\RequirePackage{fix-cm}

\documentclass[twocolumn,epjc3]{svjour3}

\smartqed  
\RequirePackage{fix-cm}

\smartqed  

\RequirePackage{graphicx}
\usepackage{xcolor}
\usepackage{subfigure}
\usepackage{amsmath,amssymb}
\usepackage{mathrsfs}

\begin{document}

\title{Shadow of novel rotating black holes in GR coupled to nonlinear electrodynamics and constraints from EHT results
}

\author{Muhammad Ali Raza\thanksref{e0,add1ab} \and Furkat Sarikulov\thanksref{e0b,addr1a,addr4a} \and Javlon Rayimbaev\thanksref{e2,addr1,addr2q,addr2r,addr2a} \and Muhammad Zubair\thanksref{e0a,add1ab} \and Bobomurat Ahmedov\thanksref{e3,addr4a,addr1,addr2} \and Zden\v{e}k Stuchl\'{i}k\thanksref{e2a,adr0}}

\thankstext{e0}{maliraza01234@gmail.com}
\thankstext{e0b}{furqatsariquloff@gmail.com}
\thankstext{e2}{javlon@astrin.uz}
\thankstext{e0a}{drmzubair@cuilahore.edu.pk}
\thankstext{e3}{ahmedov@astrin.uz}
\thankstext{e2a}{zdenek.stuchlik@physics.slu.cz}

\institute{Department of Mathematics, COMSATS University Islamabad, Lahore Campus, Lahore, Pakistan \label{add1ab}
\and
School of Mathematics and Natural Sciences, New Uzbekistan University, Mustaqillik Ave. 54, Tashkent 100007, Uzbekistan \label{addr1a}
\and
Ulugh Beg Astronomical Institute, Astronomy str. 33, Tashkent 100052, Uzbekistan \label{addr4a}
\and 
Institute of Fundamental and Applied Research, National Research University TIIAME, Kori Niyoziy 39, Tashkent 100000, Uzbekistan \label{addr1}
\and
Central Asian University, Tashkent 111221, Uzbekistan \label{addr2q} \and 
University of Tashkent for Applied Sciences, Gavhar Str. 1, Tashkent 100149, Uzbekistan \label{addr2r}
\and
Tashkent State Technical University, Tashkent 100095, Uzbekistan \label{addr2a}
\and
Institute of Theoretical Physics, National University of Uzbekistan, Tashkent 100174, Uzbekistan \label{addr2}
\and
Research Centre for Theoretical Physics and Astrophysics, Institute of Physics, Silesian University in Opava, Bezru\v covo n\' am. 13, CZ-74601 Opava, Czech Republic \label{adr0}
}

\date{Received: date / Accepted: date}

\maketitle

\begin{abstract}
We study the optical properties of spacetime around a novel regular black hole (BH) in general relativity (GR) coupled to nonlinear electrodynamics (NED), which is asymptotically flat. First, we study the angular velocity and Lyapunov exponent in unstable photon circular orbits in the novel spherically symmetric BH spacetime. Later, the rotating regular BH solution is obtained using the Newmann-Janis algorithm, and the event horizon properties of the BH are determined. We analyze the effective potential for the circular motion of photons in the spacetime of the novel rotating BH. Also, we analyze the photon sphere around the novel BH and its shadow using celestial coordinates. We obtain that an increase of the BH spin and charge as well as NED field nonlinearity parameters causes an increase in the distortion parameter of the BH shadow, while, the area of the shadow and its oblateness decrease. Moreover, we also obtain the constraint values for the BH charge and the nonlinearity parameters using Event Horizon Telescope data from shadow sizes of supermassive BHs Sgr A* and M87*. Finally, the emission rate of BH evaporation through Hawking radiation is also studied.
\end{abstract}

\maketitle

\section{Introduction}
From a theoretical point of view, astrophysical black holes (BHs) are objects that have (Arnowitt-Deser-Misner) mass and electric charge, as well as are considered in most scenarios to be rotating with spin parameters. It is also important to know the electromagnetic sources of the electric field generated by the BH charge, whether it is a linear or nonlinear electrodynamic (NED) charge. For the first time, electrically and magnetically charged BH solutions have been obtained by Reissner \cite{Reissner16} and independently by Nordstr{\"o}m \cite{Nordstrom18}. 
governed by general relativity (GR) coupled to Maxwell's electrodynamics, however, with a physical singularity. Other exact solutions for charged BHs in GR coupled to NED that avoid the singularity are called regular BH solutions \cite{Bardeen68,Ayon-Beato98,Ayon-Beato99,Ayon-Beato99a,Bronnikov01,Bambi13,Fan2016,Toshmatov18b,Toshmatov18a,Dymnikova2019a,Dymnikova2019b,Dymnikova2019c,Dymnikova2020,2023AnPhy.45469329B,Toshmatov18c,Toshmatov2019PhRvD}.

The Event Horizon Telescope (EHT) collaboration revealed in 2019 that they had captured the first photograph of a BH, a shadow image of the supermassive BH at the center of the M87 galaxy \cite{EventHorizonTelescope:2019dse}. Thanks to the information provided by this revelation about BH physics, scientists are now more interested in researching the BH shadow. In relation to BH shadow, one of the first quantitative recommendations for validating the Kerr metric with shadow analysis was done by Johannsen \& Psaltis \cite{Johannsen:2010ru}. 
Since the strong-field phenomena are the only indirect tests that can access the event horizon \cite{Falcke:1999pj}, the BH shadow plays an important role in GR \cite{Synge:1966okc}. The EHT team published the first horizon-scale image of the BH M87*. Based on the publications of EHT, a BH was rendered physically tangible, and constraints could be placed on the size of the shadow, which is the central flux depression with a factor of $\gtrsim10$ and a compact emission zone with an angular diameter of $\theta_d = 42\pm 3\mu as$ \cite{EventHorizonTelescope:2019uob,EventHorizonTelescope:2019jan,EventHorizonTelescope:2019pgp,EventHorizonTelescope:2019ggy}. 

Researchers from the EHT project later unveiled a picture of the Milky Way BH Sgr A* in 2022 based on star dynamical priors on its mass and distance \cite{EventHorizonTelescope:2022wkp,EventHorizonTelescope:2022exc,EventHorizonTelescope:2022xqj}, showing the angular shadow of diameter ($d_{sh} = 48.7\pm 7 \mu a s$). According to GR, the two BHs' recorded images, M87* and Sgr A* are consistent with the traits of a Kerr BH \cite{EventHorizonTelescope:2019dse,EventHorizonTelescope:2022wkp}. Although Kerr-like BHs arising under modified gravities are not entirely confirmed in the relative deviation of quadrupole moments and the current measurement error of spin or angular momentum, they are not entirely excluded \cite{Cardoso:2019rvt}. Furthermore, Sgr A * shows concordance with the predictions of GR in three orders of magnitude in central mass compared to the EHT results for M87* \cite{EventHorizonTelescope:2022wkp}. Therefore, one of the actual problems in astrophysics nowadays is testing modified or alternative gravity models using the data reported by the EHT Collaboration.

Recently, in Refs.\cite{2023MNRAS.523..375S} shadow of the Simpson-Vesser BHs (wormholes) was investigated and constraints on the length parameter were obtained using the image size of supermassive BHs/wormhole candidates at the center of galaxies M87, and Milky Way observed by EHT observations, and the similarity of the SV BH shadow and the shadow of the Kerr BHs was given. However, the SV wormhole (with $l>2$) with a large spin can cast a closed photon ring. Gravitational lensing and retrolensing in both weak and strong gravitational field limits, together with quasinormal spectra and gray-body factors, have been studied in Refs. \cite{2022arXiv220102971G,2022PhRvD.105h4036T,2022EPJC...82..471Z,2020AnPhy.41868181C}. Also, the strong deflection limits of the Simpson-Visser spacetime have been studied in Ref.\cite{2021PhRvD.104f4022T,2021JCAP...10..013I} and found that the photonsphere around the SV spacetime does not depend on (weakly depend on) the length parameters when $l \leq 3$. This implies that distinguishing the BH from the wormhole in the SV metric is not possible. 

Moreover, the degeneration of combined effects of electric charge of SV BH and bounce parameters was studied in Ref.\cite{2022EPJC...82..854Z} providing orbital motion around the black-bounce-Reissner-Nordstr{\"o}m BHs the same as around Schwarzschild BH. Relationships between the BH charge and bounce parameter that may break the degeneracy using the precession data from the S2 star orbits around Sgr A* detected by GRAVITY collaboration and the shadow size of Sgr A* measured by EHT. Similar tests of various BHs in gravity theories using both EHT observations have been performed in Refs. \cite{2023AnPhy.45469335R,2023CQGra..40s5003P,2023CQGra..40p5007V,2023PDU....4001178U,2023AnPhy.44869197P,2022CQGra..39b5014A,2023EPJC...83..318G}

The main focus of this work is to study the photon motion in the spacetime of the novel regular BH and its optical properties, such as the photon sphere around the BH, the BH shadow, and its distortion. Also, we obtained the constraints on the BH charge and coupling parameters using EHT data from shadow sizes of supermassive BHs SgrA* and M87*. Moreover, we study the emission rate of novel regular BH evaporation through Hawking radiation. 

The work is organized as follows: in Sect. \ref{foursbh} we give a brief explanation of the novel regular BH solution. In Section \ref{effectmetric} we analyze the effect of the interaction between the photon and NED field on the novel BH shadow using the effective metric for photon motion in the NED field. that is small enough (smaller than the error in BH shadow observations) to not consider.
Sect.\ref{AVLE} is devoted to studying the angular velocity and Lyapunov exponent in unstable photon circular orbits. In Sect.\ref{spin}, we obtain a rotating regular BH solution using the Newman-Janis algorithm (NJA). The geodesic structure around the obtained rotating BH and circular photon orbits of the BH together with the shadow cast by the BH are investigated and constraints on the BH charge and coupling parameter for the BHs are obtained using EHT observations in Sect.\ref{shdwsctn}. Moreover, in Sect.\ref{emissionrate}, we study the energy emission rate of the BH evaporation through Hawking radiation in the spacetime near the BH horizon. Finally, we summarize our findings and results in Sect.\ref{conclusion}.

Throughout this paper, we use geometrized units $c=G=1$ and $M=1$ if not mentioned.

\section{Charged Black Hole with Nonlinear Electrodynamics}
\label{foursbh}

We begin by defining the coupling of gravity through Einstein's GR with the NED model determined by the action \cite{2023AnPhy.45469329B}.
\begin{eqnarray}
S=\int dx^4\sqrt{-g}\left[\frac{R}{16 \pi G}+\mathscr{L}(F)\right], \label{action}
\end{eqnarray}
where $g$ is the determinant of the metric tensor, $R$ is Ricci scalar, $G$ is Newton's gravitational constant, and the Lagrangian defining the NED model is denoted by  $\mathscr{L}(F)$ given by \cite{2023AnPhy.45469329B}
\begin{equation}\label{Lageq}
\mathscr{L}({\cal F})=\frac{-4b {\cal F}}{\left(\sqrt{b}+\sqrt{b-2\sqrt{-\frac{{\cal F}}{2}}} \right)^2},
\end{equation}
such that $4{\cal F}=F_{\mu \nu}F^{\mu \nu}=2\left(B^2-E^2\right)$ is the Maxwell invariant and $b$ is nonlinearity parameter. If we expand the Lagrangian (\ref{Lageq}) for $b\rightarrow\infty$, we get
\begin{equation}
\mathscr{L}({\cal F})=-{\cal F}-\frac{\sqrt{-{\cal F}}{\cal F}}{\sqrt{2}b}+\frac{5{\cal F}^2}{8b^2}+\frac{7\sqrt{-{\cal F}}{\cal F}^2}{8\sqrt{2}b^3}+{\cal O}\left(\frac{1}{b^4}\right).
\end{equation}
Therefore, under the limit $b\rightarrow\infty$, we get $\mathscr{L}({\cal F})=-{\cal F}$ which is the Lagrangian for linear Maxwell electrodynamics. By considering the magnetic field $B=0$ and solving the equation $\nabla_\mu\left(F^{\mu \nu}\frac{\partial\mathscr{L}}{\partial{\cal F}}\right)=0$, the electric field $E(r)$ becomes \cite{2023AnPhy.45469329B}
\begin{equation}
E(r)=\frac{bq(2br^2+q)}{2(br^2+q)^2}, \label{EF}
\end{equation}
where $q$ is the electric charge. Note that, the electric field in Eq. (\ref{EF}) can be expressed asymptotically and at the origin, respectively as 
\begin{eqnarray}
E(r)&=&\frac{q}{r^2}-\frac{3q^2}{2br^4}+{\cal O}\left(\frac{1}{r^6}\right), \label{Einfty}\\
E(r)&=&\frac{b}{2}-\frac{b^3r^4}{2q^2}+{\cal O}\left(r^6\right). \label{Eorigin}
\end{eqnarray}
It means that the electric field $E(r)$ vanishes far from the source and unlike Maxwell's linear electrodynamics, the electric field is finite at the origin. The variation of the action (\ref{action}) and solving the Einstein's field equations for the static spherically symmetric metric
\begin{eqnarray}
ds^2=-f(r)dt^2+\frac{dr^2}{f(r)}+r^2\left(d\theta^2+\sin^2\theta d\phi^2\right) \label{GBme}
\end{eqnarray}
yields the BH solution given by
\begin{equation}
f(r)=1-\frac{2M}{r}+\frac{\pi q\sqrt{b q}}{2r}-\frac{q\sqrt{b q}}{r}\tan^{-1}{\left(\frac{\sqrt{b} r}{\sqrt{q}}\right)}, \label{meme}
\end{equation}
where $M$ is the BH mass. It is interesting to note that the asymptotic expansion of the metric function $f(r)$ can be written as
\begin{equation}
f(r)=1-\frac{2M}{r}+\frac{q^2}{r^2}+{\cal O}\left(\frac{1}{r^4}\right), \label{finf}
\end{equation}
which shows that the metric function $f(r)\rightarrow1$ as $r\rightarrow\infty$, i.e, the spacetime metric (\ref{GBme}) is asymptotically flat. Furthermore, far from the source, the BH solution mimics the Reissner-Nordstr{\" o}m BH solution. Furthermore, expanding $f(r)$ around $b=0$, we get
\begin{equation}
f(r)=1-\frac{2M}{r}+\frac{\pi q\sqrt{bq}}{2r}-bq+\frac{b^2r^2}{3}+{\cal O}\left(b^{\frac{5}{2}}\right). \label{fb0}
\end{equation}
Then, the BH metric (\ref{GBme}) reduces to the Schwarzschild metric when $b\rightarrow0$. 

\section{Effective Metric for Photons}\label{effectmetric}

In fact, there is no interaction between electromagnetic waves and electrostatic fields in linear electrodynamics. However, in NED, a photon interacts with the field due to the field's nonlinearity and does not follow null geodesics. Thus, one has to take into account considerations for photon motion spacetime of regular BHs obtained in GR coupled with NED governed by the lapse function given by Eq.~(\ref{meme}) that is determined by the "new-null" geodesics in the spacetime of the effective geometry for photons around the regular BH (introduced in \cite{Dymnikova04,Schee19ApJ,Stuchlik2019,Dima19ApJ})
\begin{eqnarray}\label{effmetric1}
\tilde{g}^{\mu \nu}&=&g^{\mu \nu}-4\frac{\mathscr{L}_{\rm FF}}{\mathscr{L}_{\rm F}}F^{\lambda}_{\mu}F^{\mu \nu} \ , \label{effmetric2} \\
\tilde{g}_{\mu \nu}&=&16\frac{\mathscr{L}_{\rm FF}F_{\mu \eta}F^{\eta}_{\nu}-(\mathscr{L}_{\rm F}+2F\mathscr{L}_{\rm FF}) g_{\mu \nu}}{F^2\mathscr{L}^2_{\rm FF}-16(\mathscr{L}_{\rm F}+F\mathscr{L}_{\rm FF})^2}\ ,
\end{eqnarray}

where

\begin{equation}
\mathscr{L}_{\rm F}=\frac{\partial \mathscr{L}(F)}{\partial F}\ , \quad  \mathscr{L}_{\rm FF}=\frac{\partial^2 \mathscr{L}(F)}{\partial F^2} \ .
\end{equation}

The eikonal equation for photons in the effective geometry related to the regular BH can be written as

\begin{equation}
\tilde{g}_{\mu \nu} k^{\mu} k^{\nu}=0,
\end{equation}
where $k^{\mu}$ is the four-wave vector related to the four-momentum of photons by $p^{\mu}=\hbar k^{\mu}$ (in Gaussian units, $\hbar = 1$).
\begin{figure*}[h!]
\centering
\includegraphics[width=0.4\textwidth]{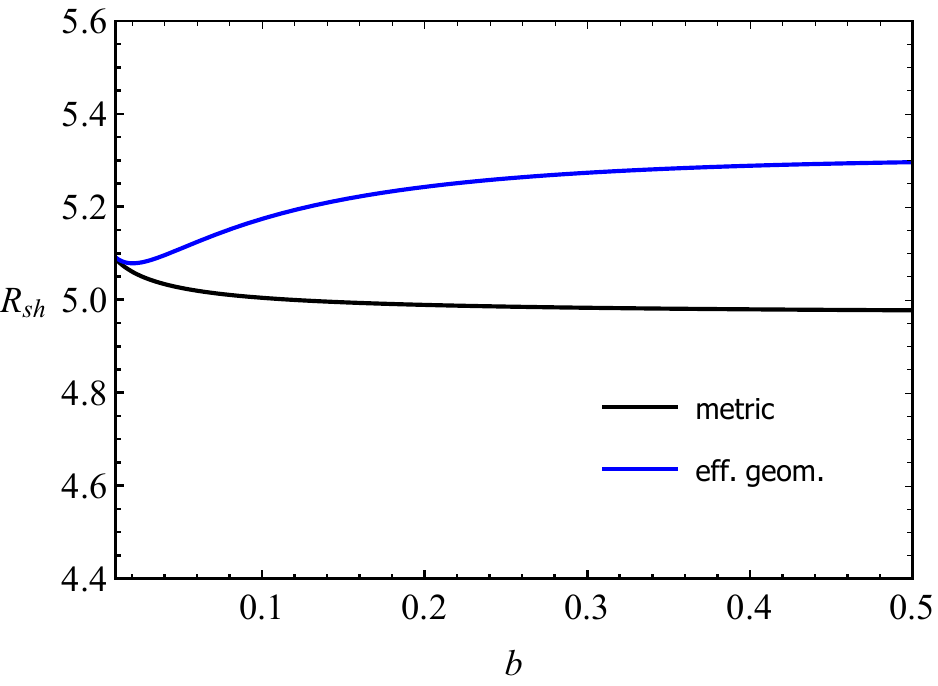}
\includegraphics[width=0.4\textwidth]{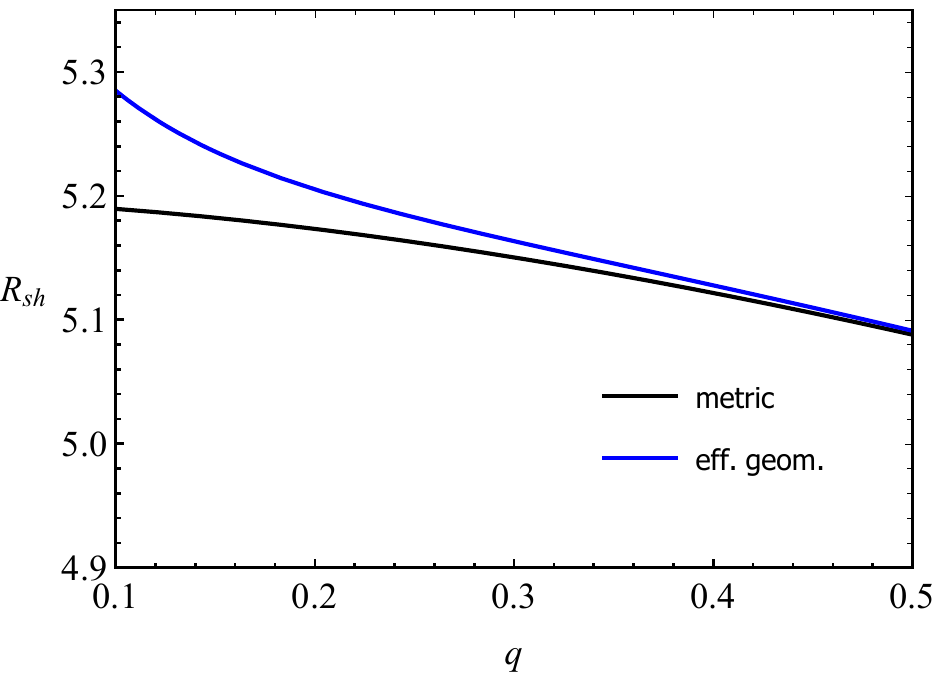}
\caption{Comparison of the shadow radii of the BH while considering (blue) and not considering (black) the nonlinear electrodynamics (NED) effect on photon motion, with $M=1$.\label{effgeom}}
\end{figure*}

Our graphical and numerical analyses have shown that the difference in shadow sizes obtained with and without taking into account the interaction between the photon and the NED field (see Fig. \ref{effgeom}, it is less than 6\%) is smaller than the error in the data of EHT observations of Sgr A* ($\sim 14\%$) and M87* (about 7.14\%). Moreover, the influence of NED on photon motion weakens due to the appearance of the spin parameter when we obtain the rotating BH for this spacetime metric. Since in this work, we mainly focus on obtaining constraints on spacetime parameters using EHT data, we will work with the metric (\ref{GBme}) for further analysis.

\section{Angular Velocity and Lyapunov Exponent}\label{AVLE}
Now, we study the quasinormal modes in terms of angular velocity $\Omega$ and Lyapunov exponent $\Gamma$ for static BH (\ref{GBme}). The angular velocity and Lyapunov exponent are respectively the real and imaginary parts of the eikonal quasinormal frequencies $\omega_n$ in dimensions $D\geq4$ and can be written as \cite{KONOPLYA2017597}
\begin{equation}
\omega_n=\Omega l_c-i\left(n+\frac{1}{2}\right)|\Gamma|,
\end{equation}
where $n$ and $l_c$ are overtones and multipole numbers, respectively. The parameters $\Omega$ and $\Gamma$ depend on the unstable circular photon trajectories in the vicinity of a static spherically symmetric asymptotically flat BH \cite{PhysRevD.79.064016,RevModPhys.83.793,Berti_2009,1999LRR.....2....2K} that are emitted by the BH in the eikonal part of its spectrum. The angular velocity and Lyapunov exponent are useful in studying the thermodynamics of the BH since these parameters allow us to construct the relation between the phase transition and quasinormal modes \cite{2023EPJC...83..407A}. The relation for angular velocity can be written as
\begin{equation}
\Omega=\frac{\dot{\phi}}{\dot{t}}=\frac{\sqrt{f(r_{ph})}}{r_{ph}}
\end{equation}
and the Lyapunov exponent is given as
\begin{equation}
\Gamma=\sqrt{-\frac{1}{2\dot{t}^2}\frac{\partial^2V_{eff}}{\partial r^2}}\bigg|_{r=r_{ph}},
\end{equation}
where $r_{ph}$ is the radius of the photon sphere determined by the roots of
\begin{equation}
\frac{d}{dr}\left(\frac{r^2}{f(r)}\right)\bigg|_{r=r_{ph}}=0
\end{equation}
and $V_{eff}$ is the effective potential in the equatorial plane $\left(\theta={\pi}/{2}\right)$ given by
\begin{equation}
V_{eff}=\frac{L^2f(r)}{r^2}-E^2,
\end{equation}
such that $L$ and $E$ are angular momentum along $z$-axis and energy of the photon, respectively.

\begin{figure*}[t]
\centering
\subfigure{\includegraphics[width=0.4\textwidth]{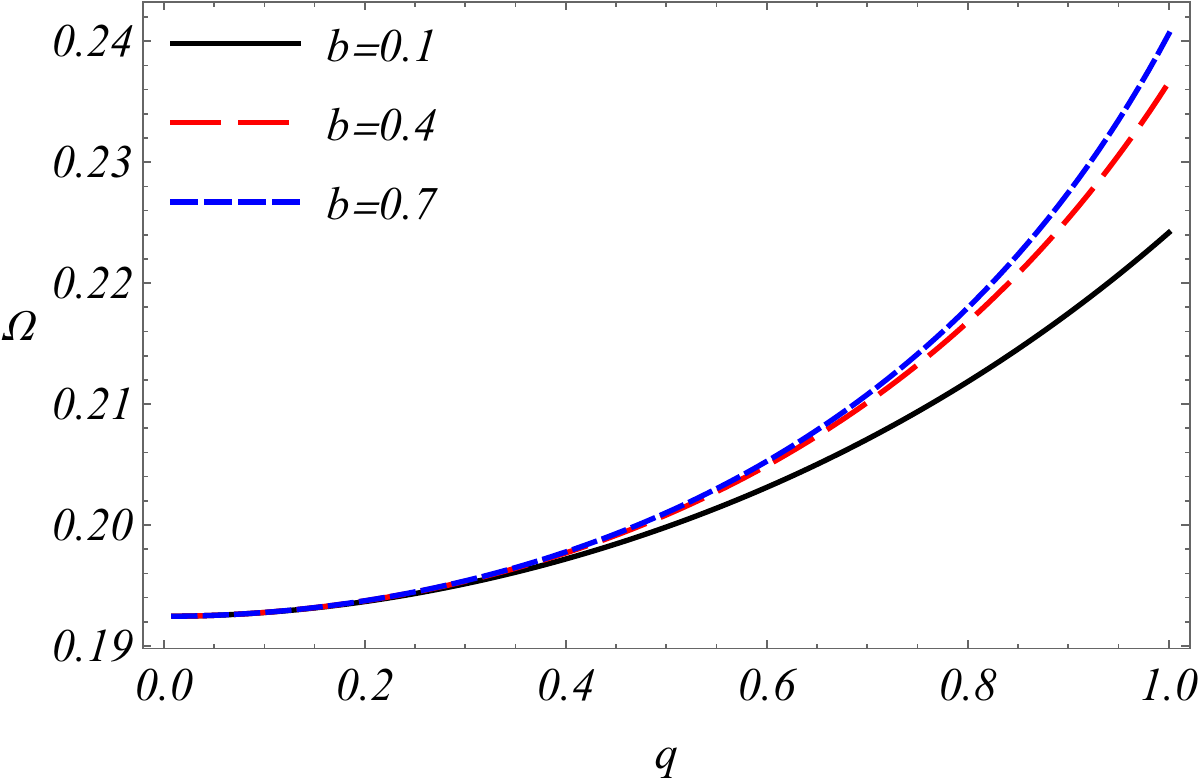}}~~~~~
\subfigure{\includegraphics[width=0.4\textwidth]{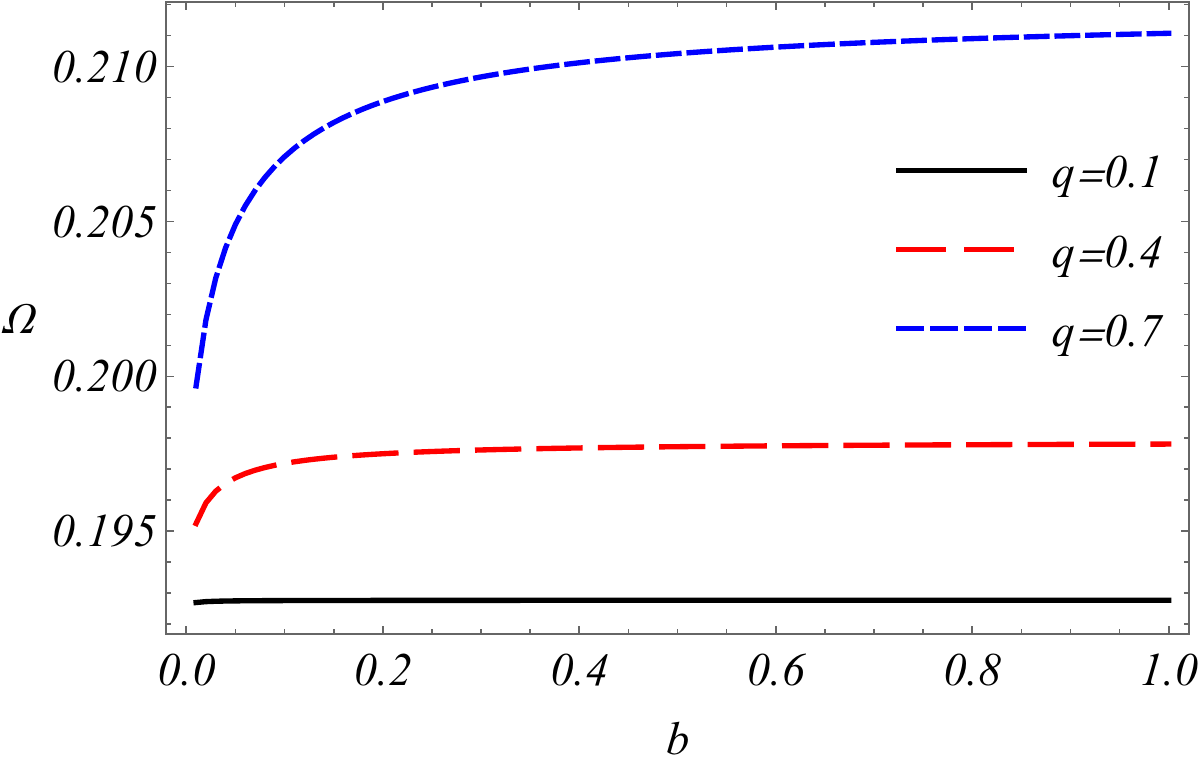}}\\
\subfigure{\includegraphics[width=0.4\textwidth]{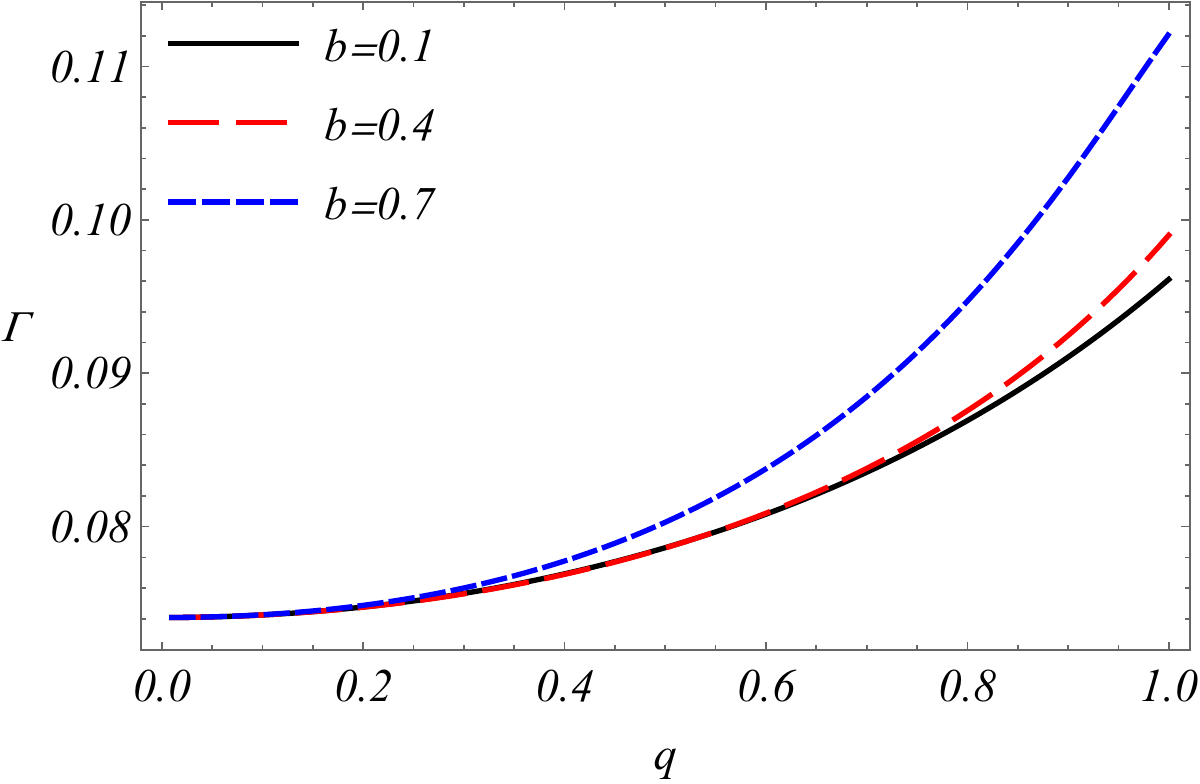}}~~~~~
\subfigure{\includegraphics[width=0.4\textwidth]{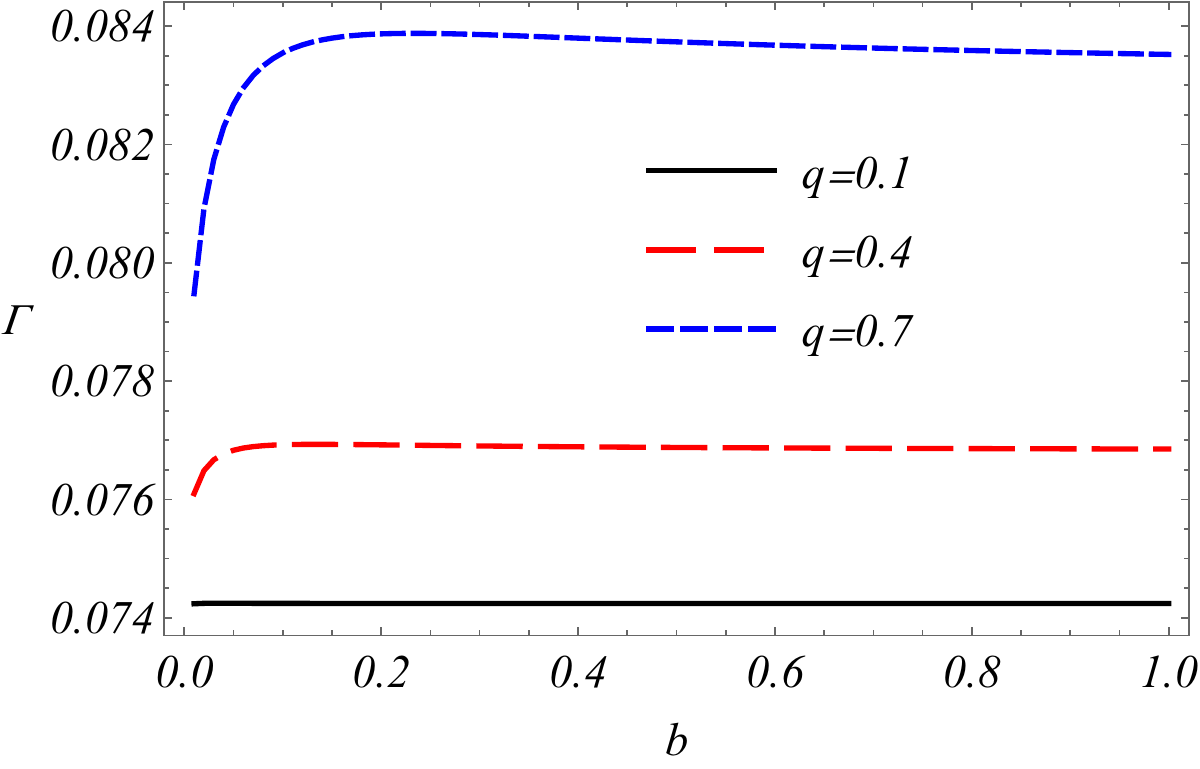}}
\caption{The behavior of angular velocity and Lyapunov exponent for different values of $b$ and $q$.} \label{QNM}
\end{figure*}

We have plotted the behavior of angular velocity $\Omega$ and Lyapunov exponent $\Gamma$ for the different values of $b$ and $q$ in Fig. \ref{QNM}. The upper panel shows the plots for angular velocity, and the lower panel corresponds to the plots for the Lyapunov exponent. In the upper left plot, the curves correspond to the different values of $b$, and the behavior of $\Omega$ is observed with respect to $q$. It shows that the angular velocity increases as $q$ increases for each curve. However, as the value of $b$ increases, the angular velocity increases more rapidly with increasing $q$. In the upper right plot, each curve corresponds to a different value of $q$, and the behavior of angular velocity is represented as a function of $b$. For a small value of $q$, the angular velocity is constant with respect to $b$. When the value of $q$ is increased, the angular velocity increases rapidly up to a small value of $b$ and then increases at such a small rate that it approaches nearly a constant value. For even a larger value of $q$, the angular velocity increases at a higher rate up to a certain value of $b$ and then increases slowly with respect to $b$. The lower left graph shows the behavior of the Lyapunov exponent with respect to $q$ for different values of $q$ corresponding to each curve. It can be seen that the behavior of the Lyapunov exponent is the same as that for the angular velocity, i.e., the value of the Lyapunov exponent increases with an increase in the value of $q$ and the increasing rate of the Lyapunov exponent also rises as the value of $b$ becomes larger. From the right plot, the Lyapunov exponent is constant with respect to $b$ for a small value of $q$. For a larger value of $q$, the Lyapunov exponent increases to a small value of $b$ and then becomes constant. When the value of $q$ increases further, the Lyapunov exponent increases rapidly to a certain value of $b$ and then decreases at a slower rate after reaching a maximum value.

\section{Rotating Black Hole}\label{spin}
A rotating BH metric is one of the simplest generalizations of a static BH with an additional spin parameter usually denoted by $a$. The behavior and properties of the spacetime structure around a rotating BH are different from its static counterpart, especially in terms of photon motion. The significance of a rotating BH can be estimated by conducting an analysis for the comparison of the BH shadow with the EHT data for supermassive BHs. Since the supermassive BHs are rotating, therefore, for a viable and meticulous comparison of the shadows, it is better to work with rotating BHs. The NJA \cite{1965JMP.....6..915N,1965JMP.....6..918N} was designed to develop the rotating counterparts of the static BH metrics within GR. The Kerr and Kerr-Newman metrics are the earliest examples of the application of this algorithm to Schwarzschild and Reissner-Nordstr\"{o}m metrics, respectively. It is well known that Schwarzschild and Kerr BHs are vacuum solutions, while the Reissner-Nordstr{\" o}m and Kerr-Newman BHs are sourced by electric charge. However, Hansen and Yunes \cite{PhysRevD.88.104020} found that some additional unknown sources arise when the NJA is applied to static BHs in non-GR gravity theories. Recently, the NJA was modified by Azreg-A\"{ı}nou \cite{PhysRevD.90.064041,2014EPJC...74.2865A} to obtain the rotating counterparts without the complexification of radial coordinates. Hence, this algorithm can easily be applied to BH metrics within GR and non-GR theories. It has successfully generated the rotating BH metrics for imperfect fluids and generic rotating regular BH metrics \cite{galaxies9020043,2022EPJC...82..547W} within GR. Therefore, we also apply this modified NJA to obtain the rotating counterpart of the static metric (\ref{GBme}). First, we introduce the Eddington-Finkelstein coordinates ($u,r,\theta,\phi$) and using the transformation
\begin{equation}
du=dt-\frac{dr}{f(r)},
\end{equation}
the static metric (\ref{GBme}) becomes
\begin{equation}
ds^2=-f(r)du^2-2dudr+r^2d\theta^2+r^2\sin^2\theta d\phi^2. \label{sur}
\end{equation}
Further, the conjugate metric tensor can be expressed as
\begin{equation}
g^{ab}=-l^an^b-n^al^b+m^a\bar{m}^b+\bar{m}^am^b
\end{equation}
with the null tetrads given by
\begin{eqnarray}
l^a&=&\delta^a_r,\\
n^a&=&\delta^a_u-\frac{f(r)}{2}\delta^a_r,\\
m^a&=&\frac{1}{\sqrt{2}r}\left(\delta^a_\theta+\frac{i}{\sin\theta}\delta^a_\phi\right),
\end{eqnarray}
where $\bar{m}^a$ is the complex conjugate of $m^a$. It is easy to find that for these null tetrads, the following relations hold:
\begin{eqnarray}
l_al^a=n_an^a=m_am^a=\bar{m}_a\bar{m}^a&=&0,\\
l_am^a=l_a\bar{m}^a=n_am^a=n_a\bar{m}^a&=&0,\\
-l_an^a=-l^an_a=m_a\bar{m}^a=m^a\bar{m}_a&=&1.
\end{eqnarray}
Now, we perform the complex coordinate transformations in ($u,r$)-plane as

\begin{eqnarray}
u'\rightarrow u-ia\cos\theta, \qquad r'\rightarrow r+ia\cos\theta,\label{rp}
\end{eqnarray}
with $a$ being a spin parameter of the BH. The Next step in NJA is complexifying the radial coordinate $r$. However, it is not necessary as shown by \cite{PhysRevD.90.064041}. The complexification process can be avoided by considering that $\delta^\mu_\nu$ transforms as a vector under (\ref{rp}). At the same time, the metric functions of the metric (\ref{sur}) transform to new undetermined functions, that is,
\begin{eqnarray}
f(r)\rightarrow F(r,a,\theta), \qquad r^2\rightarrow H(r,a,\theta),
\end{eqnarray}
such that
\begin{eqnarray}
\lim\limits_{a\rightarrow0}F(r,a,\theta)=f(r), \qquad \lim\limits_{a\rightarrow0}H(r,a,\theta)=r^2, \label{cond}
\end{eqnarray}
Using this transformation, the null tetrads become
\begin{eqnarray}
l^a&=&\delta^a_r,\\
n^a&=&\delta^a_u-\frac{F}{2}\delta^a_r,\\
m^a&=&\frac{1}{\sqrt{2H}}\left((\delta^a_u-\delta^a_r)ia\sin\theta+\delta^a_\theta
+\frac{i}{\sin\theta}\delta^a_\phi\right).
\end{eqnarray}
Making use of the new null tetrads, the rotating metric in the Eddington-Finkelstein coordinates is given by
\begin{eqnarray}
ds^2&=&-Fdu^2-2dudr+2a\sin^2\theta(F-1)dud\phi \nonumber\\
&&+2a\sin^2\theta drd\phi+Hd\theta^2 \nonumber\\
&&+\sin^2\theta\left(H+a^2\sin^2\theta(2-F)\right)d\phi^2.
\end{eqnarray}
Bringing these coordinates back to the Boyer-Lindquist coordinates, we obtain the rotating counterpart for the static BH metric (\ref{GBme}). Therefore, we introduce a global coordinate transformations
\begin{eqnarray}
du&=&dt+\lambda(r)dr,\\
d\phi&=&d\phi'+\chi(r)dr,
\end{eqnarray}
with
\begin{eqnarray}
\lambda(r)&=&-\frac{a^2+r^2}{a^2+r^2f(r)},\\
\chi(r)&=&-\frac{a}{a^2+r^2f(r)}.
\end{eqnarray}
Finally, we choose
\begin{eqnarray}
F&=&\frac{\left(r^2f(r)+a^2\cos^2\theta\right)}{H},\\
H&=&r^2+a^2\cos^2\theta
\end{eqnarray}
and the Kerr-like rotating BH metric reads
\begin{eqnarray}
ds^2&=&-\frac{\Delta(r)-a^2\sin^2\theta}{\rho^2}dt^2+\frac{\rho^2}{\Delta(r)}dr^2+\rho^2d\theta^2 \nonumber\\
&&+\frac{\sin^2\theta}{\rho^2}\left(\left(r^2+a^2\right)^2-\Delta(r)a^2\sin^2\theta\right)d\phi^2 \nonumber\\
&&-\frac{2a\sin^2\theta}{\rho^2}\left(a^2+r^2-\Delta(r)\right)dtd\phi, \label{Romet}
\end{eqnarray}
with the metric functions given as
\begin{eqnarray}
\rho^2&=&r^2+a^2\cos^2\theta, \label{rho}\\
\Delta(r)&=&a^2+r^2f(r)=r^2-2Mr+a^2+\frac{\pi rq\sqrt{bq}}{2} \nonumber\\
&&-rq\sqrt{bq}\tan^{-1}{\left(\frac{\sqrt{b}r}{\sqrt{q}}\right)}. \label{DDm}
\end{eqnarray}

The metric (\ref{Romet}) with metric functions (\ref{rho}) and (\ref{DDm}) describe the novel charged rotating BH in NED. It is obvious that under the limit $a\rightarrow0$, the metric (\ref{Romet}) reduces to the static metric (\ref{GBme}), and the conditions (\ref{cond}) are also satisfied. Note that the asymptotic expansion of the metric function $\Delta(r)$ can be written as
\begin{eqnarray}
\Delta(r)&=&r^2-2Mr+a^2+q^2+{\cal O}\left(\frac{1}{r^2}\right),
\end{eqnarray}
which shows that far from the source, the rotating BH solution (\ref{Romet}) mimics the Kerr-Newman BH solution. Moreover, expanding $\Delta(r)$ around $b=0$, we get
\begin{eqnarray}
\Delta(r)&=&r^2-2Mr+a^2+\frac{\pi q^{\frac{3}{2}}\sqrt{b}r}{2}-bqr^2 \nonumber\\
&&+{\cal O}\left(b^{\frac{3}{2}}\right).
\end{eqnarray}
Therefore, under the limit $b\rightarrow0$, The rotating metric (\ref{Romet}) reduces to the Kerr metric.

\begin{figure*}[h!]
\begin{center}
\subfigure{\includegraphics[width=0.324\textwidth]{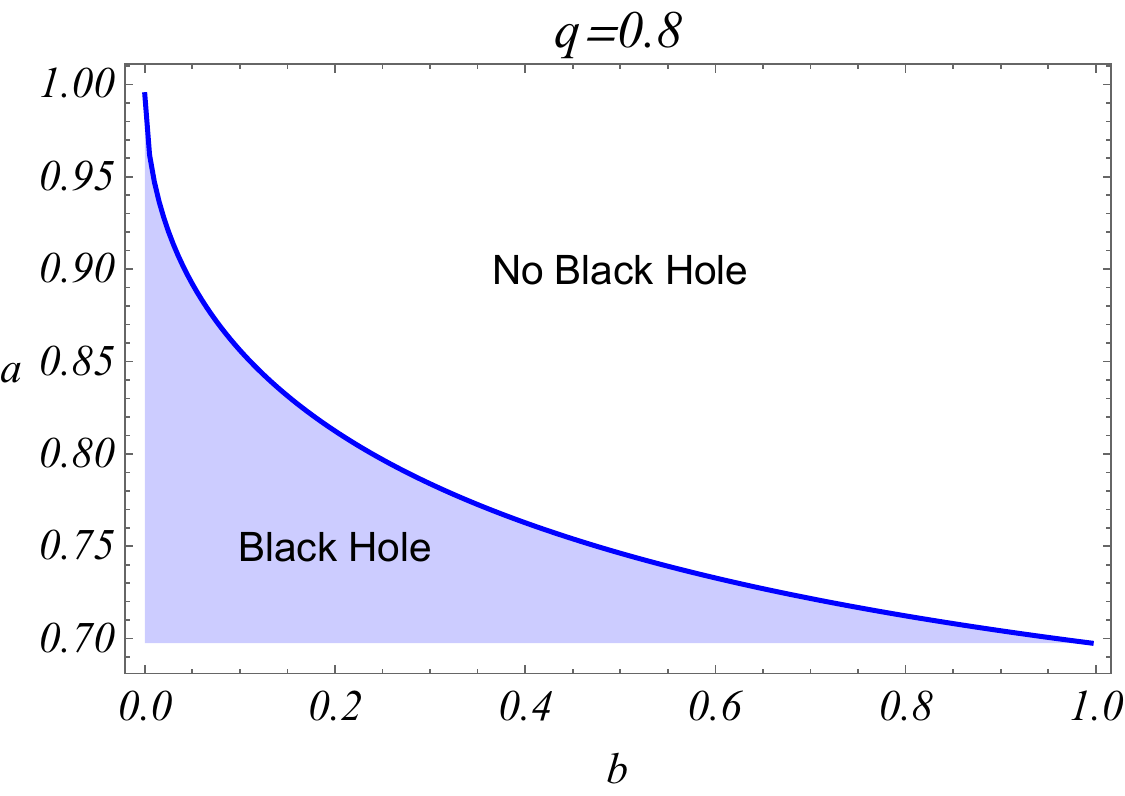}}~~
\subfigure{\includegraphics[width=0.324\textwidth]{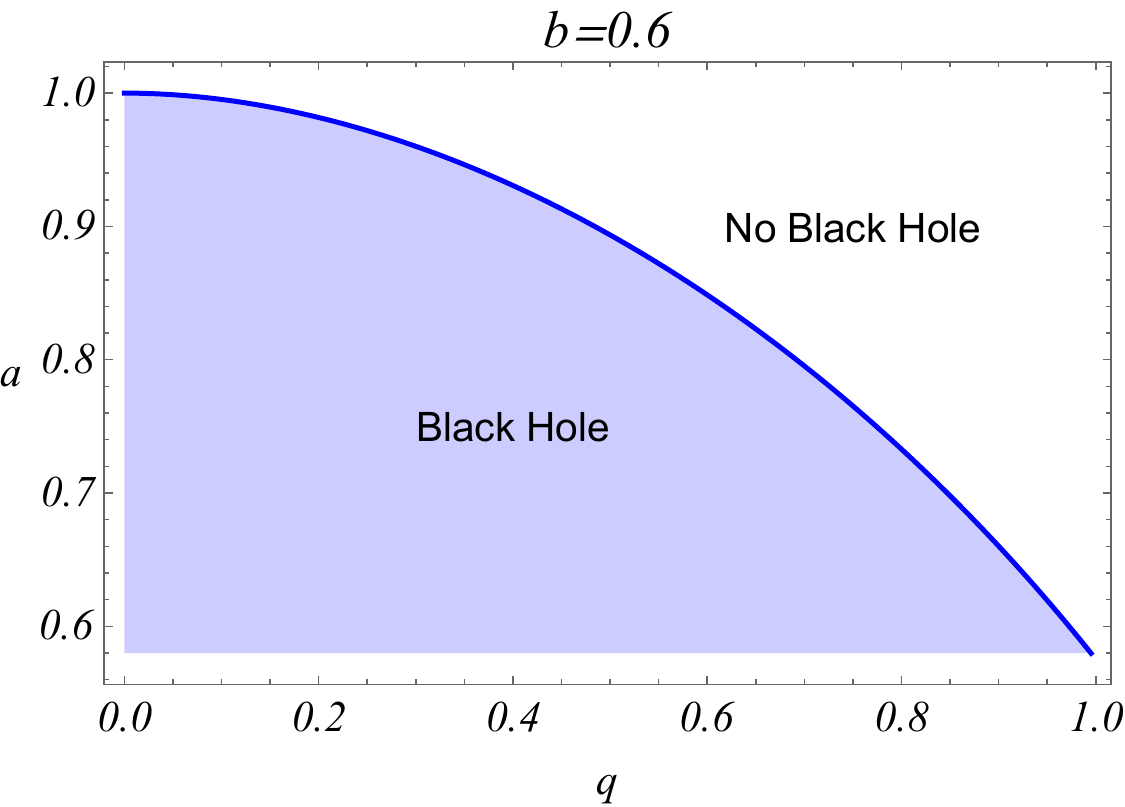}}
\subfigure{\includegraphics[width=0.324\textwidth]{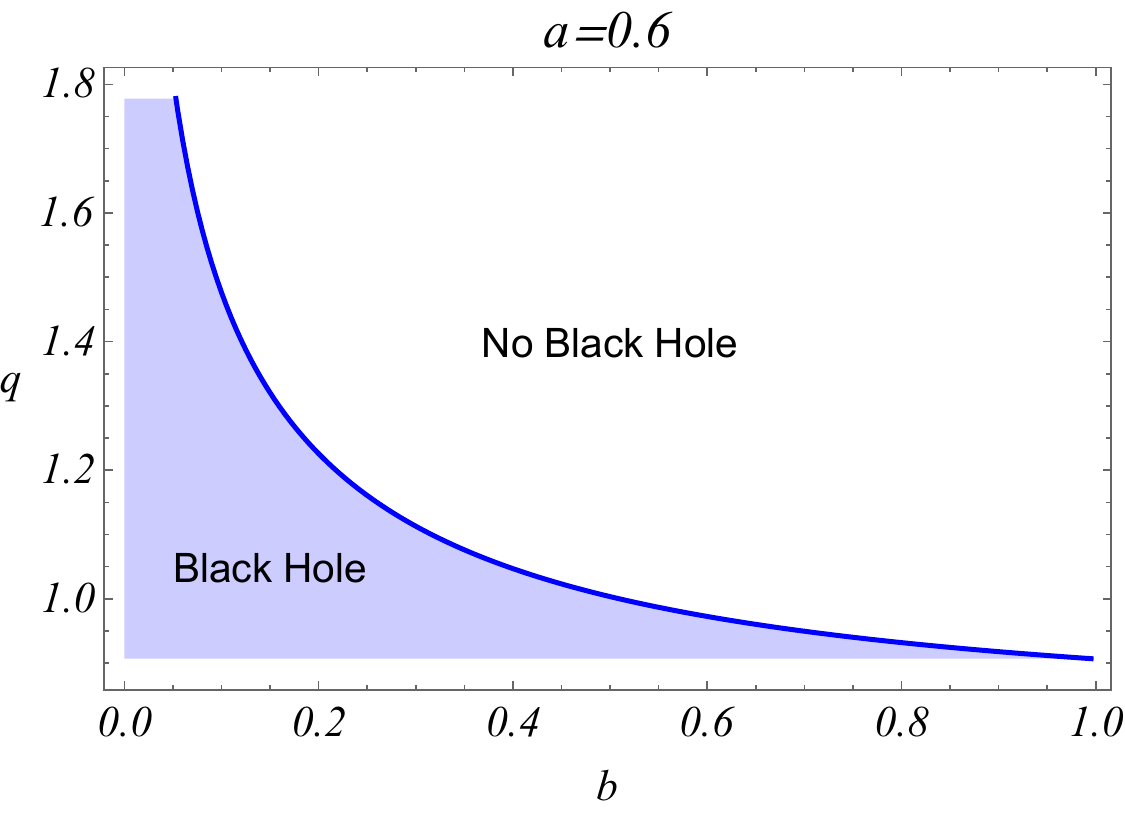}}
\end{center}
\caption{Plots showing the limits of BH parameters for which event horizon exists. The left panel is for $q=0.8$, the middle $b=0.6$ and the right panel is for $a=0.6$ cases. \label{BNB}}
\end{figure*}

In Fig. \ref{BNB}, we have plotted the limits of BH parameters as parametric spaces for which the event horizon exits. In particular, these parametric spaces are $a$-$b$, $a$-$q$, and $b$-$q$ spaces for which the corresponding third parameter has been kept fixed. The highlighted curves correspond to the extremal BHs. It shows that for a fixed $q$, when $b$ is increased, the value of spin $a$ decreases drastically and gradually slows down as $ b\rightarrow l$. The same behavior is observed when $a$ is fixed and the event horizon is plotted for space $b$-$q$. However, for a fixed $b$, $a$ decreases slowly as $q$ increases and further the decrement in $a$ becomes faster as $q\rightarrow1$.

\begin{figure*}[ht!]
\begin{center}
\subfigure{\includegraphics[width=0.37\textwidth]{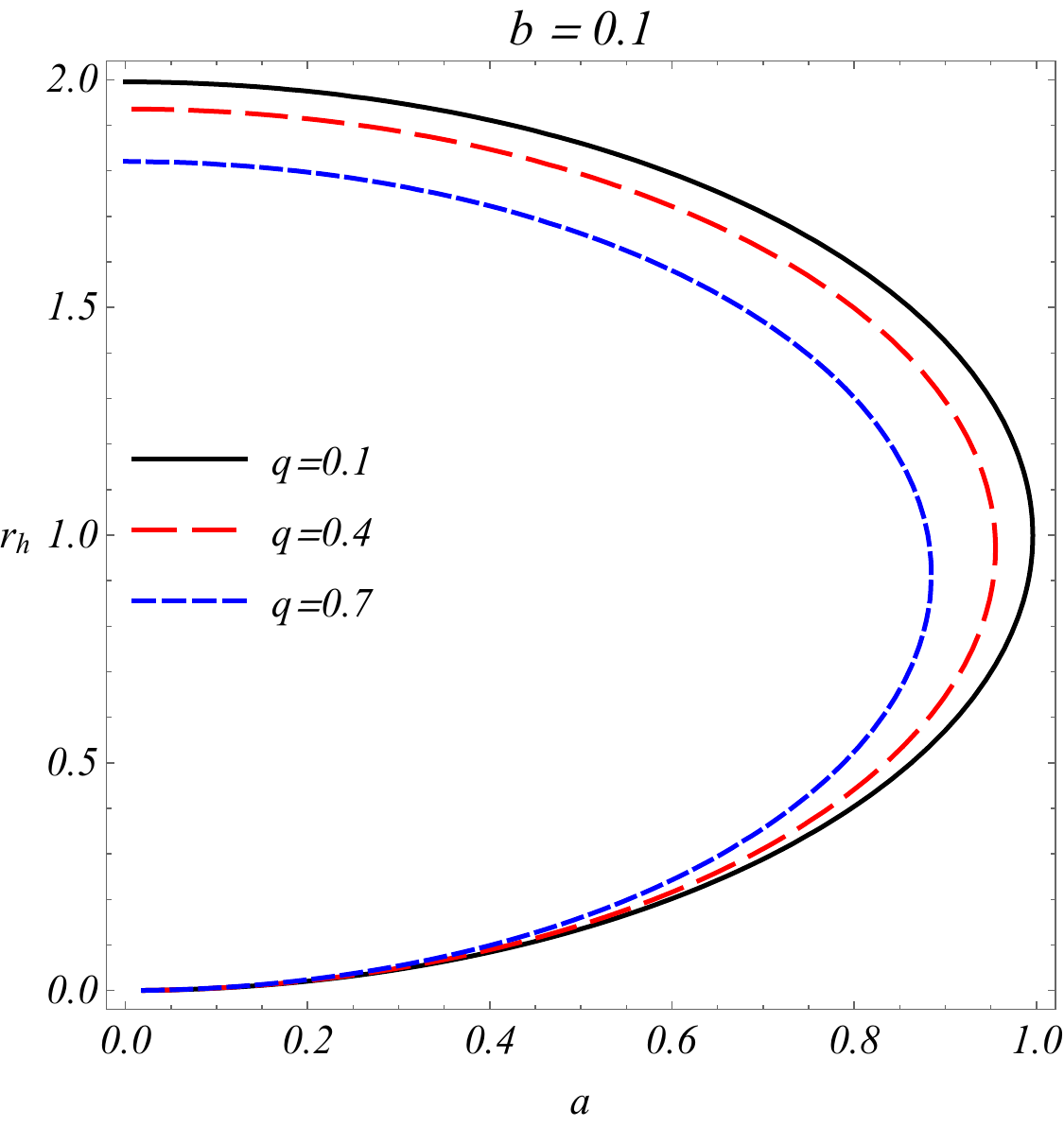}}~~~~~~~
\subfigure{\includegraphics[width=0.37\textwidth]{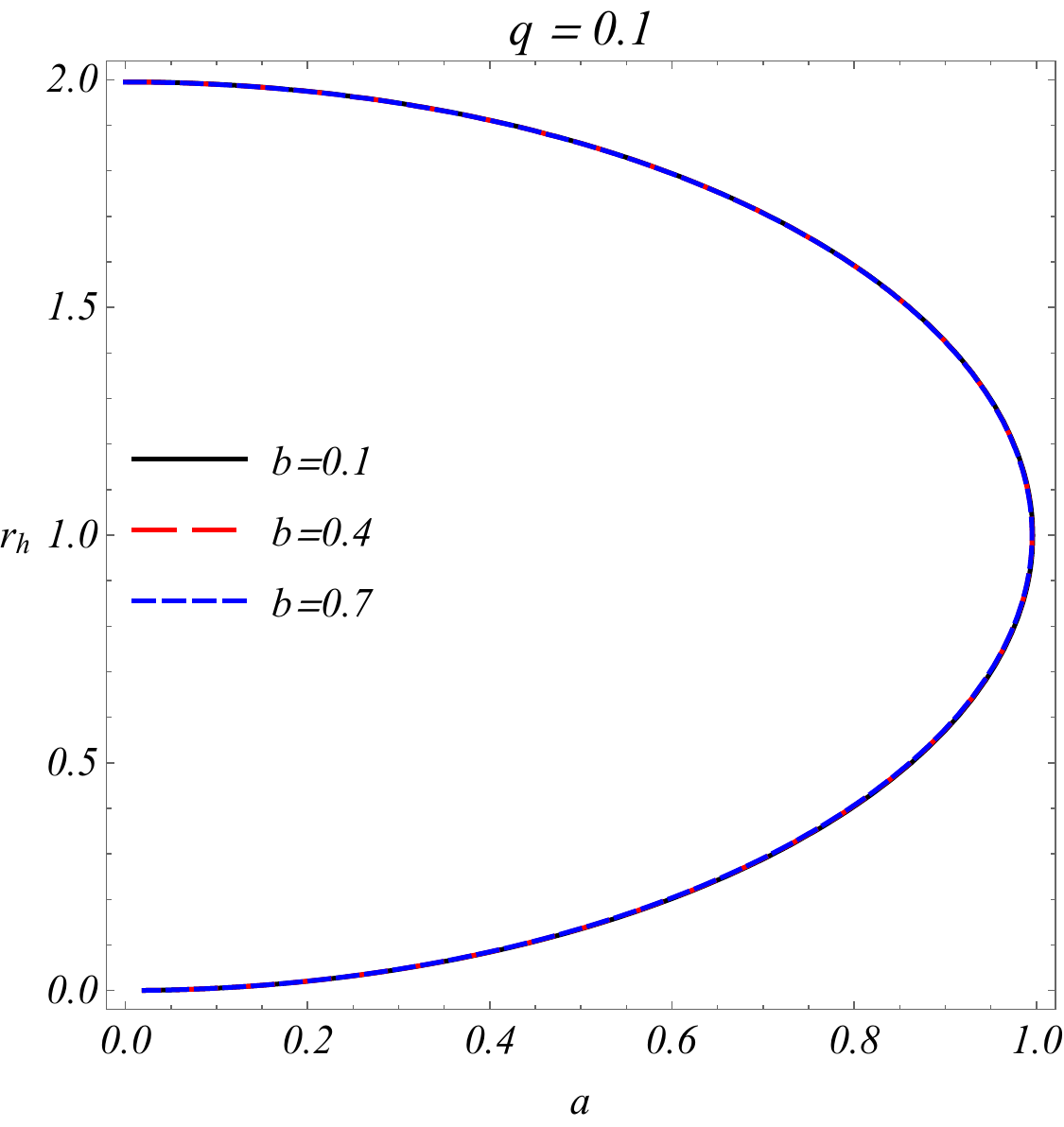}}\\
\subfigure{\includegraphics[width=0.37\textwidth]{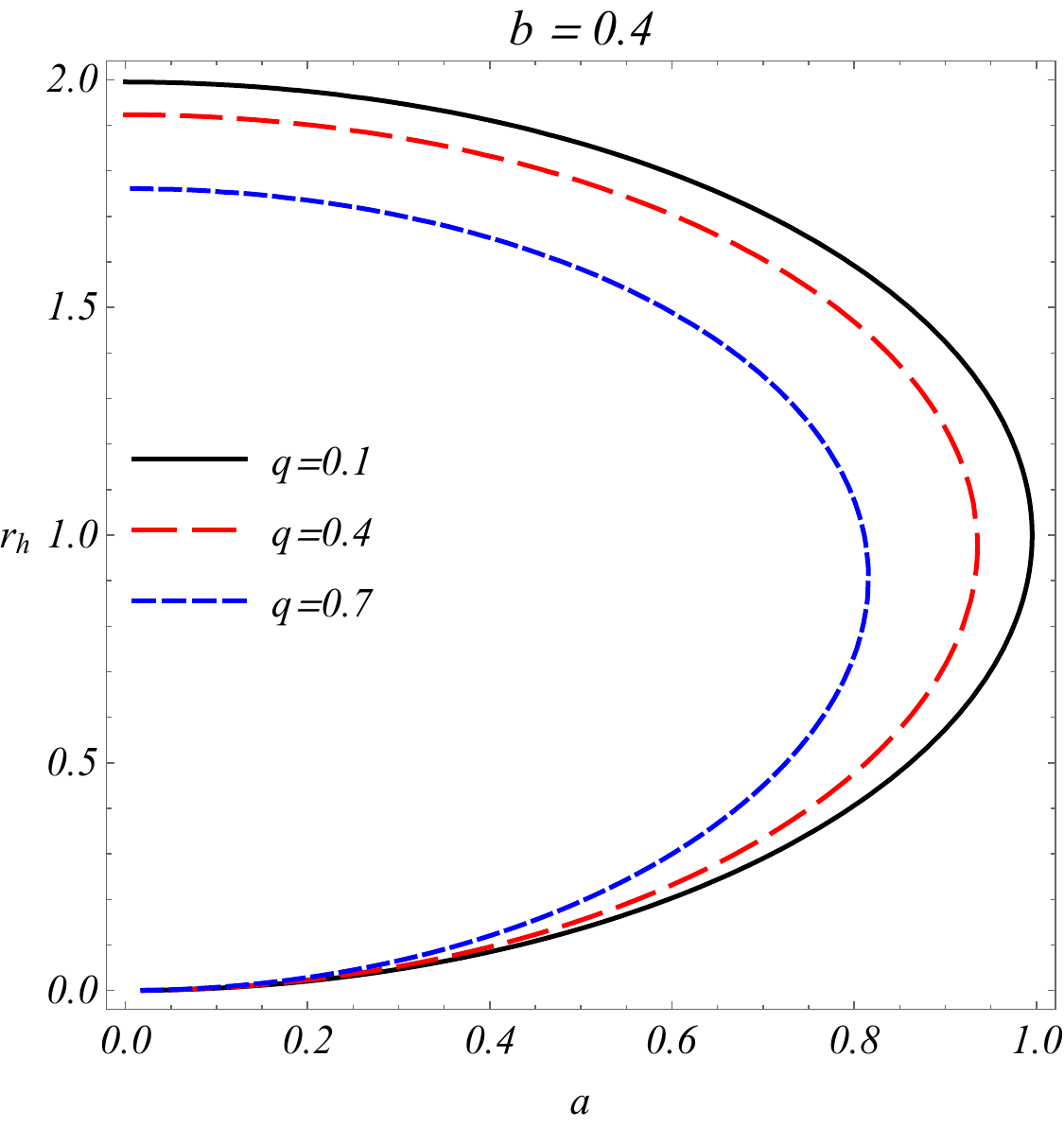}}~~~~~~~
\subfigure{\includegraphics[width=0.37\textwidth]{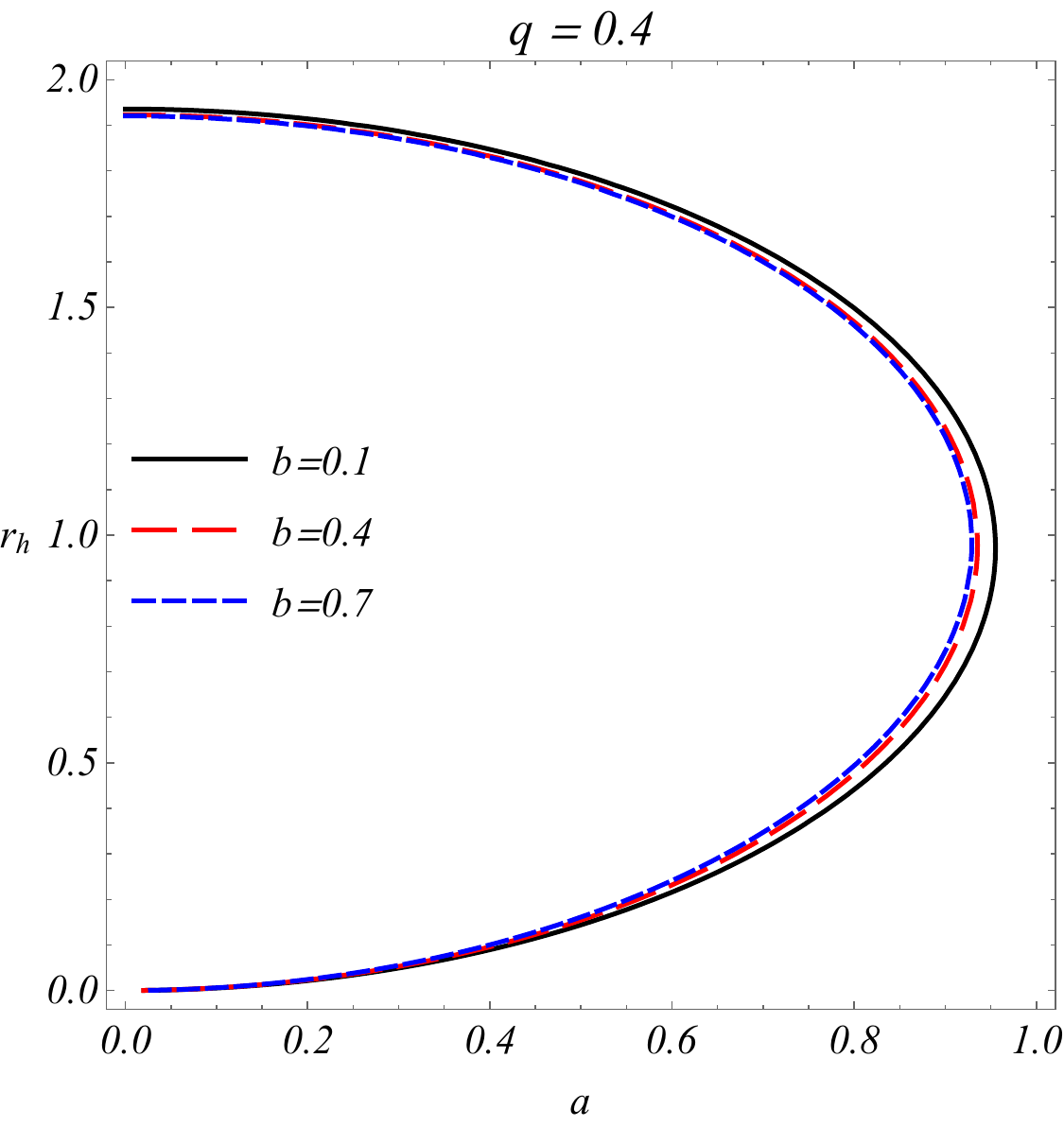}}\\
\subfigure{\includegraphics[width=0.37\textwidth]{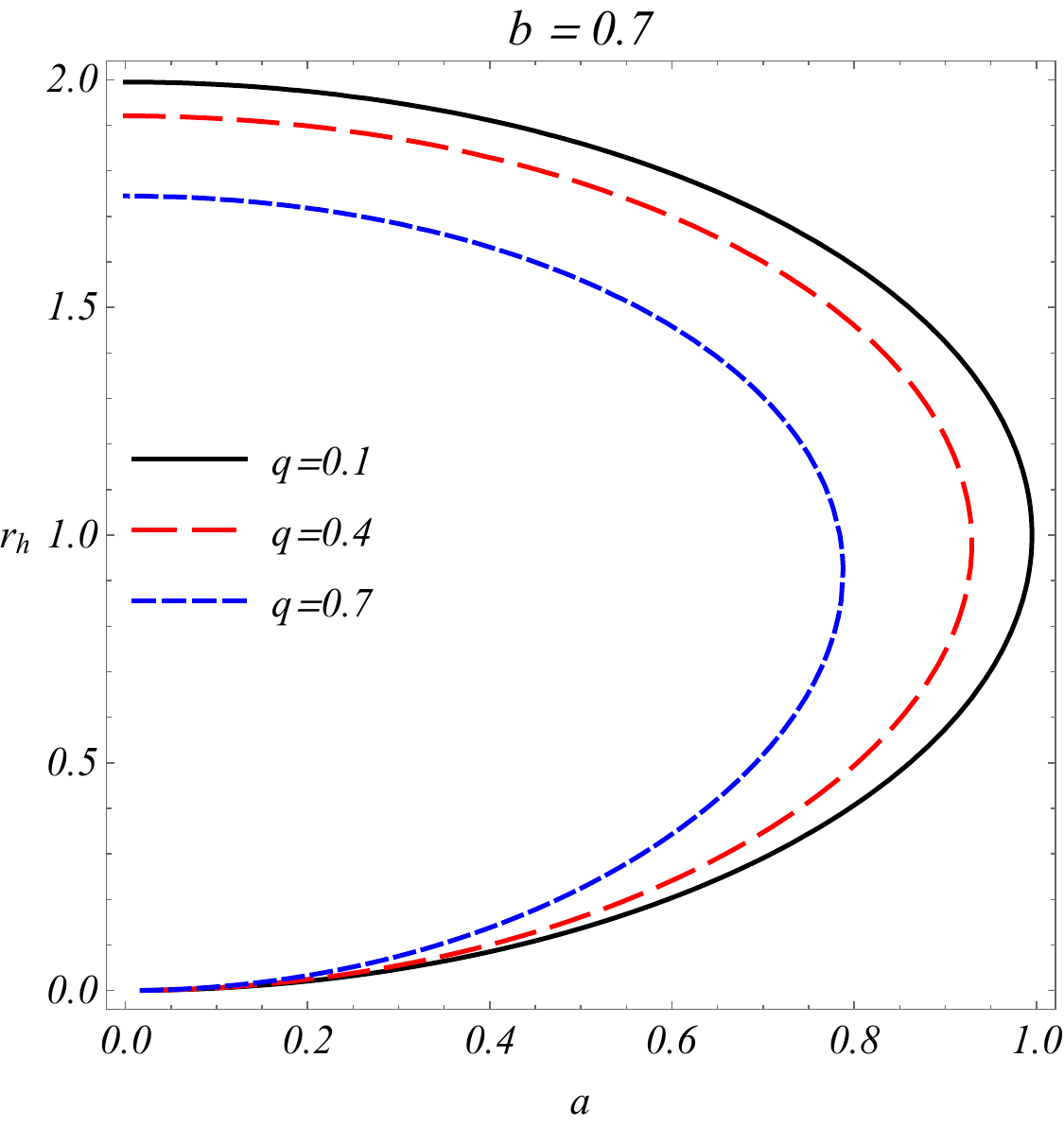}}~~~~~~~
\subfigure{\includegraphics[width=0.37\textwidth]{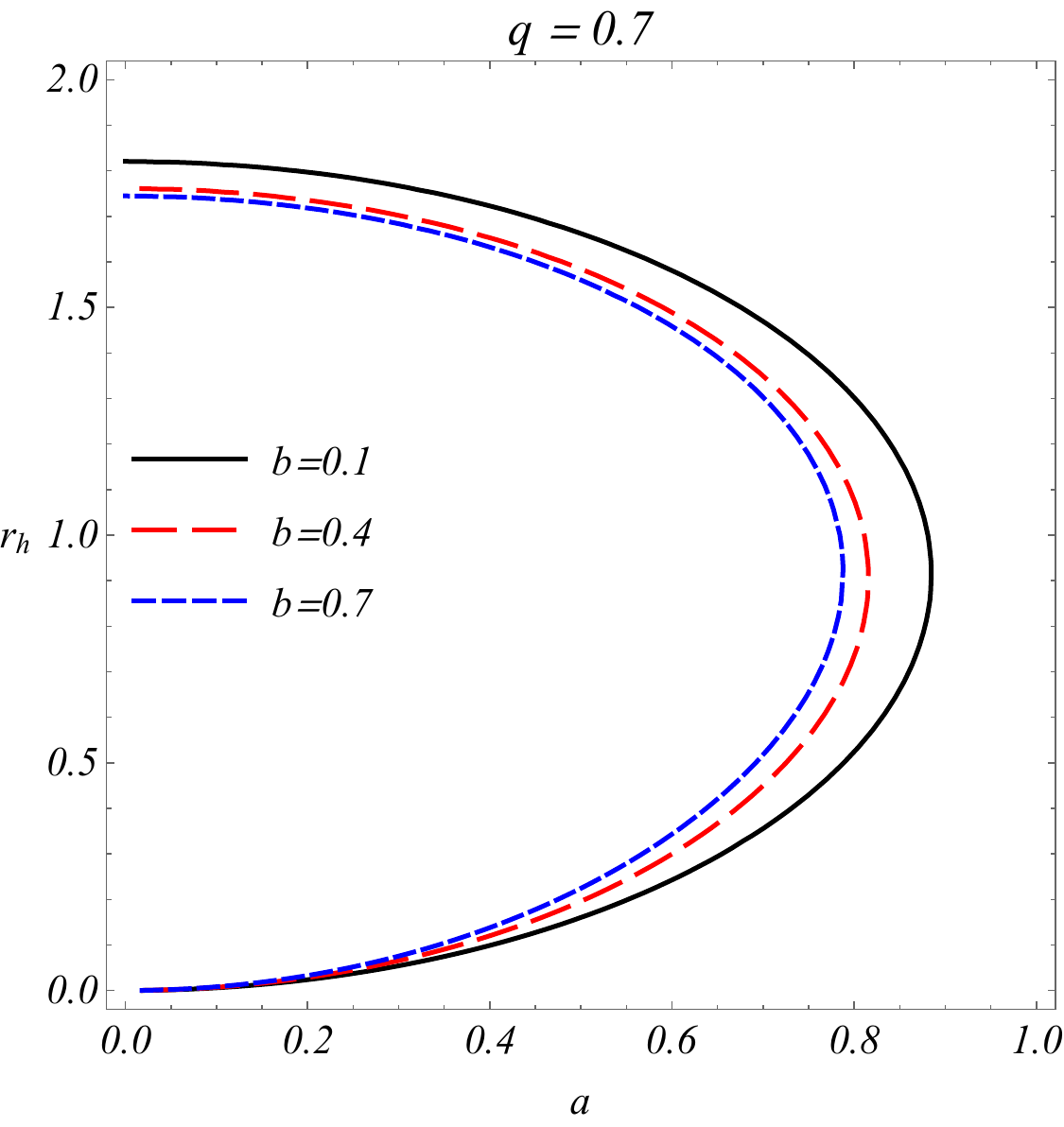}}
\end{center}
\caption{Plots showing the effects of $a$, $b$ and $q$ on horizon radius $r_h$ for the charged rotating BH in NED. The left column is for fixed values of $b=0.1$, $b=0.4$ and $b=0.4$. The right column is for fixed values of $q=0.1$, $q=0.4$, and $1=0.7$.} \label{HR}
\end{figure*}

Next, we study the horizon structure of the charged rotating BH in NED defined by the metric (\ref{Romet}). We employ a numerical technique to solve the equation $g^{rr}=0$ for its roots to obtain the radius of the horizon. We plotted the horizon radius $r_h$ with respect to $a$ for various values of $b$ and $q$ in Fig. \ref{HR}. For all curves, it is quite usual to observe that the event horizon decreases and the Cauchy horizon increases with an increase in $a$. For all cases, the extremal value of $a$ decreases. Moreover, when $q$ increases, the event horizon decreases for all values of $b$. Conversely, when $b$ increases, the event horizon does not change significantly for small values of $q$. However, for large values of $q$, the event horizon decreases with an increase in $b$.

\section{Photon motion and shadow of the BH}\label{shdwsctn}
\begin{figure*}[t]
\begin{center}
\subfigure{
\includegraphics[width=0.42\textwidth]{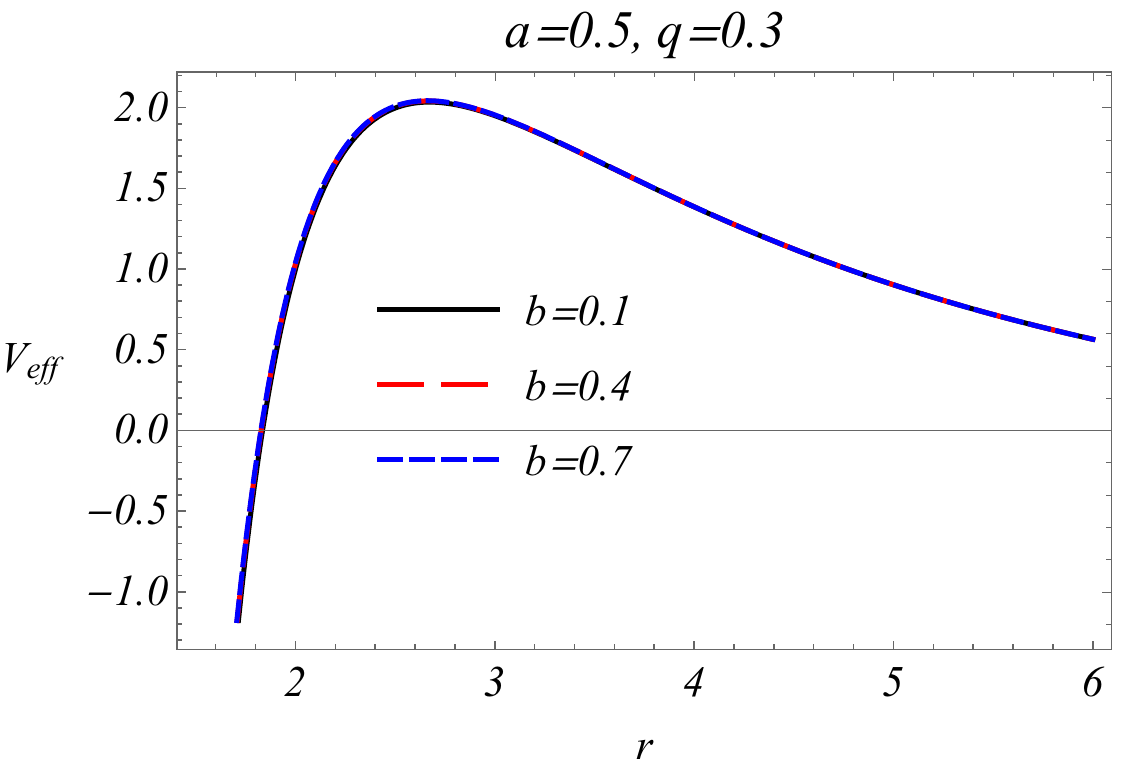}}~~~~
\subfigure{
\includegraphics[width=0.42\textwidth]{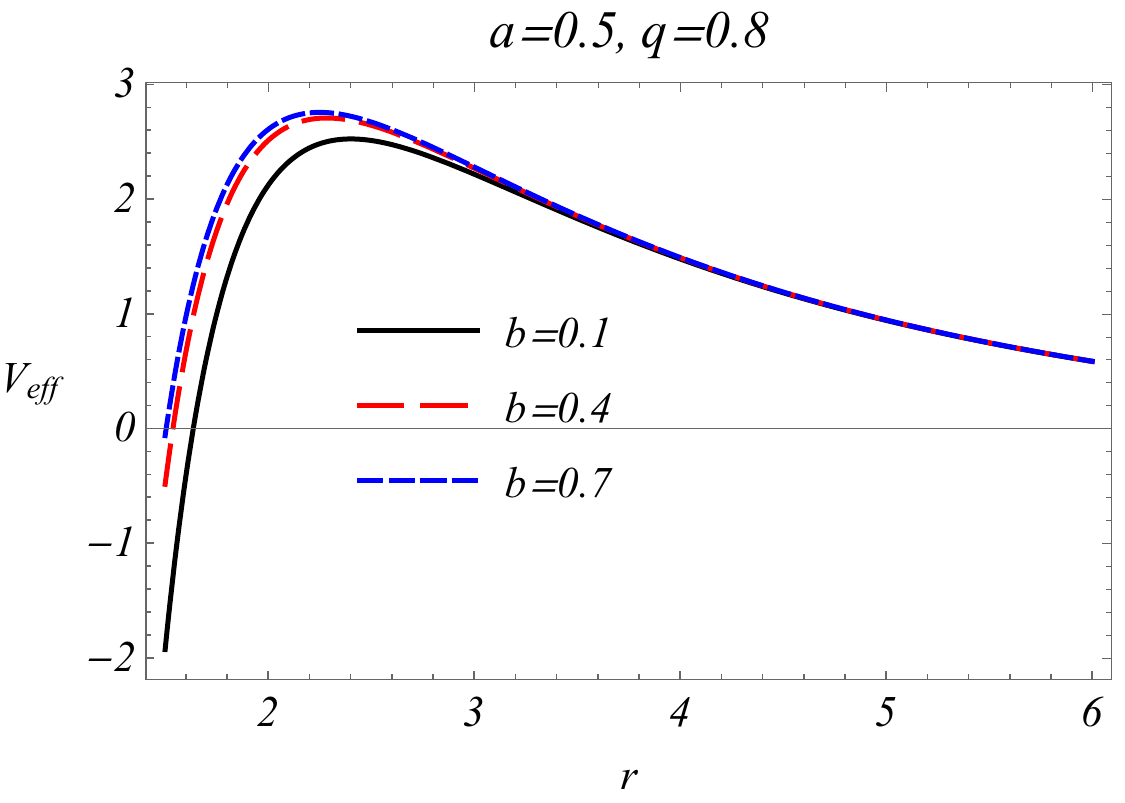}}\\
\subfigure{
\includegraphics[width=0.42\textwidth]{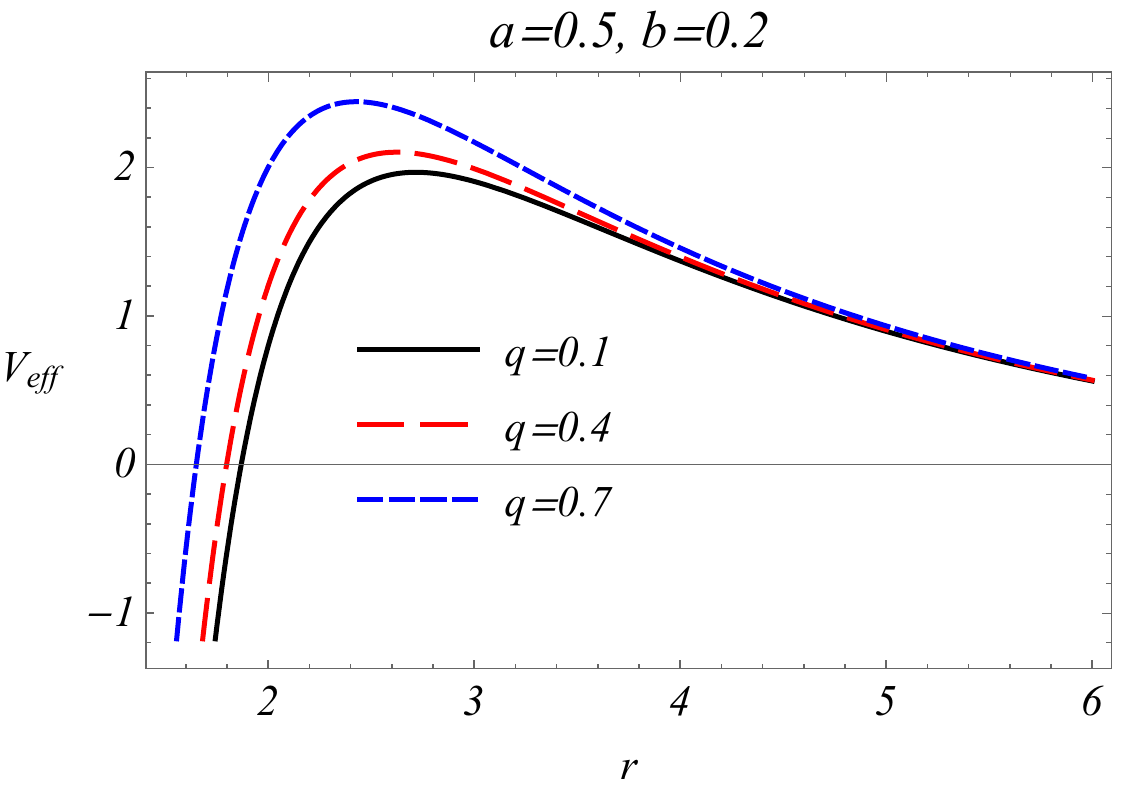}}~~~~
\subfigure{
\includegraphics[width=0.42\textwidth]{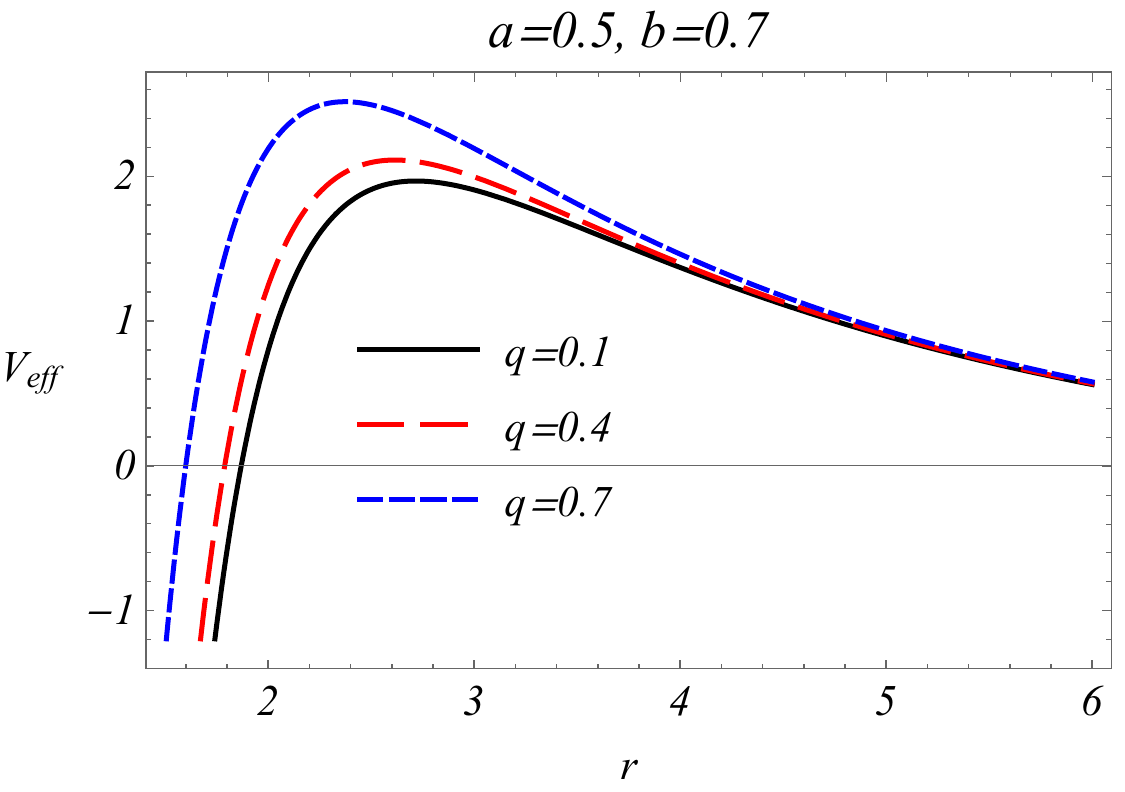}}\\
\subfigure{
\includegraphics[width=0.42\textwidth]{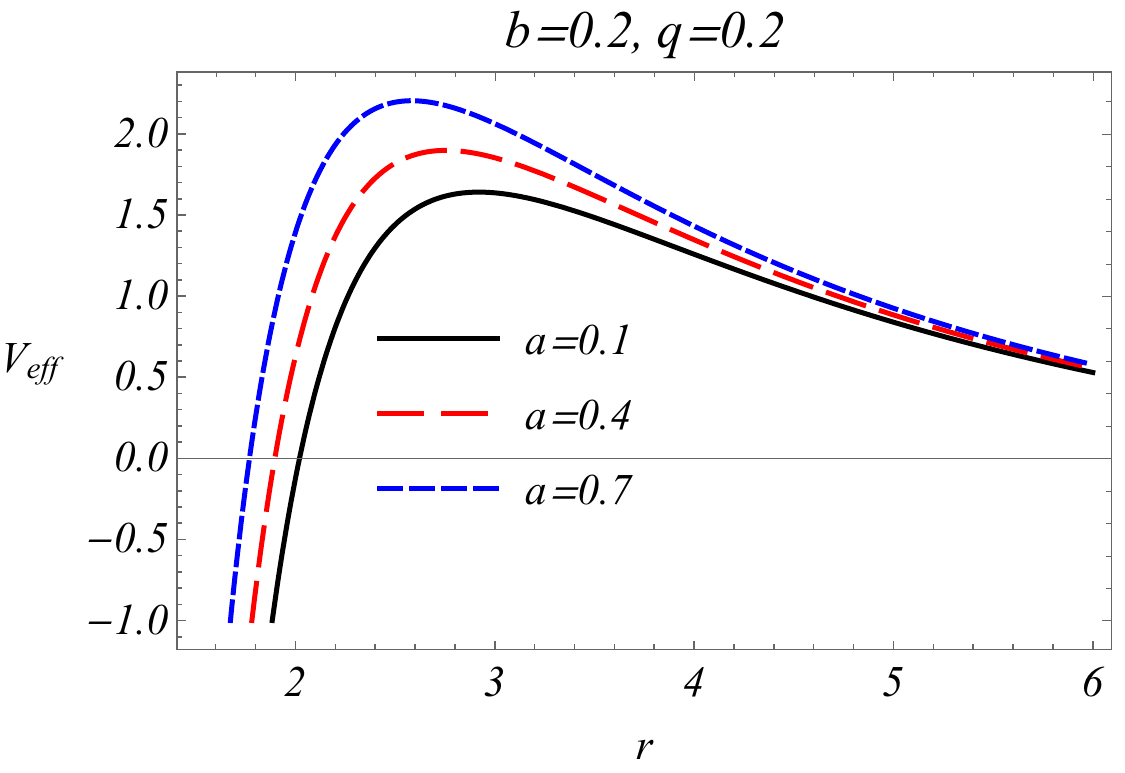}}~~~~
\subfigure{
\includegraphics[width=0.42\textwidth]{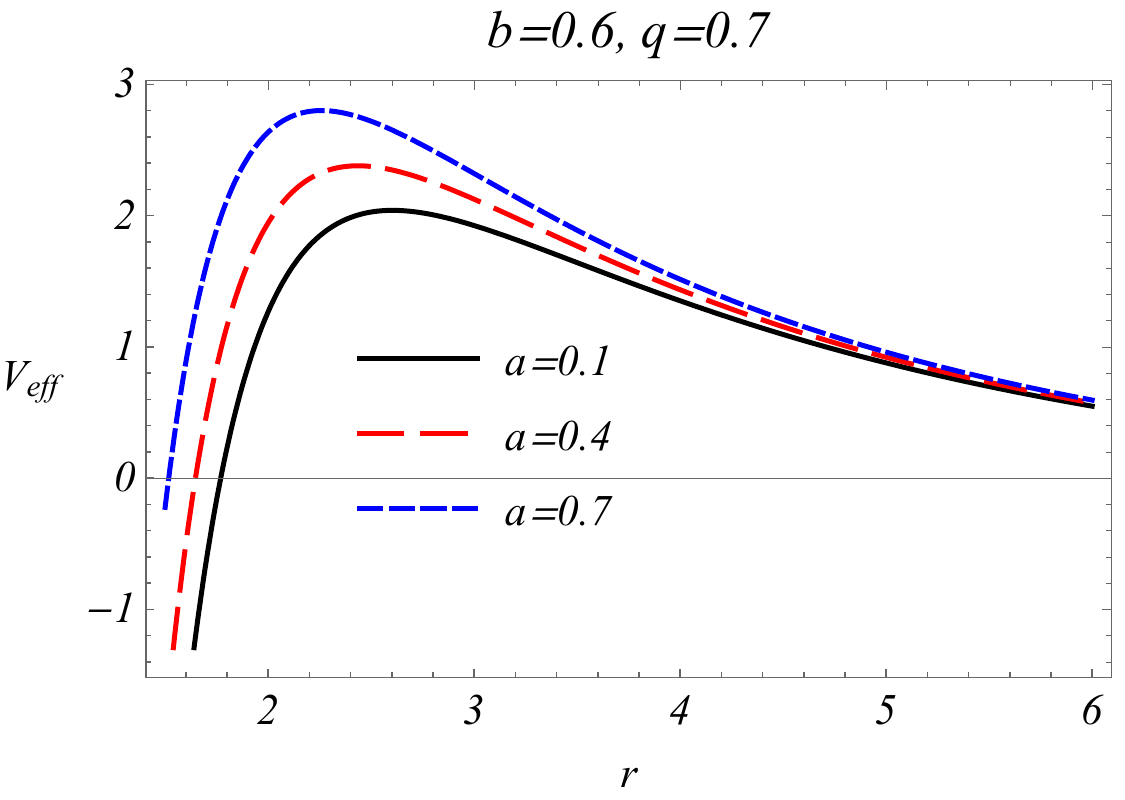}}
\end{center}
\caption{Plots showing the behaviour of effective potential $V_{eff}$ versus $r$ for different values of $a$, $b$ and $q$.} \label{EP}
\end{figure*}

In this section, we would like to investigate the geodesic structure around the rotating BH (\ref{Romet}). We will study the circular photon orbits of the BH characterized by the effective potential, which is a key quantity for examining the shadow cast by the novel charged rotating BH in NED. The geodesics of a particle moving in this background can be obtained by solving the geodesic equations. To derive geodesic equations, one can adopt the Hamilton-Jacobi (HJ) approach. The HJ equation describing the motion of a particle is given by
\begin{equation}
\frac{\partial S}{\partial\tau}=-\frac{1}{2}g^{\mu\nu}\frac{\partial S}{\partial x^{\mu}}\frac{\partial S}{\partial x^{\nu}}, \label{sequation}
\end{equation}
where $\tau$ is the affine parameter. For the BH metric under consideration, there exist two Killing vector fields $\partial_t$ and $\partial_\phi$ as translational and rotational invariance of time, which generate two constants; the particle energy $E$ and orbital angular momentum $L$ along the $z$ axis given as
\begin{eqnarray}
-E=g_{t\mu}\dot{x}^{\mu}&=&p_t,\\
L=g_{\phi\mu}\dot{x}^{\mu}&=&p_\phi,
\end{eqnarray}
where $p_t$ and $p_\phi$ are generalized momenta in respective directions. Then we can assume the Jacobi action in the form
\begin{equation}
S=\frac{1}{2}m_p^2\tau-Et+L\phi+S_r(r)+S_\theta(\theta),\label{jaction}
\end{equation}
where $m_p$ is the rest mass of the particle and is the third constant of motion. The functions $S_r(r)$ and $S_\theta(\theta)$, respectively, depend only on $r$ and $\theta$ and are arbitrary functions yet to be determined. Substituting the Jacobi action (\ref{jaction}) into the HJ equation (\ref{sequation}) and considering photon mass $m_p=0$, one obtains
\begin{eqnarray}
S_r(r)&=&\int^r\frac{\sqrt{\mathcal{R}(r)}}{\Delta(r)}dr, \\
S_\theta(\theta)&=&\int^\theta\sqrt{\Theta(\theta)}d\theta,
\end{eqnarray}
with
\begin{eqnarray}
\mathcal{R}(r)&=&\left(\left(r^2+a^2\right)E-aL\right)^2 \nonumber\\
&&-\Delta(r)\left(\mathcal{Z}+(L-aE)^2\right), \\
\Theta(\theta)&=&\mathcal{Z}+\cos^2\theta\left(a^2E^2-L^2\csc^2\theta\right),
\end{eqnarray}
where $\mathcal{Z}$ is the Carter constant related to the Killing-Yano tensor field and is the fourth constant of the geodesics. Since we have four equations corresponding to four coordinate variables, and the system of equations is completely integrable, we have four constants. Further, taking derivatives of Jacobi action with respect to these four constants and solving by setting the equations equal to zero, we obtain the following geodesic equations
\begin{eqnarray}
\rho^2\frac{dt}{d\tau}&=&\frac{r^2+a^2}{\Delta(r)}\left(E\left(r^2+a^2\right)-aL\right) \nonumber\\
&&+a\left(L-aE\sin^2\theta\right), \label{rhot}\\
\rho^2\frac{dr}{d\tau}&=&\pm\sqrt{\mathcal{R}(r)}, \label{Rad}\\
\rho^2\frac{d\theta}{d\tau}&=&\pm\sqrt{\Theta(\theta)}, \label{theta}\\
\rho^2\frac{d\phi}{d\tau}&=&\frac{a}{\Delta(r)}\left(E\left(r^2+a^2\right)-aL\right) \nonumber\\
&&+\left(l\csc^2\theta-aE\right). \label{rhophi}
\end{eqnarray}
Now we focus on the circular photon orbits by analyzing the radial motion. The radial geodesic equation (\ref{Rad}) can also be expressed as
\begin{equation}
\frac{1}{2}\left(\rho^2\frac{dr}{d\tau}\right)^2+V_{eff}=0,
\end{equation}
where $V_{eff}$ is the effective potential and in the equatorial plane, it takes the form
\begin{eqnarray}\label{veff}
V_{eff}&=&-\frac{\mathcal{R}(r)}{2r^4}.
\end{eqnarray}
The behavior of effective potential $V_{eff}$ is plotted with respect to $r$ for various values of $b$ and $q$ in Fig. \ref{EP}. We focus only on the peak of the curves in the plots that correspond to the unstable circular null orbits. In the top panel, we can see that for a large value of $q$ and a fixed value of $a$, the unstable circular null orbits decrease in size as $b$ increases. However, no significant change has been observed for a small value of $q$. In the middle panel, the unstable circular null orbits reduce in size with the increase in $q$ for both cases. Similarly, as the value of $a$ increases, the unstable circular null orbits shrink. The circular photon orbits satisfy the following conditions
\begin{equation}
V_{eff}=0\ , \quad \frac{\partial V_{eff}}{\partial r}=0\ , \label{qqq}
\end{equation}
and the condition 
\begin{equation}
\frac{\partial^2 V_{eff}}{\partial r^2}<0\ , \label{qqq1}
\end{equation}
ensures the orbits are unstable. Now, supposing the new definitions $\xi=\frac{L}{E}$ and $\eta=\frac{\mathcal{Z}}{E^2}$ and solving both equations in (\ref{qqq}), we obtain the following.
\begin{eqnarray}
\xi(r_p)&=&\frac{\left(a^2+r_p^2\right)\Delta'(r_p)-4\Delta(r_p)r_p}{a\Delta'(r_p)}, \label{xx1}\\
\eta(r_p)&=&\frac{16r_p^2\Delta(r_p)\left(a^2-\Delta(r_p)\right)}{a^2\Delta'(r_p)^2}-\frac{r_p^4}{a^2} +\frac{8r_p^3\Delta(r_p)}{a^2\Delta'(r_p)}, \label{xx2}
\end{eqnarray}
where the prime denotes the derivative with respect to $r$ and $r_p$ is the radius of the photon sphere. Solving for the unstable orbits, the condition (\ref{qqq1}) yields
\begin{eqnarray}
r+2\frac{\Delta(r)}{\Delta'(r)^2}\left(\Delta'(r)-r\Delta''(r)\right)\bigg|_{r=r_p}>0. \label{ccdc}
\end{eqnarray}

\subsection{The rotating black Hole in NED shadow}
Now, we would like to study the shadow cast by the rotating charged BH in NED. It is worth pointing out that, in our case, all light sources are located at infinity and distributed uniformly in all directions. Moreover, there is no light source between the BH and the observer, and the observer is located at infinity. In order to develop the shadow images under the above-mentioned assumption, one needs to introduce two celestial coordinates \cite{1973blho.conf..215B}
\begin{eqnarray}
\alpha&=&-\lim_{r_0\rightarrow\infty}\left(r_0^2\sin\theta_0\frac{d\phi}{dr}\bigg|_{\theta\rightarrow\theta_0,r\rightarrow r_0}\right) \nonumber\\
&=&-\xi(r_p)\csc\theta_0, \label{alpha}\\
\beta&=&\lim_{r_0\rightarrow\infty}\left(r_0^2\frac{d\theta}{dr}\bigg|_{\theta\rightarrow\theta_0,r\rightarrow r_0}\right) \nonumber\\
&=&\pm\sqrt{\eta(r_p)+a^2\cos^2\theta_0-\xi(r_p)^2\cot^2\theta_0}, \label{beta}
\end{eqnarray}
where $\theta_0$ is the inclination angle of the observer. If the observer is located on the equatorial plane, these celestial coordinates are simplified to
\begin{eqnarray}
\alpha&=&-\xi(r_p), \\
\beta&=&\pm\sqrt{\eta(r_p)}.
\end{eqnarray}

\begin{figure*}[t]
\begin{center}
\subfigure{\includegraphics[width=0.315\textwidth]{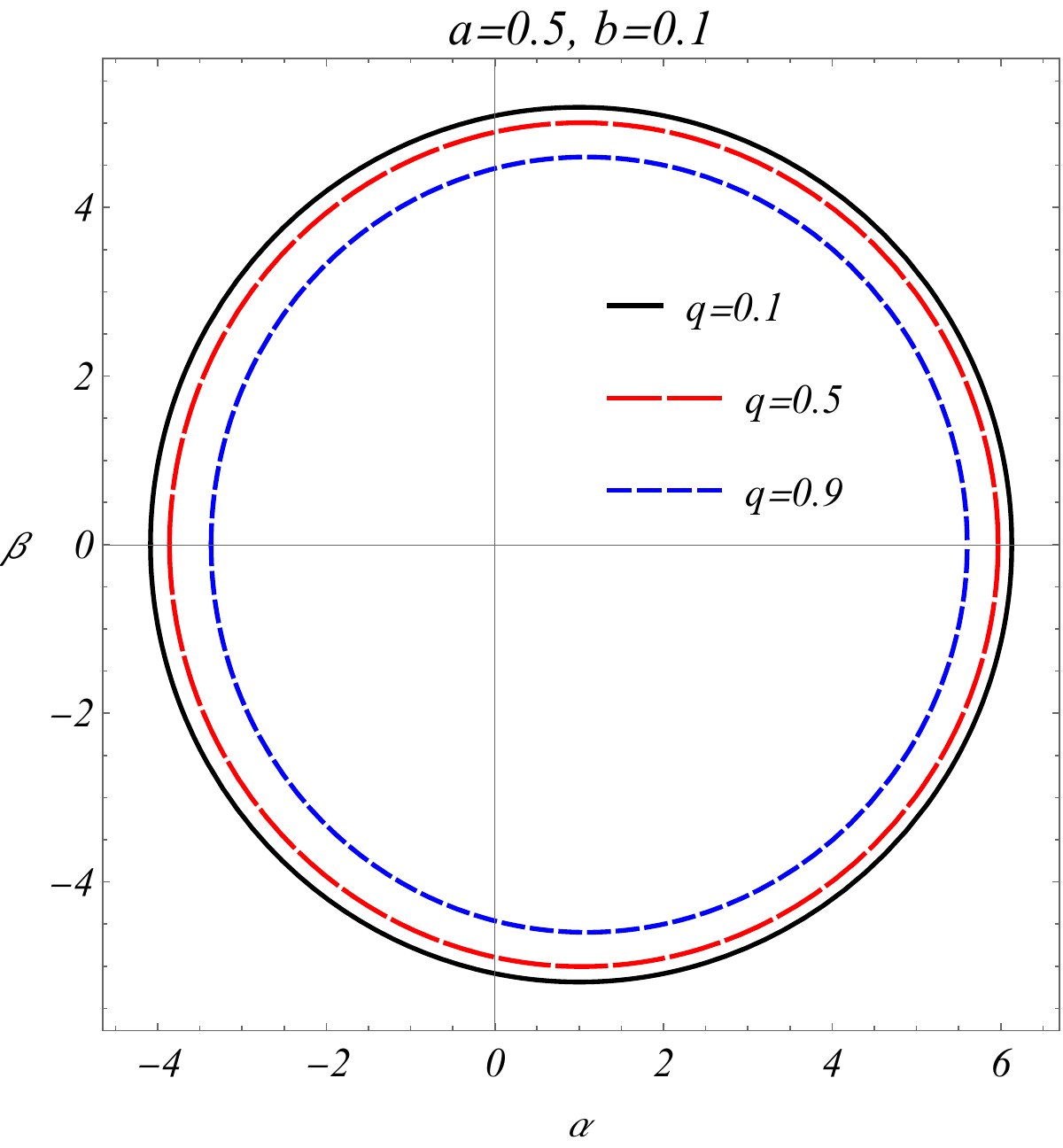}}~~
\subfigure{\includegraphics[width=0.315\textwidth]{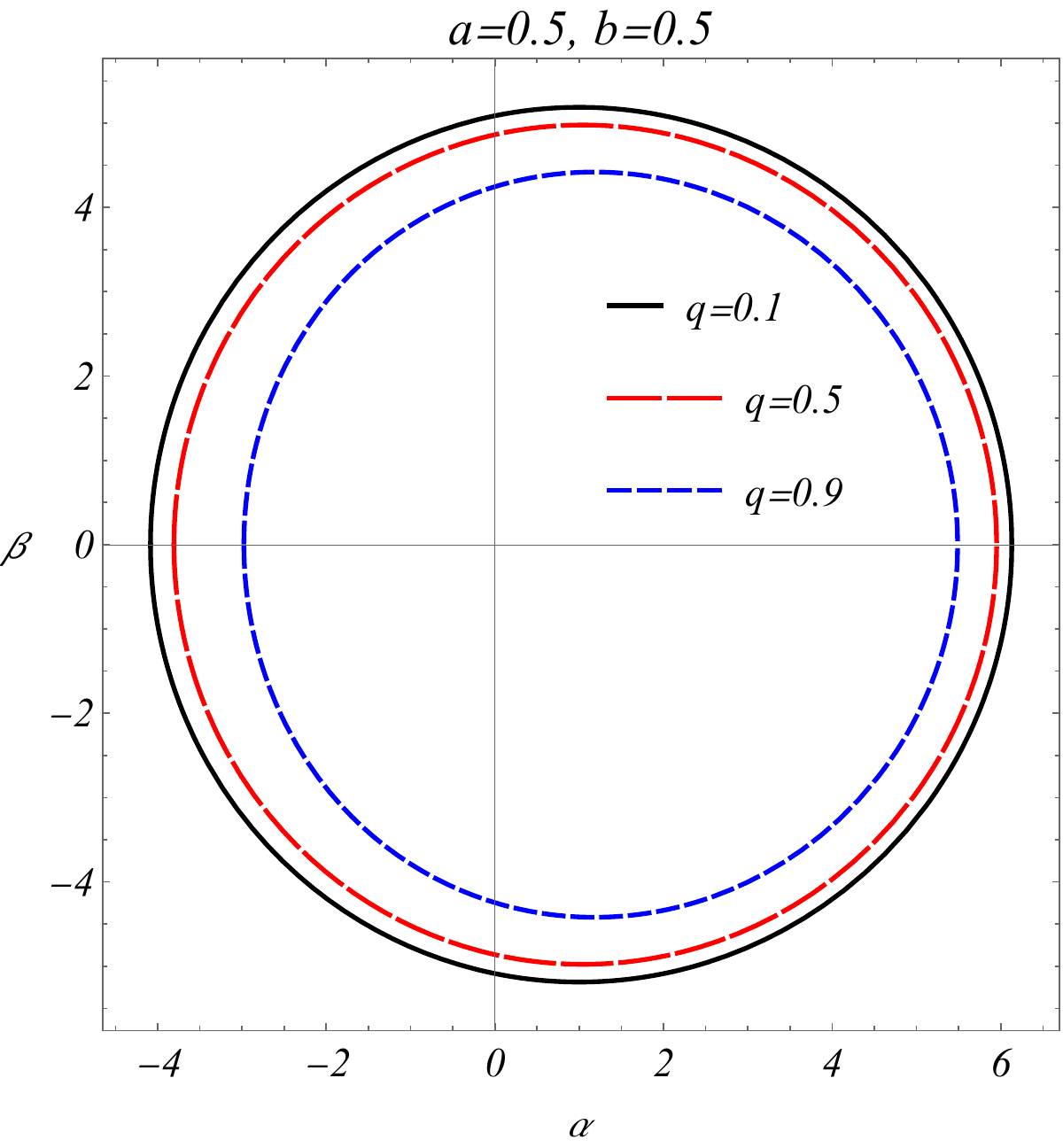}}~~
\subfigure{\includegraphics[width=0.315\textwidth]{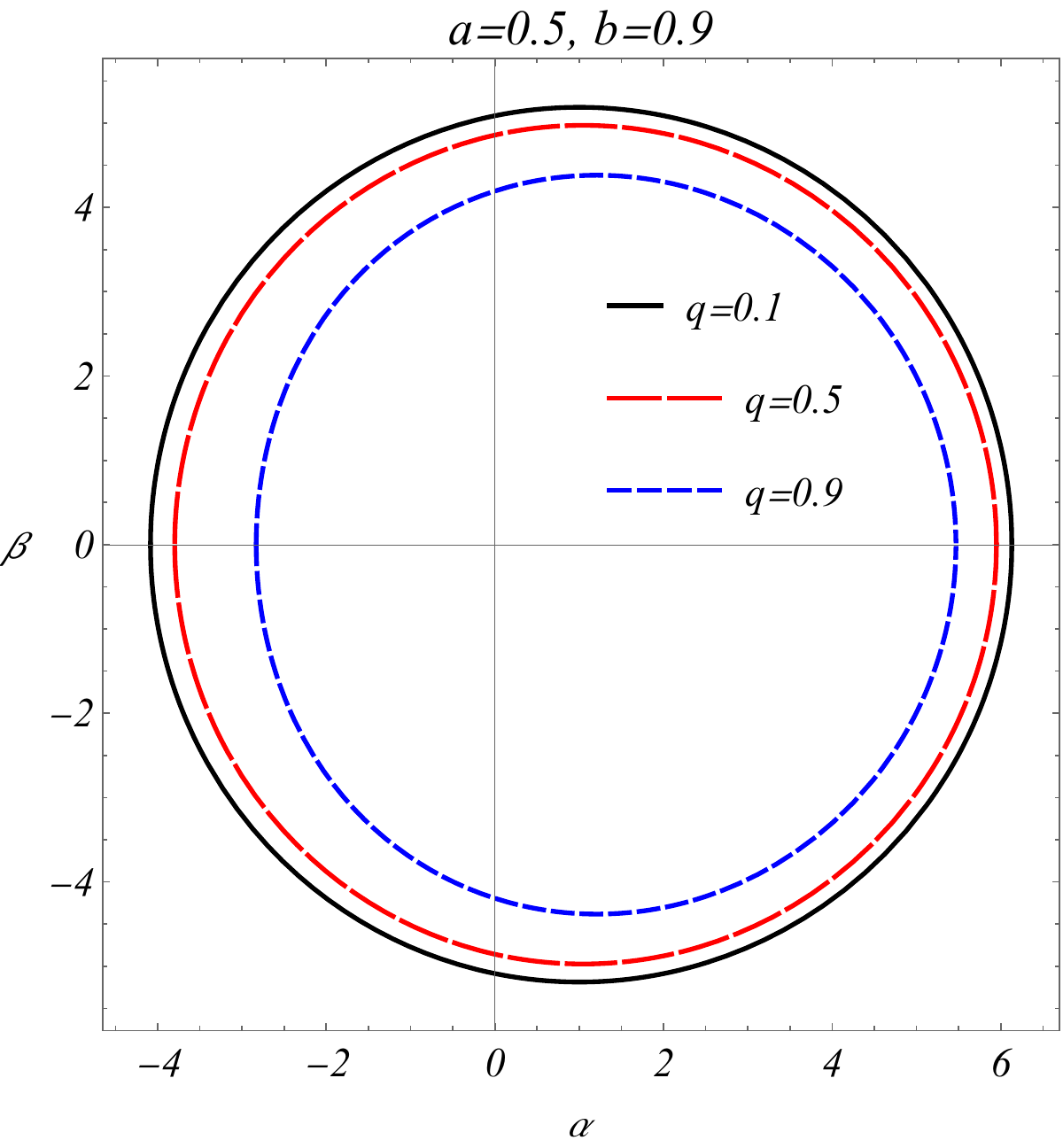}}\\
\subfigure{\includegraphics[width=0.315\textwidth]{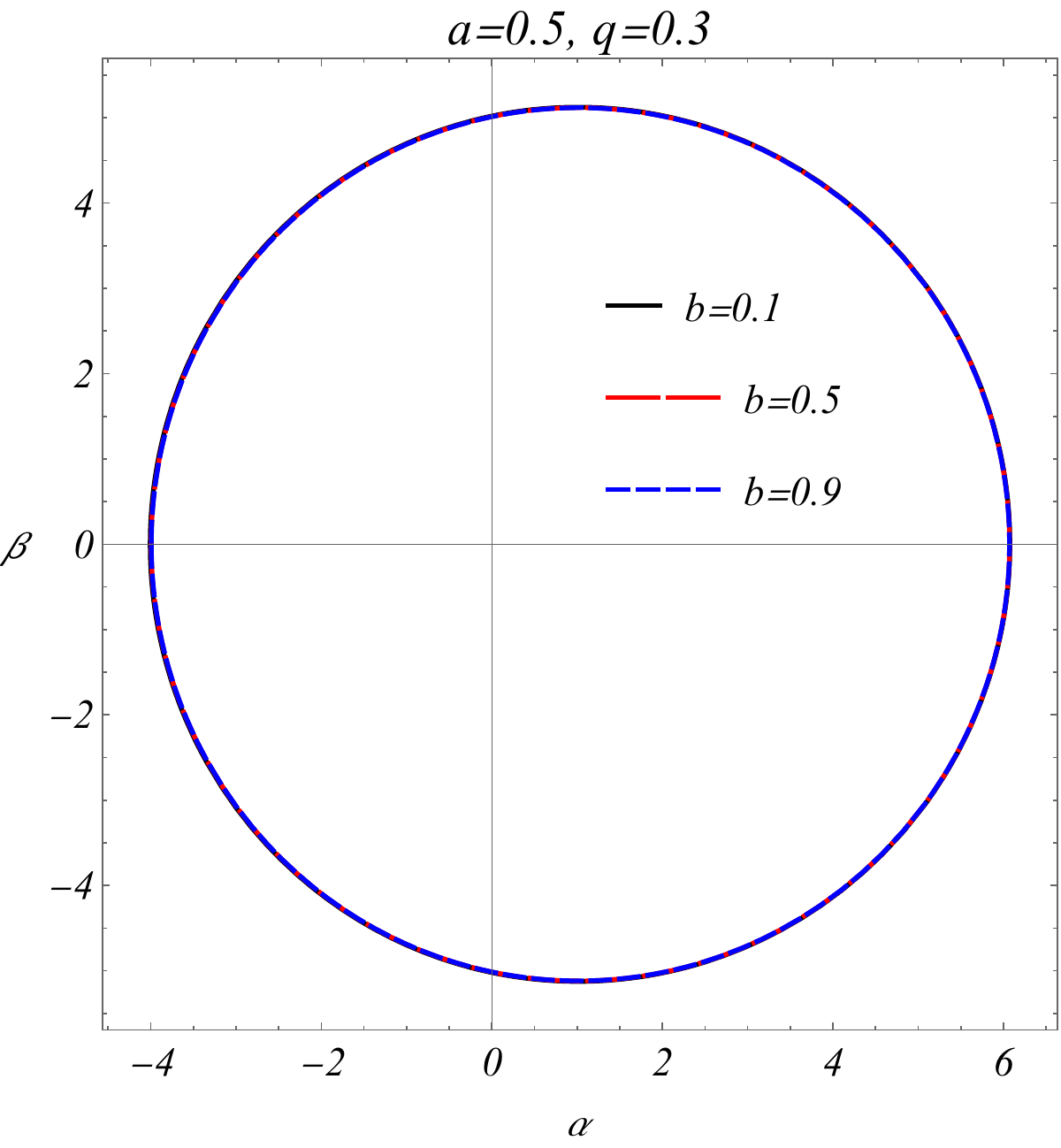}}~~
\subfigure{\includegraphics[width=0.315\textwidth]{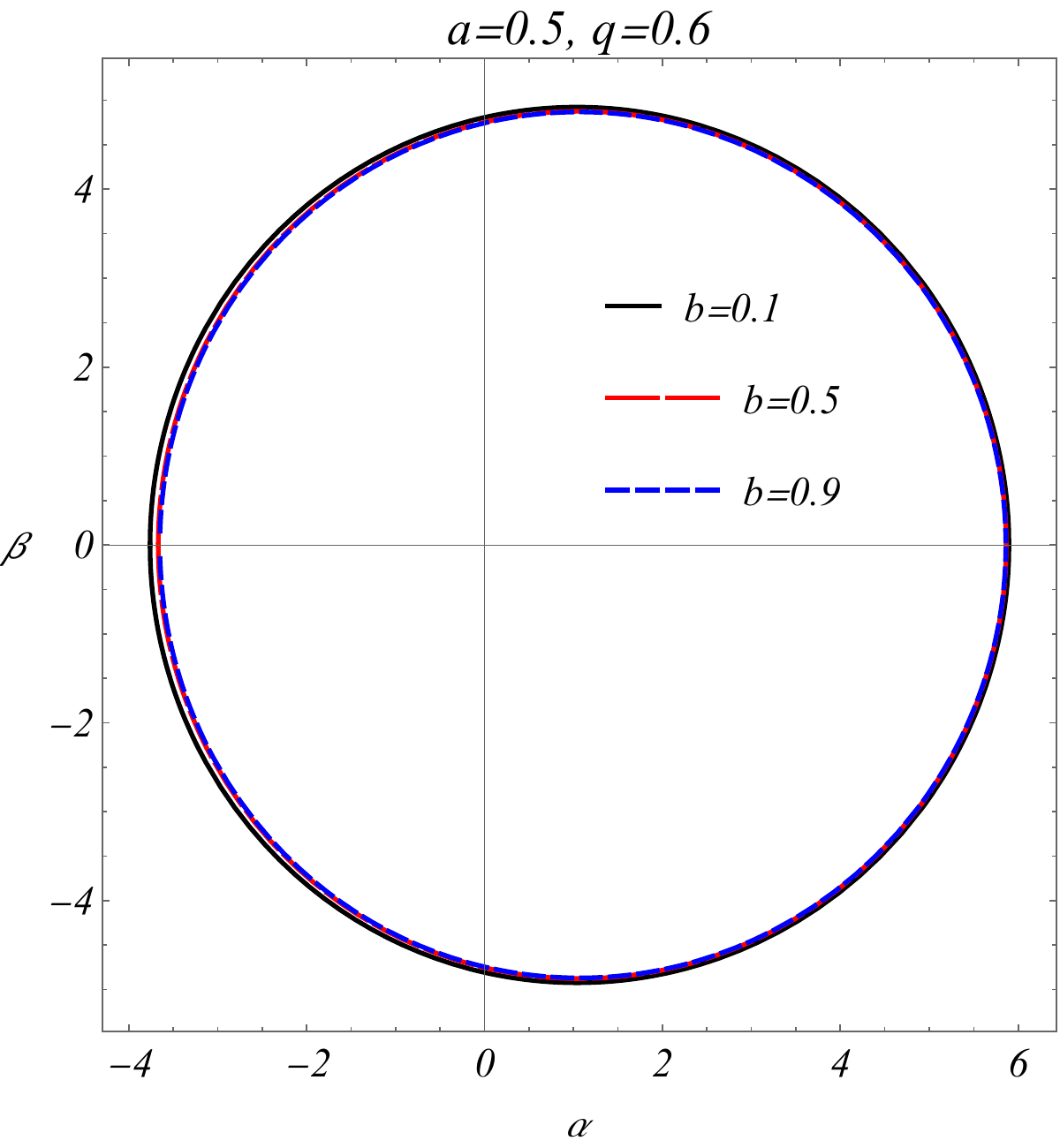}}~~
\subfigure{\includegraphics[width=0.315\textwidth]{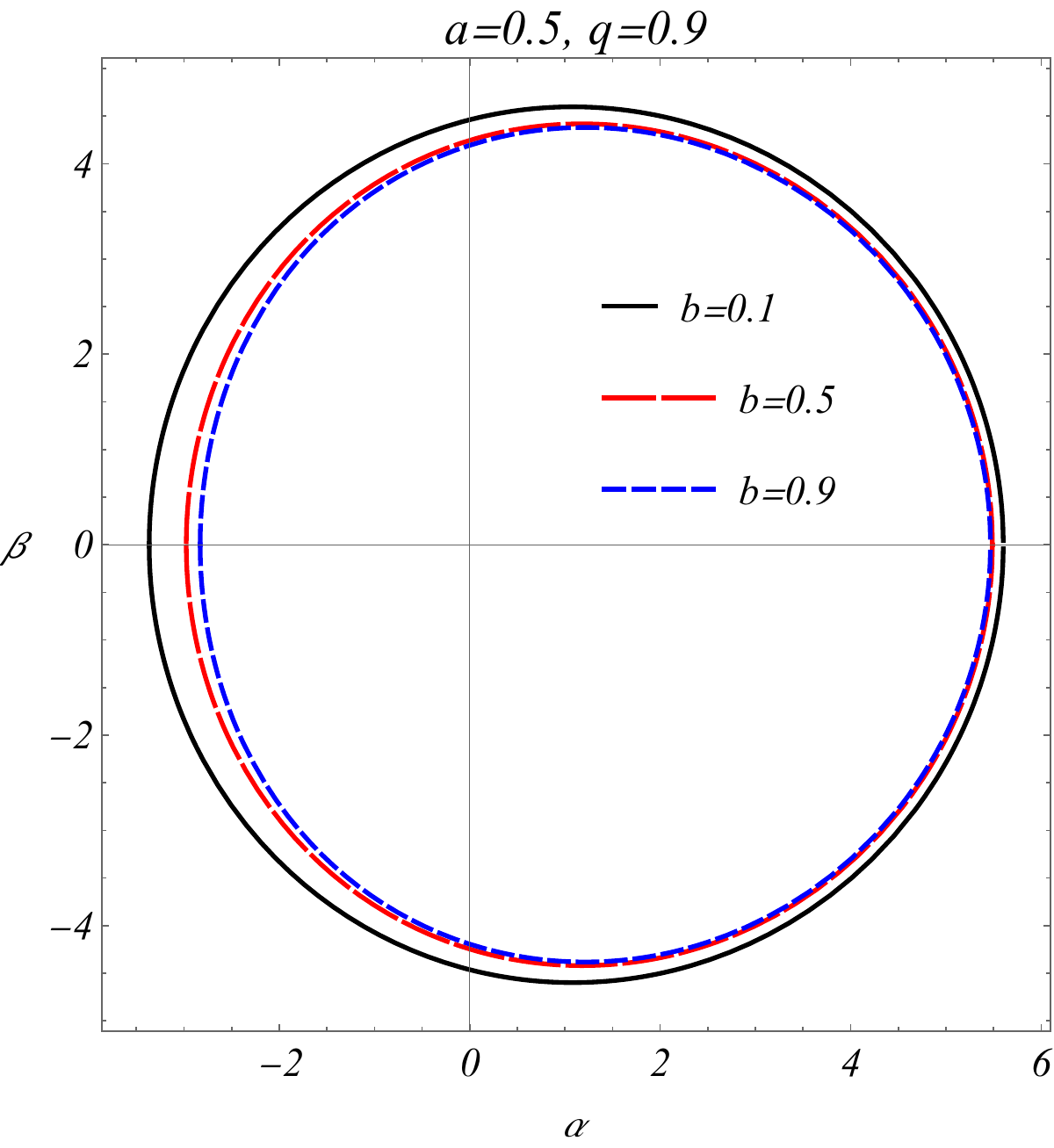}}\\
\subfigure{\includegraphics[width=0.315\textwidth]{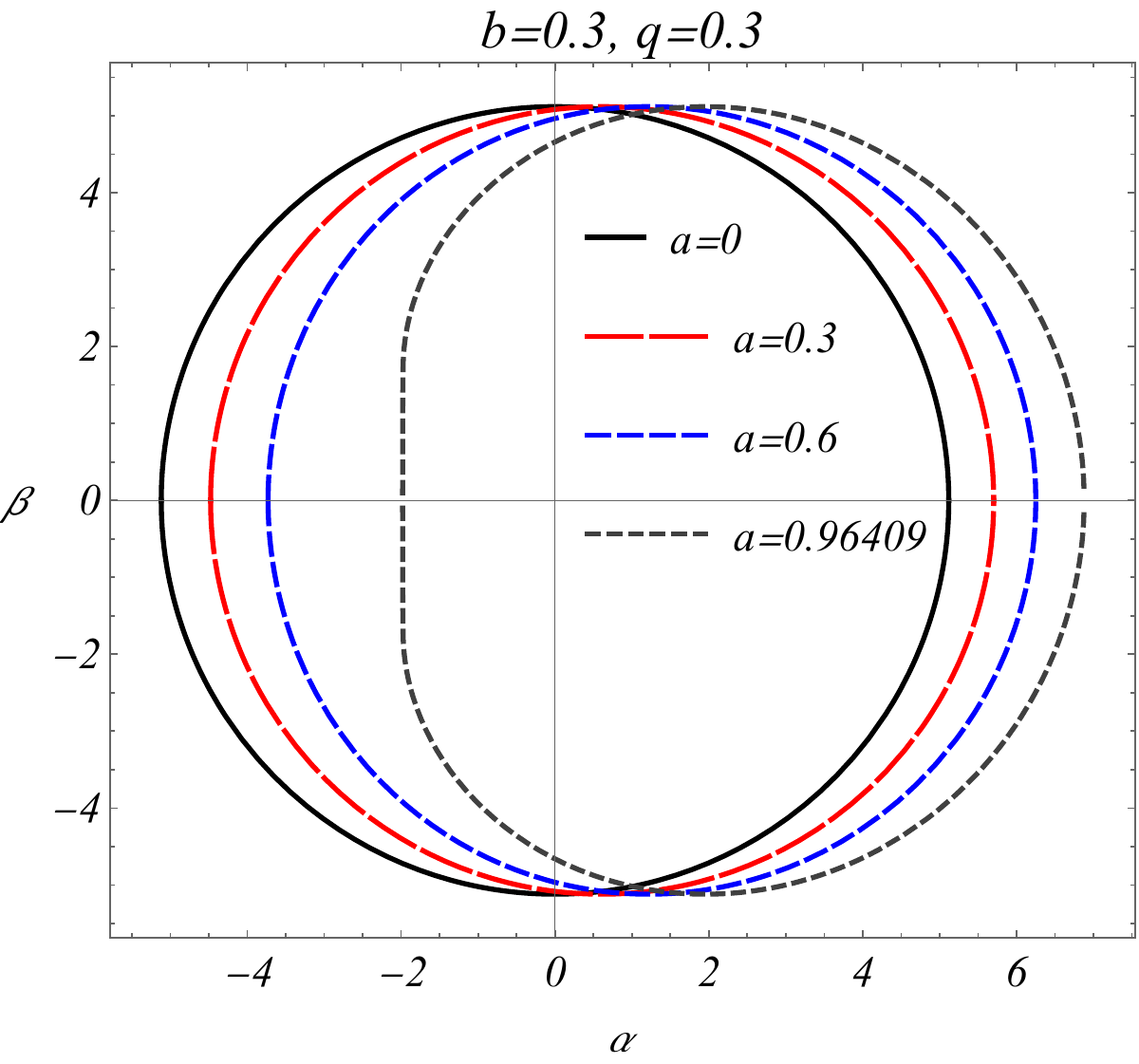}}~~
\subfigure{\includegraphics[width=0.315\textwidth]{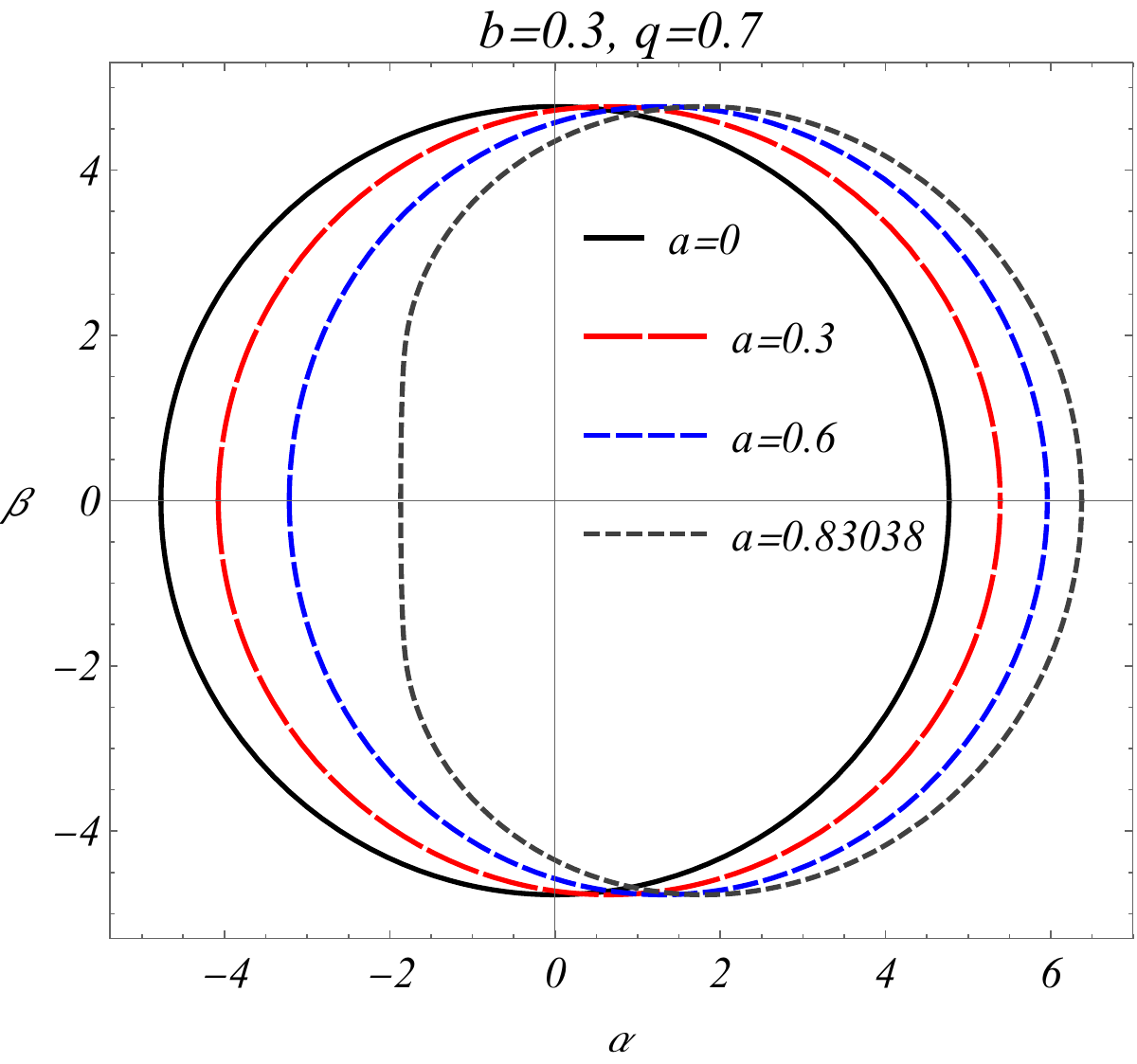}}~~
\subfigure{\includegraphics[width=0.315\textwidth]{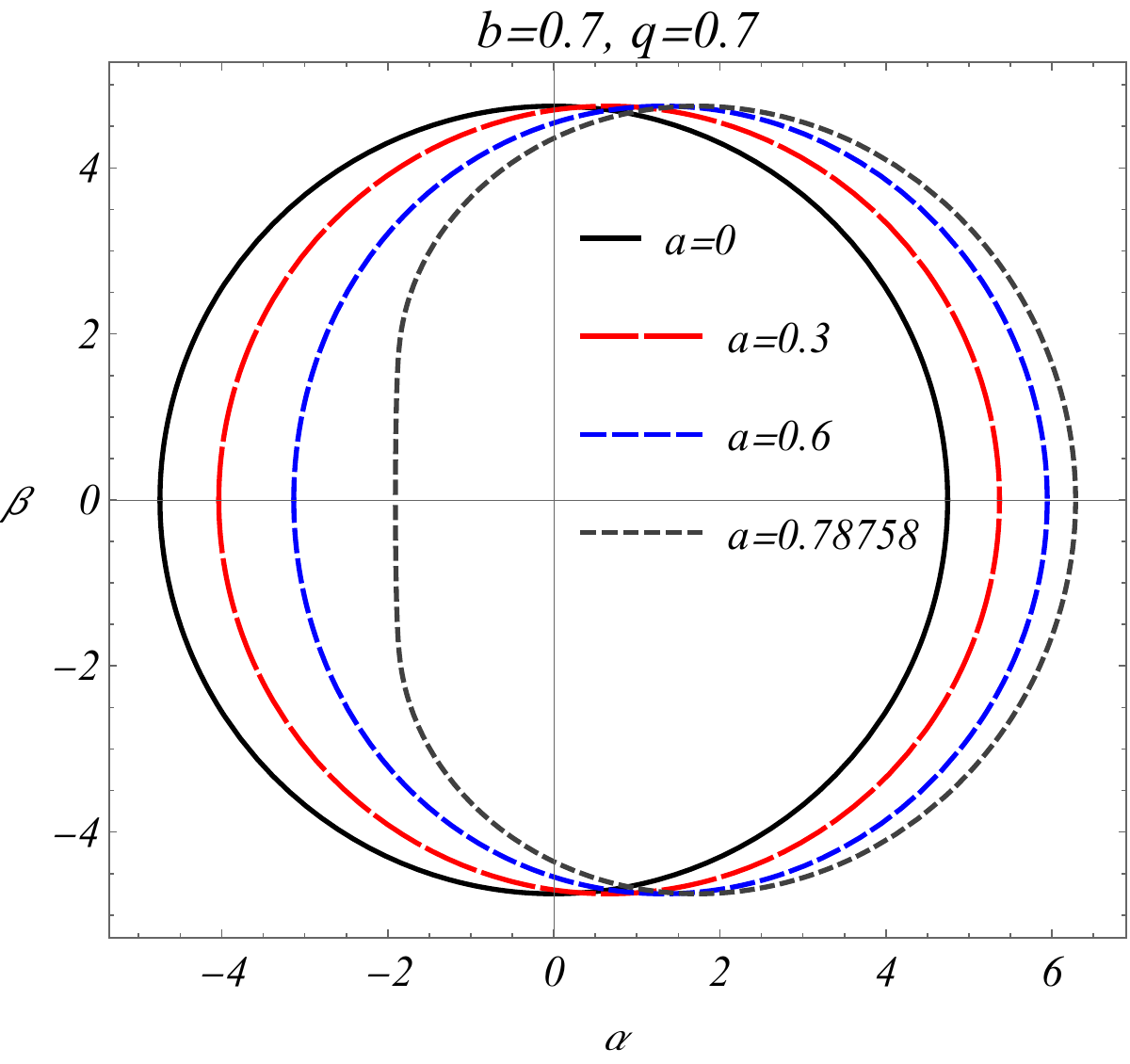}}
\end{center}
\caption{Shadow plots for various parametric values for rotating and static charged BH in NED.} \label{Sh}
\end{figure*}

The shadow can be obtained by producing a parametric plot in the $\alpha$-$\beta$ celestial plane consistent with Eqs. (\ref{xx1}) and (\ref{xx2}), where the parameter governing the plot is $r_p$. Such a region is actually not illuminated by the bright photon sources. The boundary of the shadow can be determined by the radius of the circular photon orbits. The shadow curves have been plotted in Fig. \ref{Sh} for various cases of parametric values of $a$, $b$, and $q$. The top panel shows the fixed value of $a$ while $b$ varies in the panel from left to right and each curve corresponds to a different value of $q$. It is clear that with an increase in $q$, the shadow size reduces, whereas with an increase in $b$ in the panel, the variation in the shadow size increases. In the middle panel, again $a$ has been kept fixed and the value of $q$ varies from left to right in the panel while the different values of $b$ correspond to each curve. For the small value of $q$, there is no significant variation in the size of the shadow with respect to the increase in $b$. Some variation is observed in the shadow size with an increase in $b$ when $q$ increases further in the middle plot. In the right plot for the largest value of $q$, the variation in shadow size is more prominent, and therefore we can specify that the shadow size is reduced with an increase in $b$. In the lower panel, we mainly study the role of spin parameters on the variation in shadow size. Therefore, each curve corresponds to a different value of $a$. These plots show that for the static case, the shadows are purely circular, and as the value of $a$ increases, the shadows shift towards the right, and the flatness on one side increases. The maximum flatness is also plotted for the extremal spin values for each combination of the values of $b$ and $q$.

\subsection{Distortion}
The distortion measures the difference in the shape of the rotating BH shadow. It is characterized only for the rotating BHs, because all static BHs are circular. Therefore, the distortion is studied for shadow comparison for rotating BHs and static BHs as well as for those that have the least distorted shadows. The distortion is measured in terms of an observable known as the linear radius of the shadow \cite{PhysRevD.94.024054,PhysRevD.80.024042}, defined as
\begin{equation}
R_{sh}=\frac{(\alpha_t-\alpha_r)^2+\beta_t^2}{2|\alpha_t-\alpha_r|}, \label{rad}
\end{equation}
where $R_{sh}$ is the radius of a hypothetical circle that can be assumed to touch the shadow curve at the points $(\alpha_t,\beta_t)$, $(\alpha_b,\beta_b)$ and $(\alpha_r,0)$ lying at the top, bottom and right most point on the shadow curve, respectively. The shadow curve is described by the points in the space $(\alpha,\beta)$ such that the points are characterized by the subscripts $t$, $b$, and $r$ corresponding to the top, bottom, and the rightmost point, respectively. For further details, we refer to Fig. \textbf{9} in \cite{PhysRevD.94.024054}. Eq. (\ref{rad}) is valid only for rotating BHs because the static BHs have circular shadows and can be expressed by coordinates of the curve on any coordinate axes. Then, the distortion can be obtained by the relation.
\begin{equation}
\delta_s=\frac{|\bar{\alpha}_l-\alpha_l|}{R_{sh}}, \label{dis}
\end{equation}
where $(\alpha_l,0)$ and $(\bar{\alpha}_l,0)$ intersect the $-\alpha$-axis and the points reside on the shadow and hypothetical circle, respectively. The points on the shadow are characterized by the subscript $l$ on the left side of $\beta$-axis, whereas, the bar denotes the points on the imaginary circle.

\begin{figure*}[t]
\begin{center}
\subfigure{\includegraphics[width=0.32\textwidth]{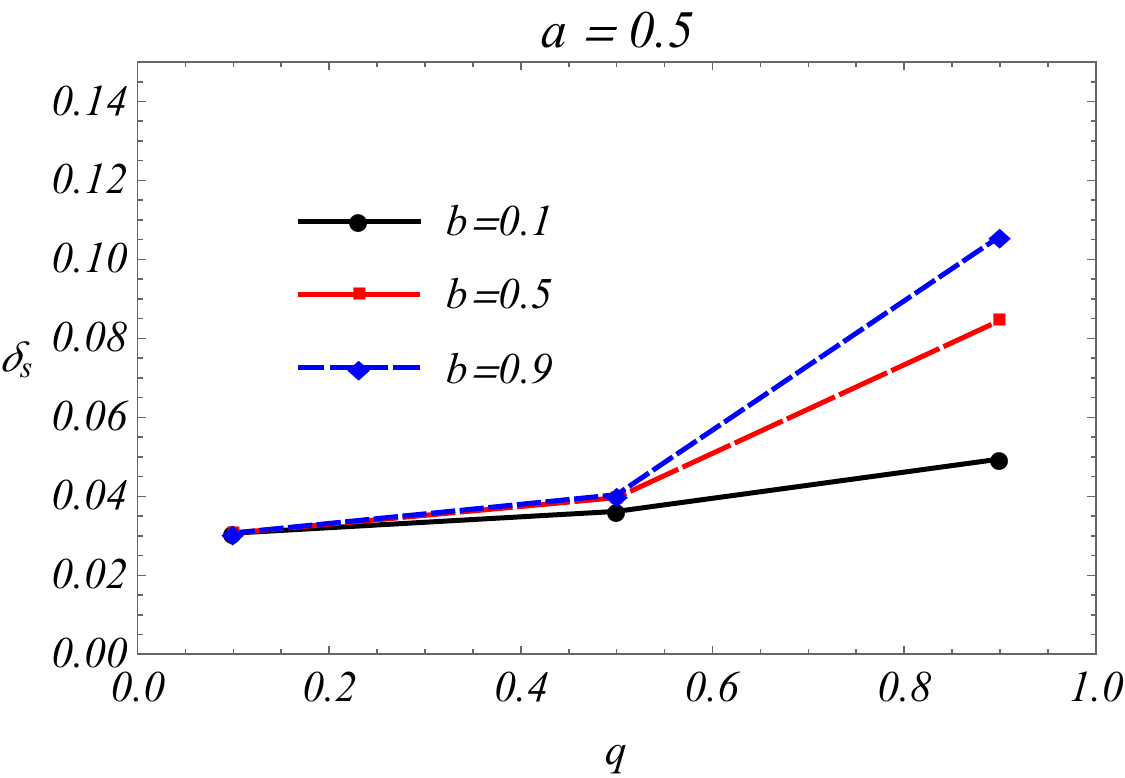}}~~
\subfigure{\includegraphics[width=0.32\textwidth]{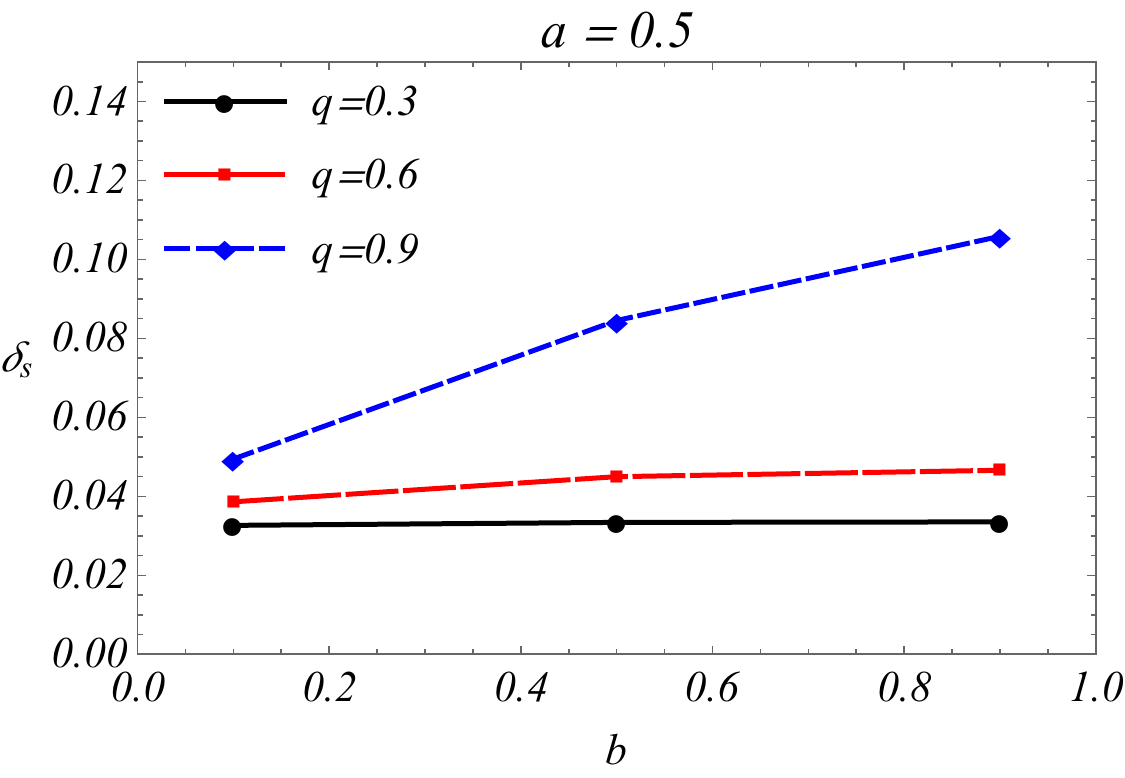}}
\subfigure{\includegraphics[width=0.32\textwidth]{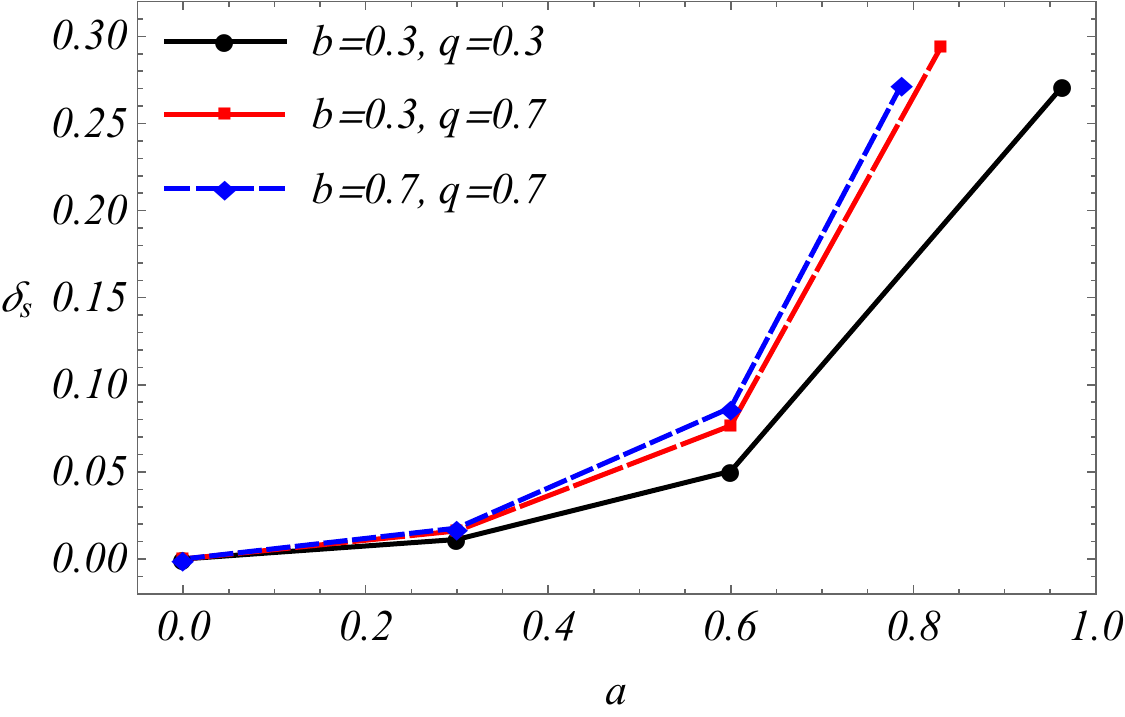}}
\end{center}
\caption{Plots showing the distortion with respect to $b$, $q$ and $a$ for the BH shadows in Fig. \ref{Sh}.} \label{Dis}
\end{figure*}

The variation of distortion is plotted with respect to the BH parameters $b$, $q$, and $a$ in Fig. \ref{Dis} corresponding to the shadows in Fig. \ref{Sh}. The value of $a$ has been fixed for both plots in the upper panel, whereas no parameter has been fixed for the plot in the lower panel. For the left plot in the upper panel, the distortion $\delta_s$ has been plotted vs $q$ and different values of $b$ correspond to the curves. We found that the distortion increases with an increase in $q$ and with an increase in $b$, the distortion increases more rapidly with respect to the increase in $q$. The right plot shows the variation of distortion with respect to $b$. It can be seen that for the smallest value of $q$, the distortion is almost constant with respect to $b$. When the value of $q$ is increased, some variation in distortion is observed for the values of $b$ up to $0.5$. The distortion becomes constant for further values of $b$. For the largest value of $q$, the variation in distortion is more obvious and increases with an increase in $b$ but at a slower rate. The lower plot shows the behavior of distortion with respect to $a$ and the curves correspond to the different values of $b$ and $q$. It is found that the distortion increases at an accelerating rate with an increase in $a$ for all cases and certainly, the distortion is maximum for the extremal cases.

\subsection{Parameter Estimation}
The shadow of the BH characterizes the features of background spacetime in its shape and size. Thus, it can serve as a useful method to test new gravity theories and constrain BH parameters \cite{Jusufi:2021fek,Okyay:2021nnh,Afrin:2021imp,Chen:2022nbb,Afrin:2022ztr,Sarikulov:2022atq,Zahid:2023csk,Zubair:2023cep}.
One can get estimations of BH parameters by using the shadow observables. To do so, we need to define the observable parameters that describe the size and shape of the BH shadow. In addition to the method by Hioki and Maeda, a straightforward method for estimating the BH parameter can be employed, which relies on the coordinate independent formalism~\cite{Abdujabbarov:2015xqa,Kumar:2018ple} and makes use of the shadow observables, shadow area, and oblateness. One can introduce the area of BH shadow $A$ and oblateness $D$ being associated with the deformation of the BH shadow in the forms \cite{Kumar:2018ple,Sarikulov:2022atq}
\begin{eqnarray}
A&=&2\int_{r_-}^{r_+}\beta(r)\frac{d\alpha(r)}{dr}dr, \label{eq:Area} \\
D&=&\frac{\Delta\alpha}{\Delta\beta}, \label{eq:oblat}
\end{eqnarray}
where $r_\pm$ are radii of stable circular orbits, which are obtained by solving the equation $\eta_c=0$.

\begin{figure*}
\subfigure{\includegraphics[width=0.45\textwidth]{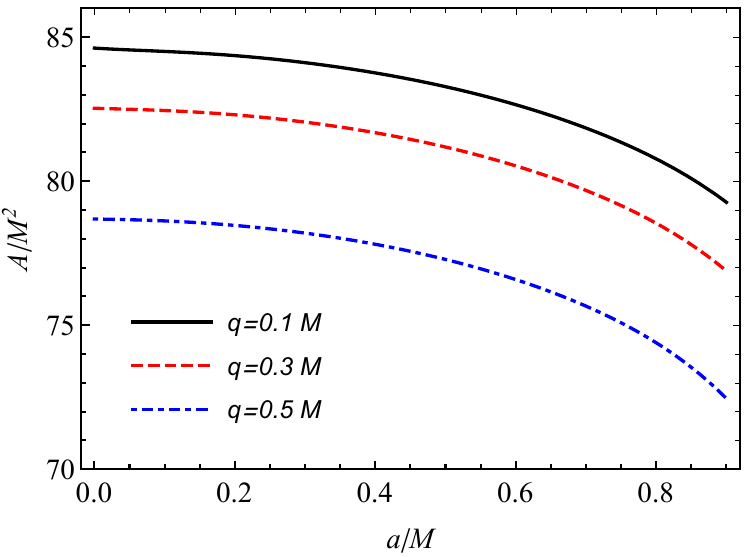}}~~~~~~
\subfigure{\includegraphics[width=0.45\textwidth]{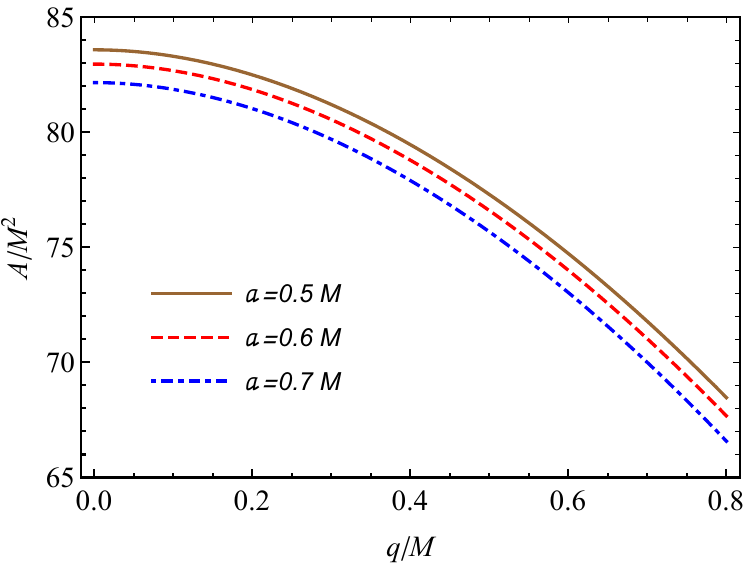}}\\
\subfigure{\includegraphics[width=0.45\textwidth]{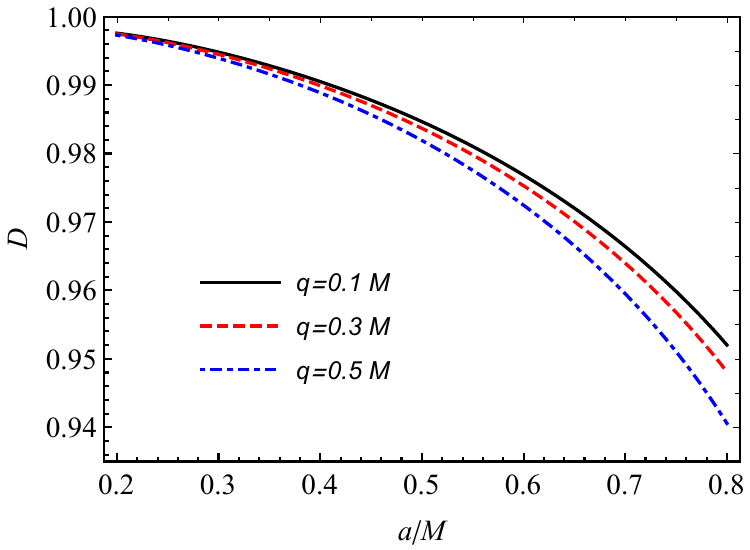}}~~~~~~
\subfigure{\includegraphics[width=0.45\textwidth]{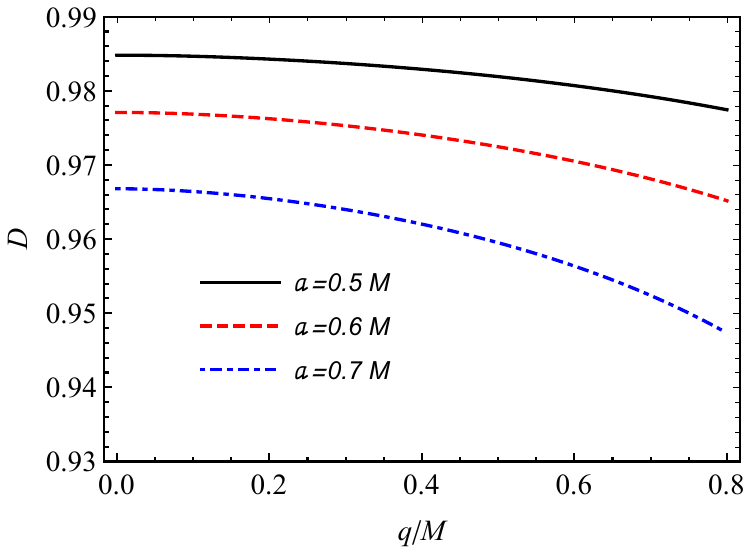}}
\caption{Area $A$ and oblateness $D$ of the BH shadow as functions of BH parameters $a$ and $q$ for the fixed value of $b=0.1M^{-1}$ \label{case1AD}}
\end{figure*}

\begin{figure*}
\subfigure{\includegraphics[width=0.45\textwidth]{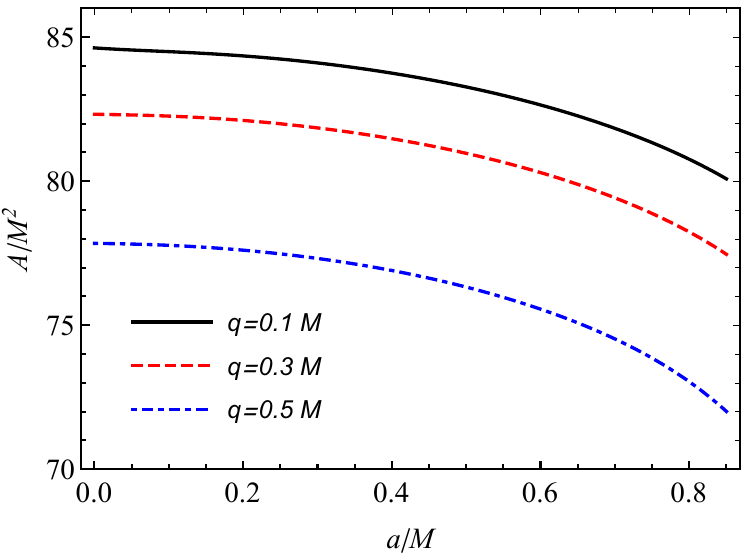}}~~~~~~
\subfigure{\includegraphics[width=0.45\textwidth]{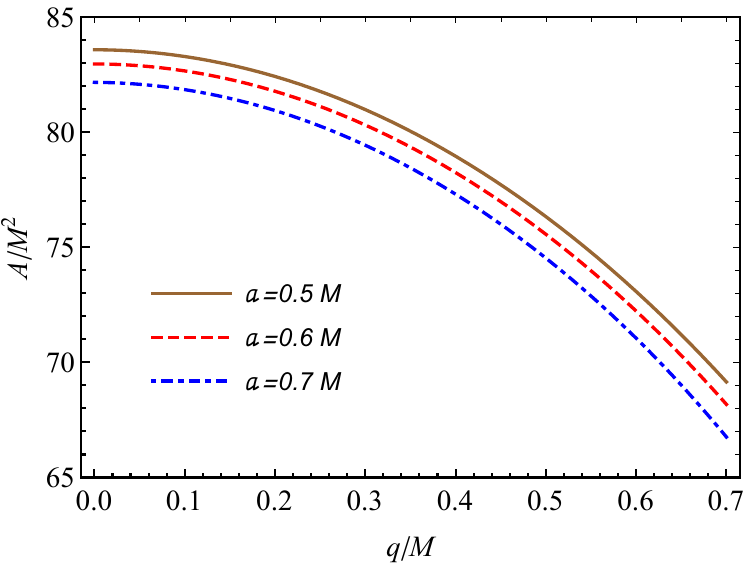}}\\
\subfigure{\includegraphics[width=0.45\textwidth]{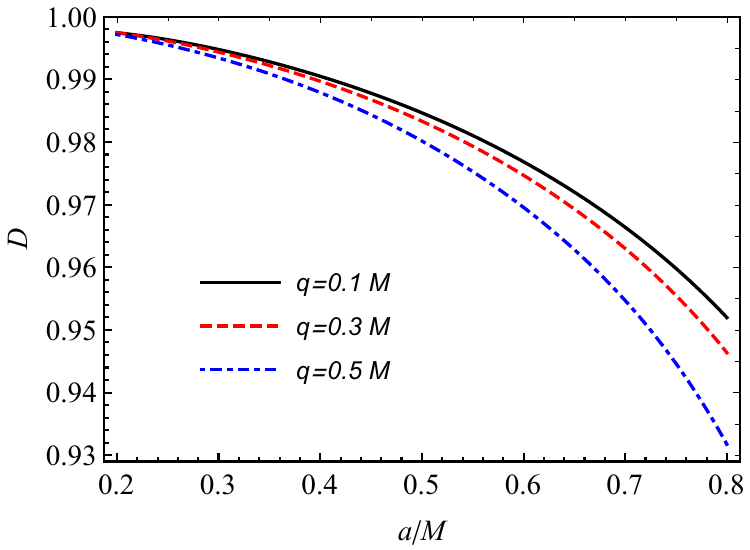}}~~~~~~
\subfigure{\includegraphics[width=0.45\textwidth]{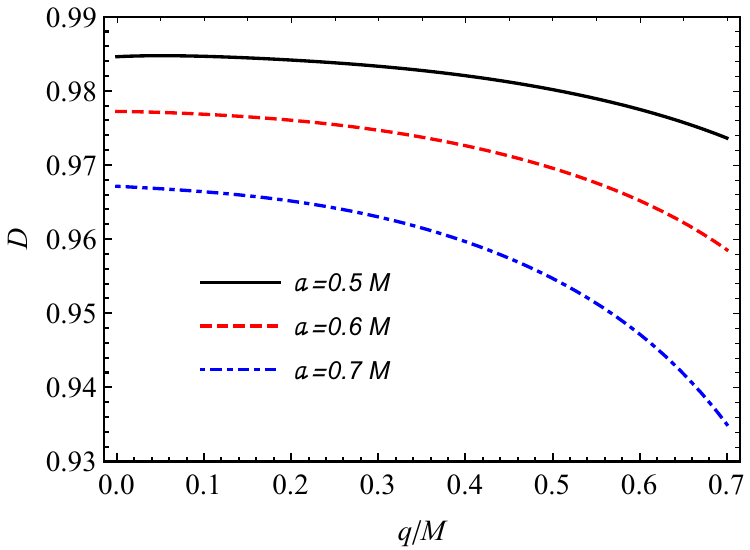}}
\caption{Area $A$ and oblateness $D$ of the BH shadow as functions of BH parameters $a$ and $q$ for the fixed value of $b=0.5M^{-1}$ \label{case2AD}}
\end{figure*}

In Figs.~\ref{case1AD} and \ref{case2AD}, we illustrate the dependencies of the shadow area $A$ and the oblateness $D$ on the parameters of rotating charged BH in NED with the help of numerical calculations. Here we studied two cases by fixing parameter $b$, for the cases of $b = 0.1 M^{-1}$ and $b = 0.5 M^{-1}$. From these plots, we can conclude that the increase of all BH parameters causes a decrease in the values of both the area and oblateness of the BH. By comparing Figs.~\ref{case1AD} and \ref{case2AD}, one can notice that there is only a small change in observable parameters with the increase of parameter $b$. Moreover, the dependence of shadow observables on $q$ is more dramatic in the larger values as $q$ compared to lower values. It is useful to note that there is a considerable decrease in the values of oblateness with an increase in the spin parameter of BH.

\begin{figure*}
\subfigure{\includegraphics[width=0.45\textwidth]{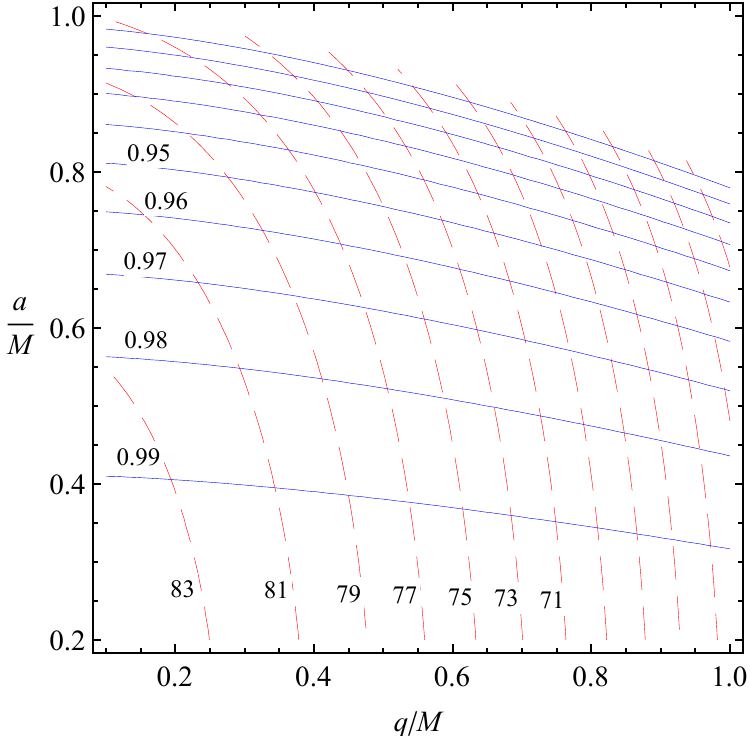}}~~~~~~
\subfigure{\includegraphics[width=0.45\textwidth]{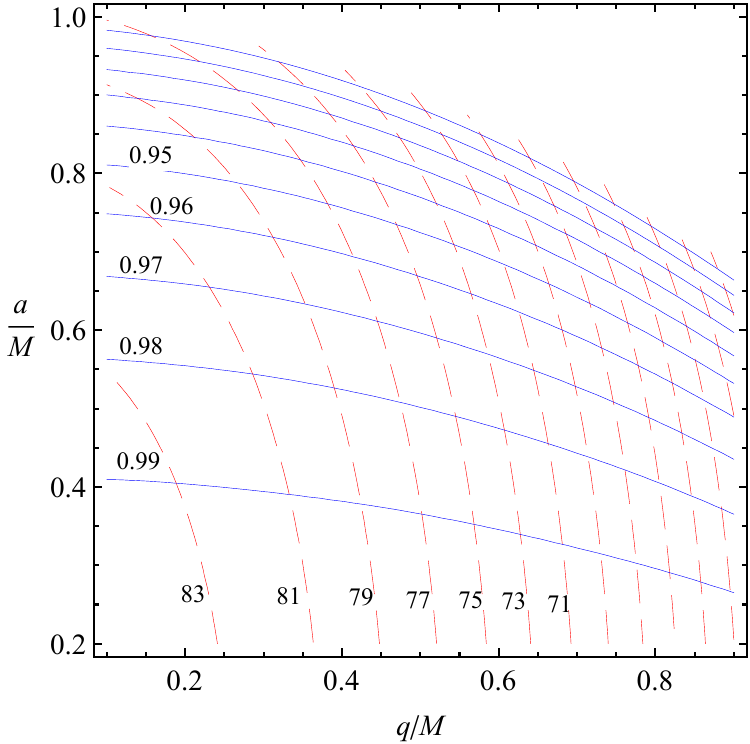}}
\caption{Contour plots of the observables $A/M^2$
and $D$ in the parameter space
($a/M$, $q/M$) of rotating charged BH in NED for the fixed value of $b=0.1 M^{-1}$ (left) and $b=0.5 M^{-1}$ (right).  The dashed red curves and the blue solid curves correspond to the contours of $A/M^2$ and $D$, respectively. \label{estimation}}
\end{figure*}

As we mentioned above, one can estimate two parameters of BH by shadow observables using Eqs.~(\ref{eq:Area})-(\ref{eq:oblat}). We present the cross-section contour plots of the area and oblateness in $a$-$q$ space for the fixed values of $b = 0.1$ and $b = 0.5$ with $M=1$ (Fig.~\ref{estimation}). It can be seen from these figures that the coordinates of the intersections uniquely determine the two BH parameters $a$ and $q$. One can determine the values of BH parameters from its shadow by applying this estimation method. These results can be a useful tool to get information about the BH based on its shadow parameters, which can be measured from observations.

\section{Constraining with EHT Observations}
Observational data of two supermassive BHs (Sgr A* and M87*) shadow images that are obtained by EHT collaboration can motivate us to study BH shadows with great scientific interest. By using this data, one may estimate the BH parameters in the framework of different modified or alternative gravity models to see if the results of the model really fit with observations. In our analysis, we take constraints for the parameters of the rotated charged BH in NED from the EHT results of M87* and Sgr A*.

We used the angular diameters of these two BHs measured by the EHT collaboration to obtain constraints. The angular diameter of the shadow image for an observer at distance $d$ from BH can be expressed as \cite{Afrin:2021imp,Zahid:2023csk}
\begin{eqnarray}
    \theta_d =2 \frac{R_a}{d}, \qquad R_a = \sqrt{\frac{A}{\pi}}
    \label{angdia}
\end{eqnarray}
where $R_a$ is the areal shadow radius. If we consider Eq.~(\ref{eq:Area}), the angular diameter of the shadow depends on the parameters of the BH and the observation angle. Also, it implicitly depends on the mass of the BH. We now consider supermassive BHs M87* (and Sgr A*) as rotating charged BHs in NED to compare our theoretical results of shadow analysis with the shadow images of M87* (and Sgr A*) from EHT data.

\begin{figure*}[t]
\begin{center}
\subfigure{\includegraphics[width=0.48\textwidth]{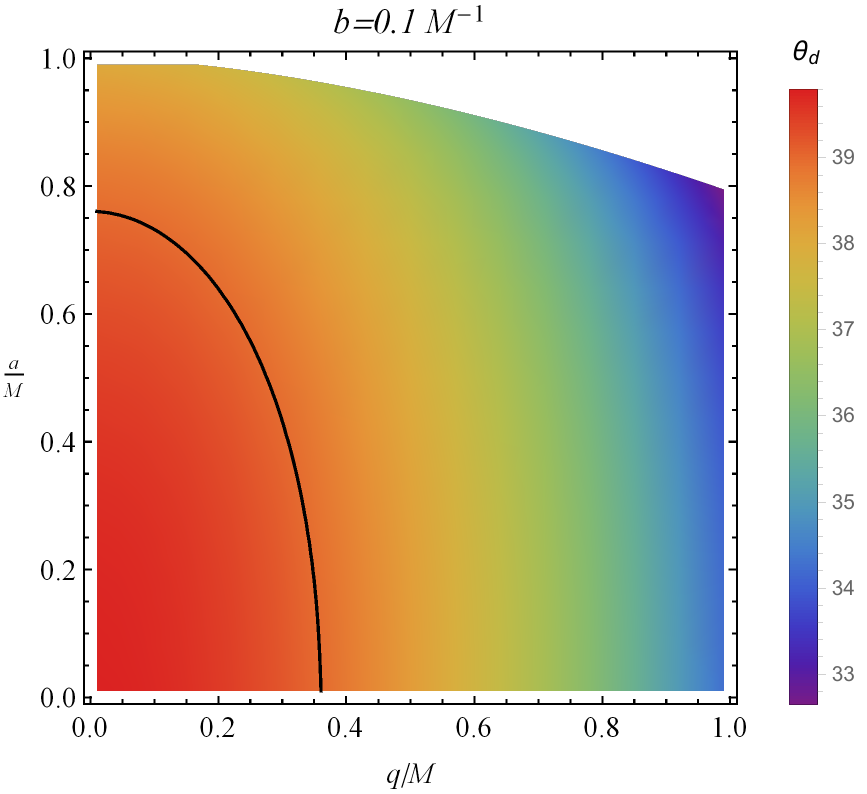}}~~~
\subfigure{\includegraphics[width=0.48\textwidth]{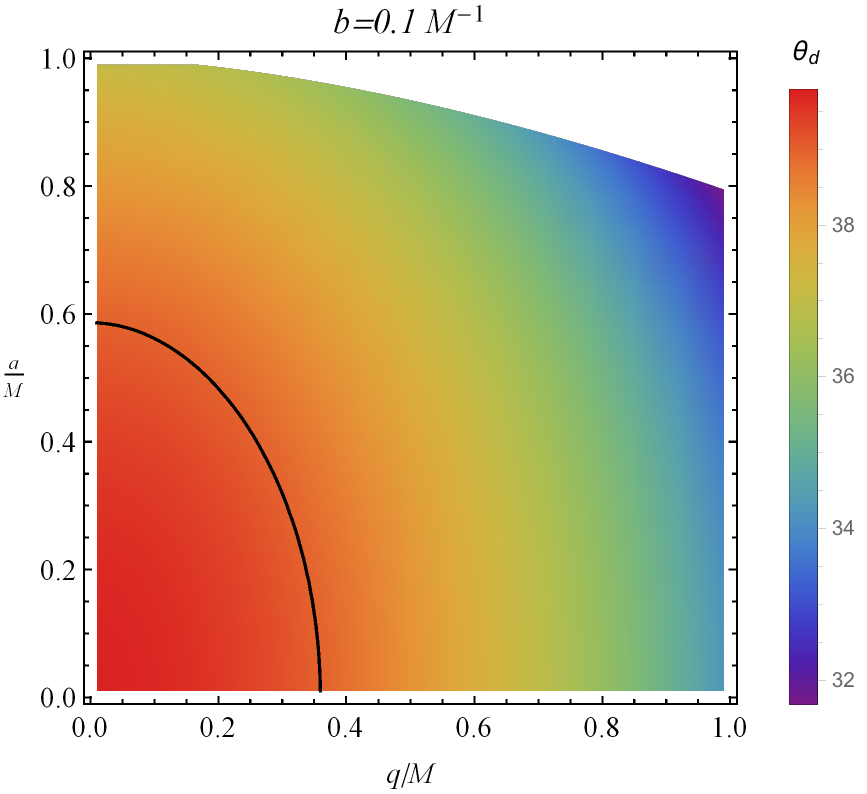}}
\subfigure{\includegraphics[width=0.48\textwidth]{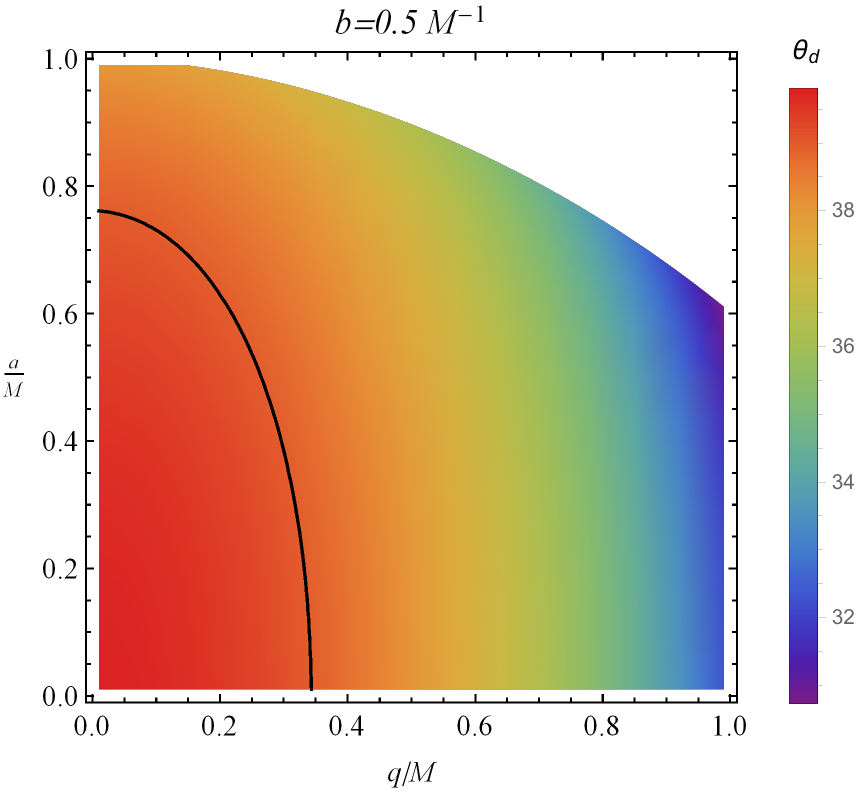}}~~~
\subfigure{\includegraphics[width=0.48\textwidth]{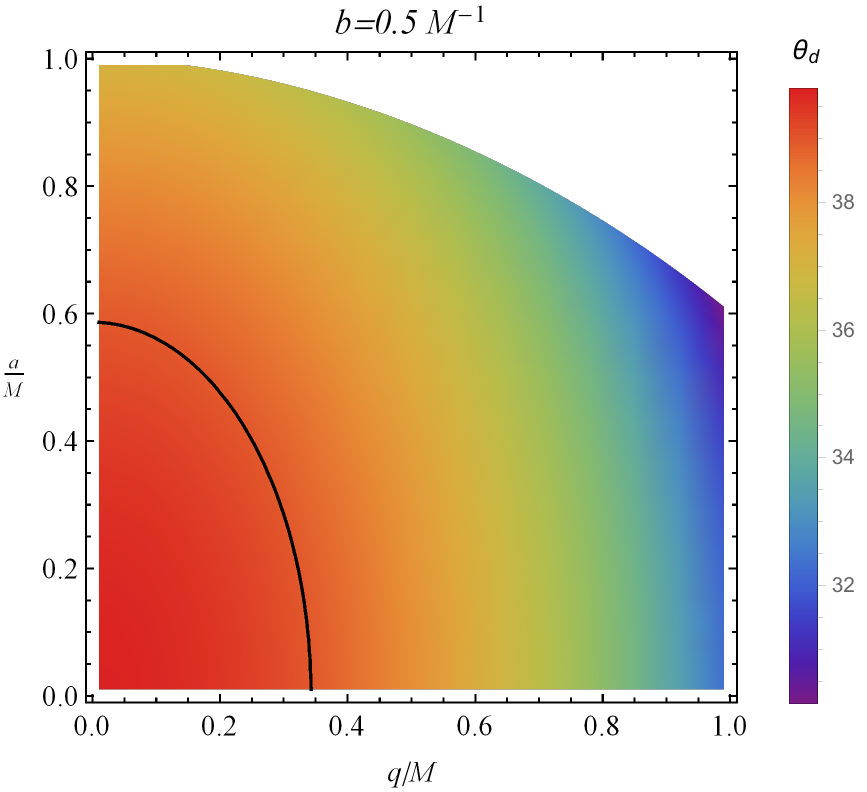}}
\end{center}
\caption{
Angular diameter observable $\theta_d$ for the BH shadows as a function of parameters $a/M$ and $q/M$ at inclinations $90^\circ$ (left panels) and $17^\circ$ (right panels). The black curves describe the borders of the image size $\theta_d=42\pm3\mu as$ measured angular diameter of the SMBH M87* reported by the EHT.
} \label{M87}
\end{figure*}

The mass of M87* and the distance from Earth can be considered as $M=6.5\times10^9M_\odot$ and $d=16.8Mpc$, respectively \cite{EventHorizonTelescope:2019pgp,EventHorizonTelescope:2019ggy}. For simplicity, we do not take into account the uncertainties of the mass and distance measurements of the SMBHs. The angular diameter of the image of the SMBH M87* is $\theta_d=42\pm3\mu as$ \cite{EventHorizonTelescope:2019dse} in the 1-$\sigma$ confidential level. In Fig. \ref{M87}, the constraints for M87* are presented, as are the density plots of the angular diameter $\theta_d$ in the $a$-$q$ space at inclination angles $90^\circ$ (left panel) and $17^\circ$ (right panel) for the fixed value's parameter $b$. Here, the black curves correspond to the lower borders of the measured angular diameter of the SMBH M87* shadow. From these plots, one can study the dependence of the angular diameter of the BH shadow on the parameters of charged BH in NED in particular cases. The maximum limit of charge parameter $q$ is more than $0.35 M$ in the constraint part for the first case, $b=0.1 M^{-1}$, while it is decreased to less than $0.35 M$ when $b=0.5 M^{-1}$. Furthermore, there is a noticeable decrease in the upper limit of spin parameter $a$ (around $0.6 M$) when we see the plots for inclination angle $17^\circ$ compared to the left panels. 

\begin{figure*}[ht!]
\begin{center}
\subfigure{\includegraphics[width=0.48\textwidth]{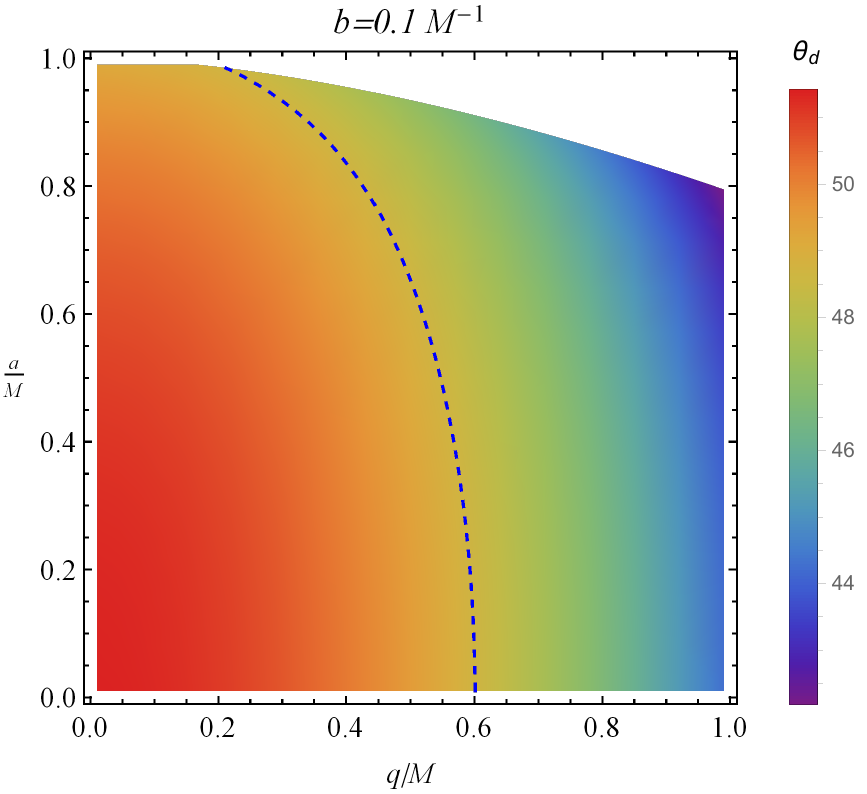}}~~~
\subfigure{\includegraphics[width=0.48\textwidth]{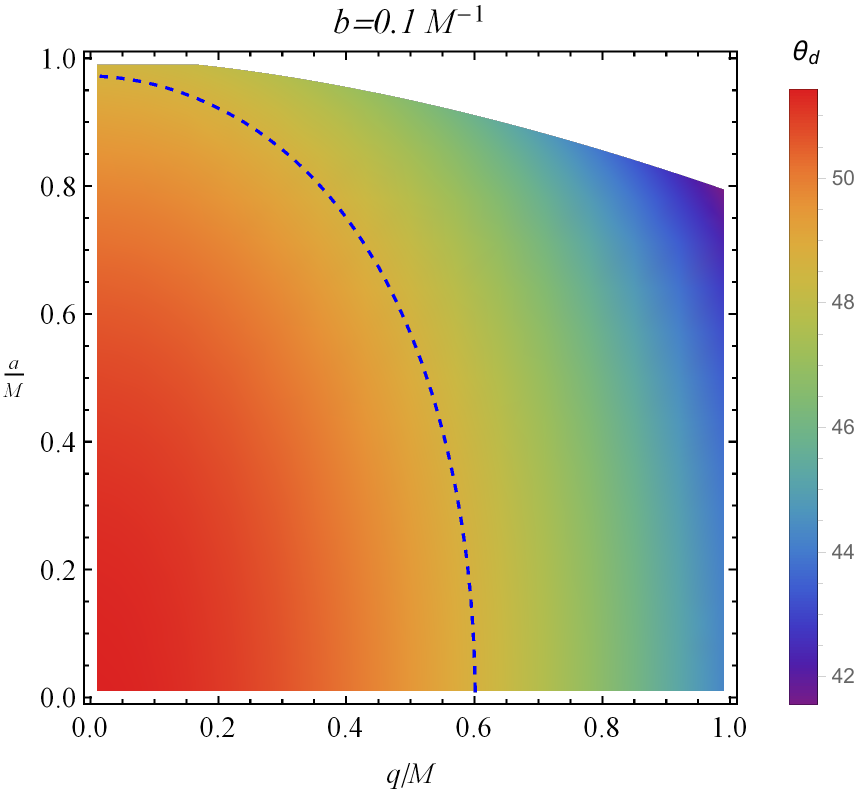}}
\subfigure{\includegraphics[width=0.48\textwidth]{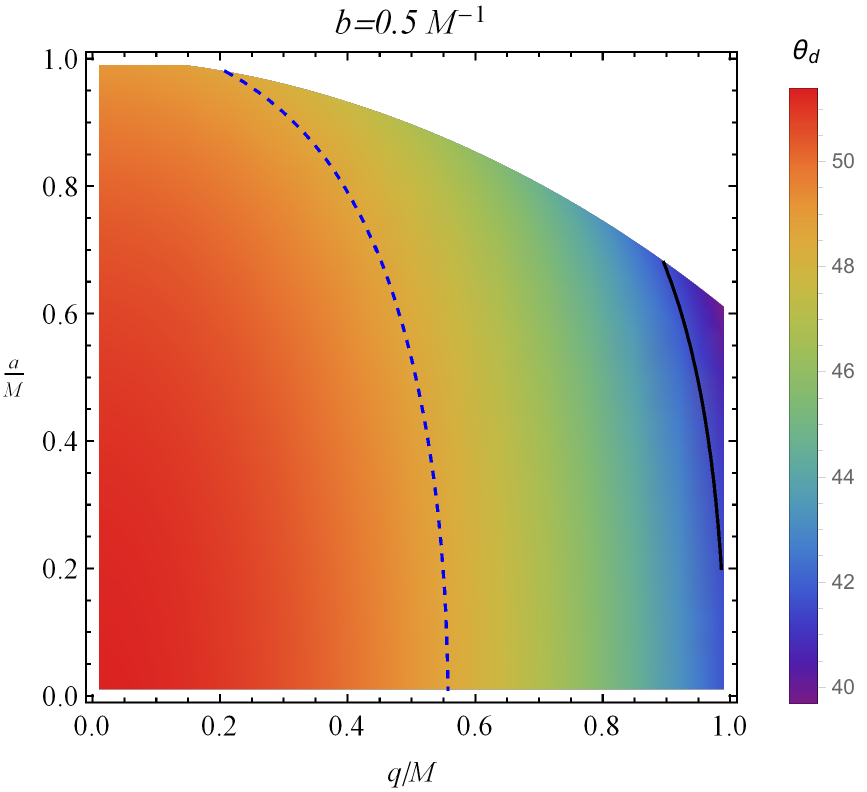}}~~~
\subfigure{\includegraphics[width=0.48\textwidth]{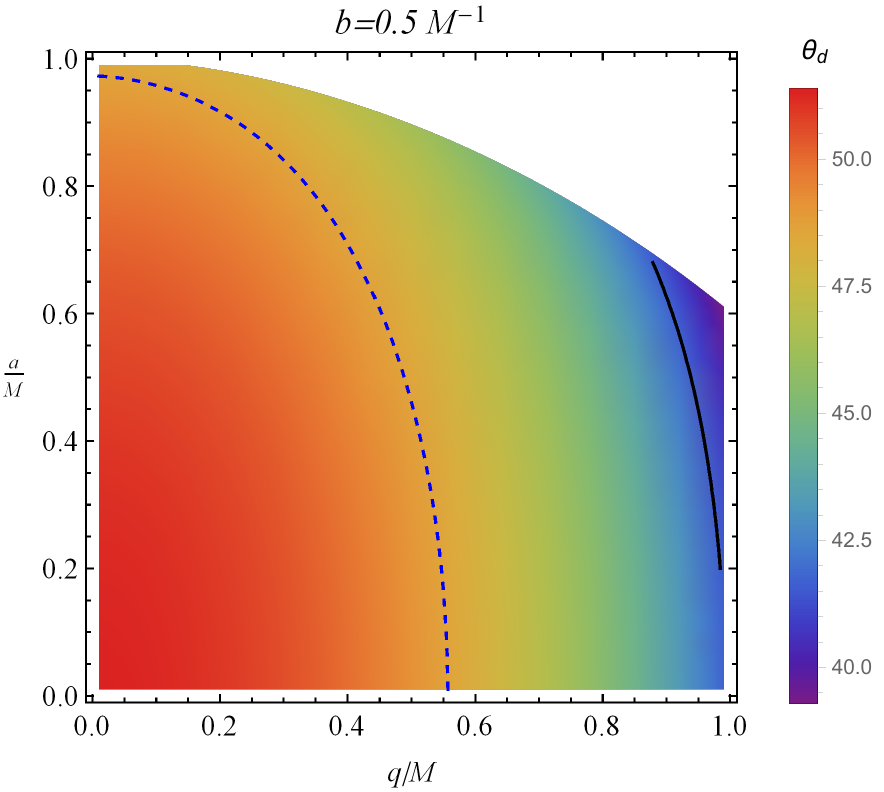}}
\end{center}
\caption{Angular diameter observable $\theta_d$ for the BH shadows as a function of parameters $a/M$ and $q/M$ at inclinations $90^\circ$ (left panels) and $50^\circ$ (right panels). The black curves correspond to $\theta_d=41.7\mu as$ within the measured angular diameter, $\theta_d=48.7\pm7\mu as$ of the Sgr A* BH reported by the EHT. The blue dashed curves correspond to $48.7\mu as$.} \label{Sgr}
\end{figure*}

Similarly, we can get constraints for the shadow image of Sgr A* from EHT observations. The angular diameter of the shadow of the SMBH Sgr A* is $\theta_d=48.7\pm7\mu as$ \cite{EventHorizonTelescope:2022wkp}. The mass of Sgr A* and the distance from the solar system be considered as $M\simeq4\times10^6M_\odot$ and $d\simeq 8$ kpc, respectively \cite{EventHorizonTelescope:2022exc,EventHorizonTelescope:2022xqj}. In Fig.~\ref{Sgr}, the constraints from Sgr A* are described for two cases we have considered. The fitted density plots of the angular diameter of BH lie inside the measured angular diameter of Sgr A*, $\theta_d = 48.7\pm7 \mu$as. Therefore, we can claim that our theoretical results of rotating charged BH in NED can be compared with the supermassive BH at the center of the Milky Way Galaxy. Only in the second case did we manage to show the upper limit of the measured angular diameter of the Sgr A* shadow with black curves, while blue dashed curves correspond to the mean value of the measured angular diameter. What can be observed from the figure is that the blue dashed curves for the limits of the observable parameters shift slightly to the left, and there is a decrease in the upper limit of the spin parameter $a$ (from $0.6 M$ to more than $0.55 M$) when we increase the fixed value of the parameter $b$. Moreover, the change in the inclination angle influences the distribution of angular diameter throughout the $a$-$q$ space.

\section{Energy Emission Rate}\label{emissionrate}
Classically, an object is lost forever if it enters a BH, whereas it is believed that a BH does emit quantum mechanically. Inside the BH horizon, the quantum fluctuations cause the creation and annihilation of particles. The particles with positive energy may escape out of the horizon due to a process known as tunneling. These escaping particles carry energy that helps the BH to evaporate. An absorption process is measured in terms of the probability of absorption cross-section. Far from the gravitational effect of BH, the absorption cross-section is related to the shadow of BH. In the high-energy regime, this absorption cross-section is obtained by a constant value denoted by $\sigma_{lim}$. The value of $\sigma_{lim}$ is approximately equal to the area of BH shadow as \cite{2013JCAP...11..063W,PhysRevD.7.2807,PhysRevD.83.044032}
\begin{equation}
\sigma_{lim}\approx\pi R_{sh}^2.
\end{equation}
The energy emission rate can be expressed as
\begin{equation}
\mathcal{E}_{\omega t}:=\frac{d^2\mathcal{E}(\omega)}{d\omega dt} =\frac{2\pi^2\sigma_{lim}\omega^3}{e^{\omega/T_H}-1}\approx\frac{2\pi^3R_{sh}^2\omega^3}{e^{\omega/T_H}-1},
\end{equation}
where $\omega$ denotes the angular frequency, $T_H=\kappa/2\pi$ is the Hawking temperature and
\begin{equation}
\kappa=\frac{\Delta'(r)}{2(a^2+r^2)}\bigg|_{r=r_h}
\end{equation}
is the surface gravity at the event horizon for rotating BH. Surface gravity for the static case becomes
\begin{equation}
\kappa=\frac{1}{2}f'(r)\bigg|_{r=r_h}.
\end{equation}

\begin{figure*}[ht!]
\begin{center}
\subfigure{\includegraphics[width=0.315\textwidth]{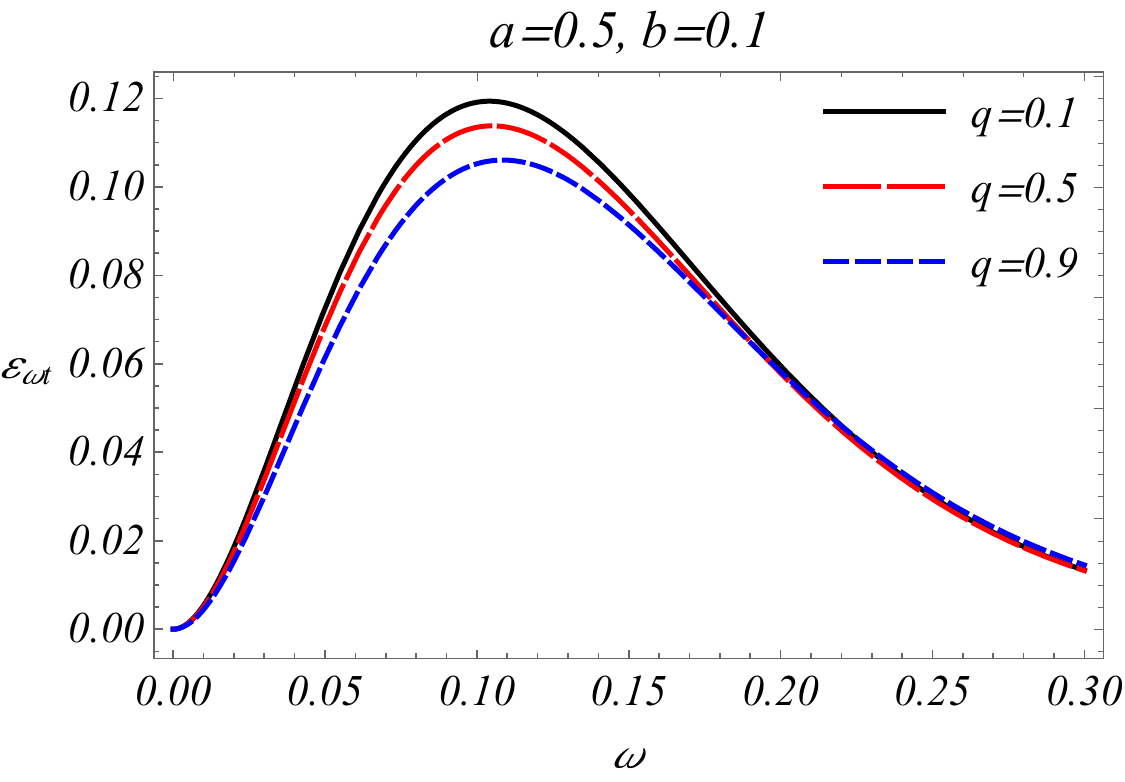}}~~
\subfigure{\includegraphics[width=0.315\textwidth]{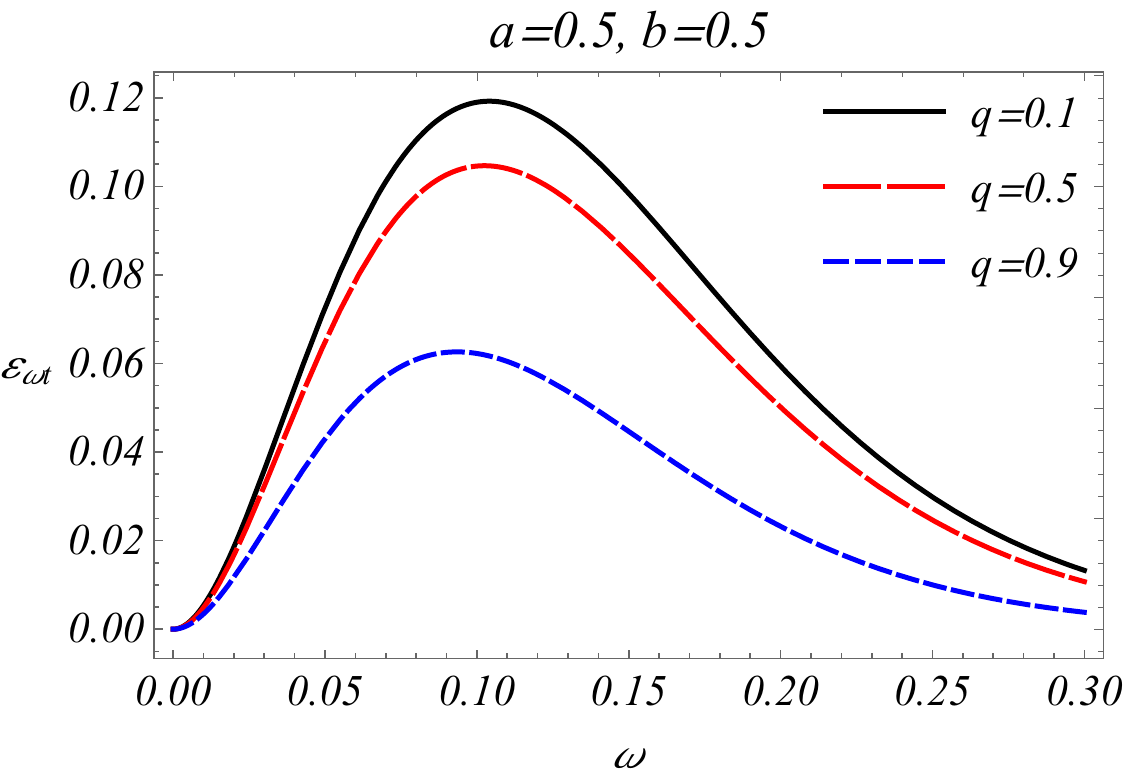}}~~
\subfigure{\includegraphics[width=0.315\textwidth]{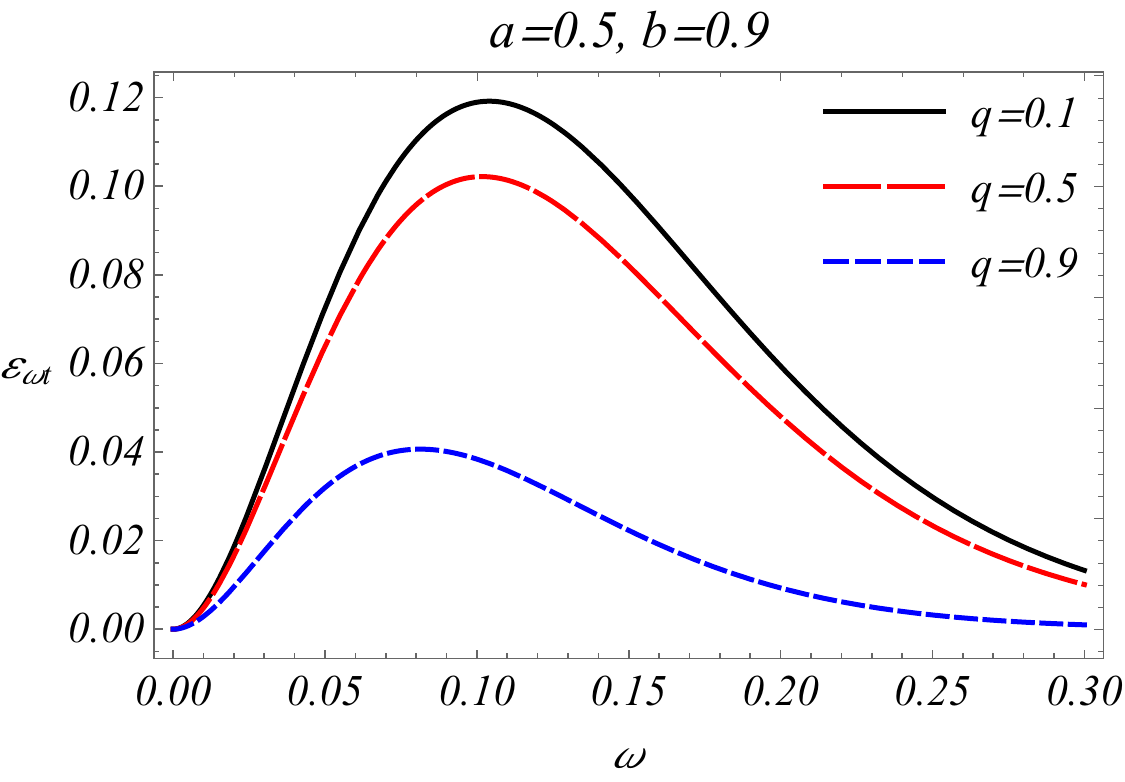}}\\
\subfigure{\includegraphics[width=0.315\textwidth]{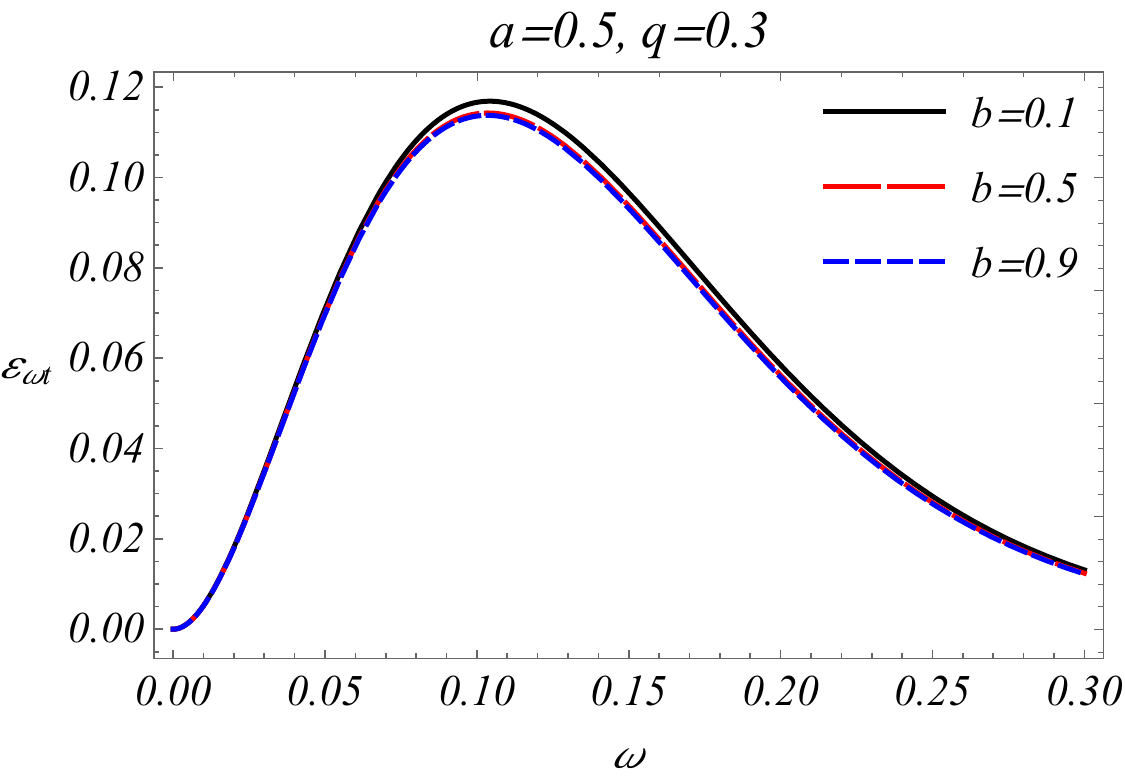}}~~
\subfigure{\includegraphics[width=0.315\textwidth]{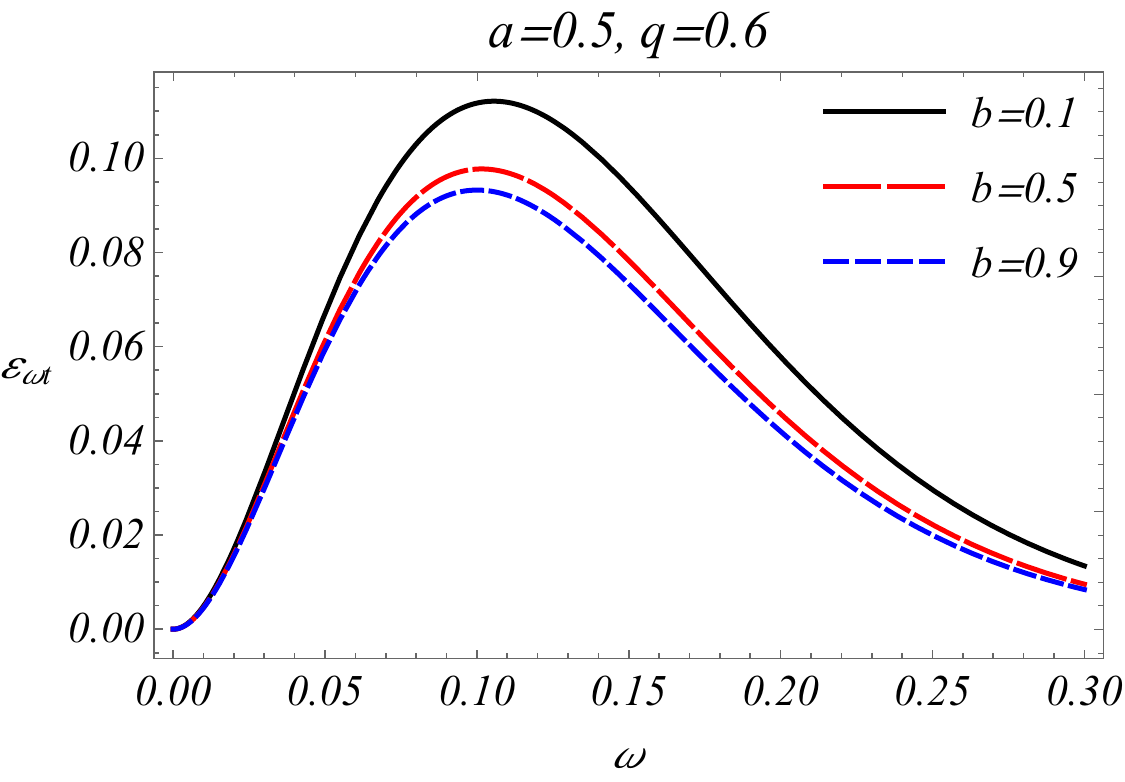}}~~
\subfigure{\includegraphics[width=0.315\textwidth]{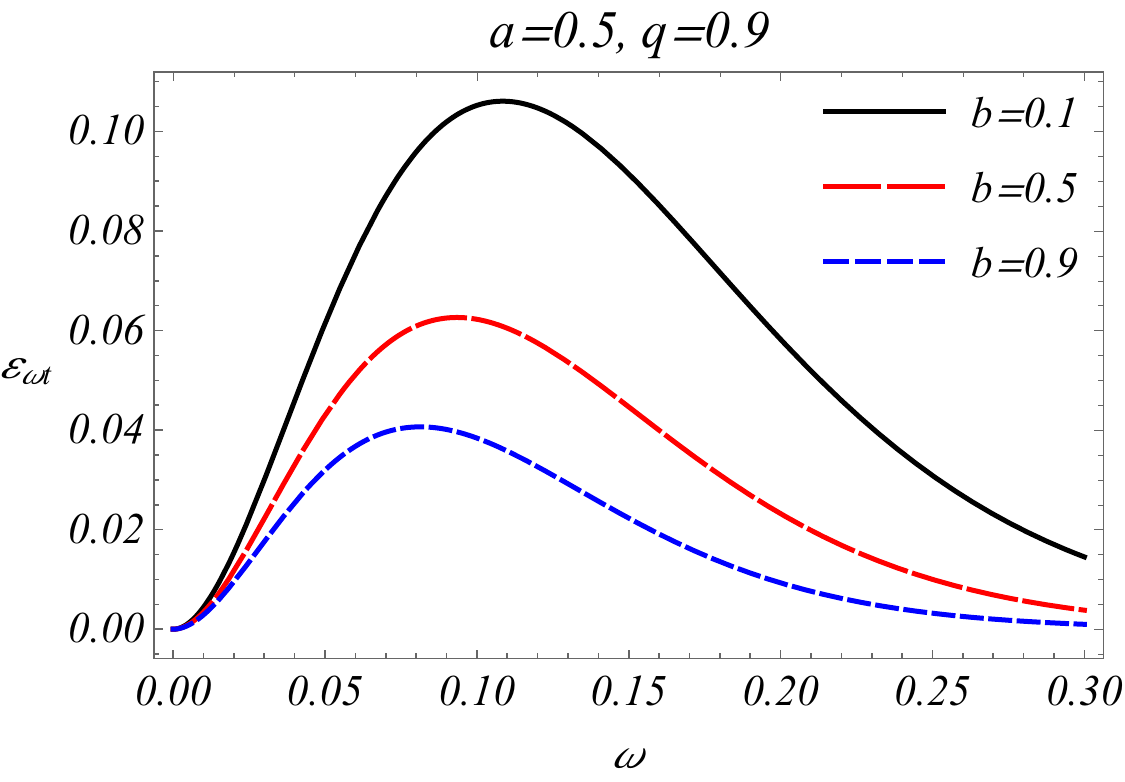}}\\
\subfigure{\includegraphics[width=0.315\textwidth]{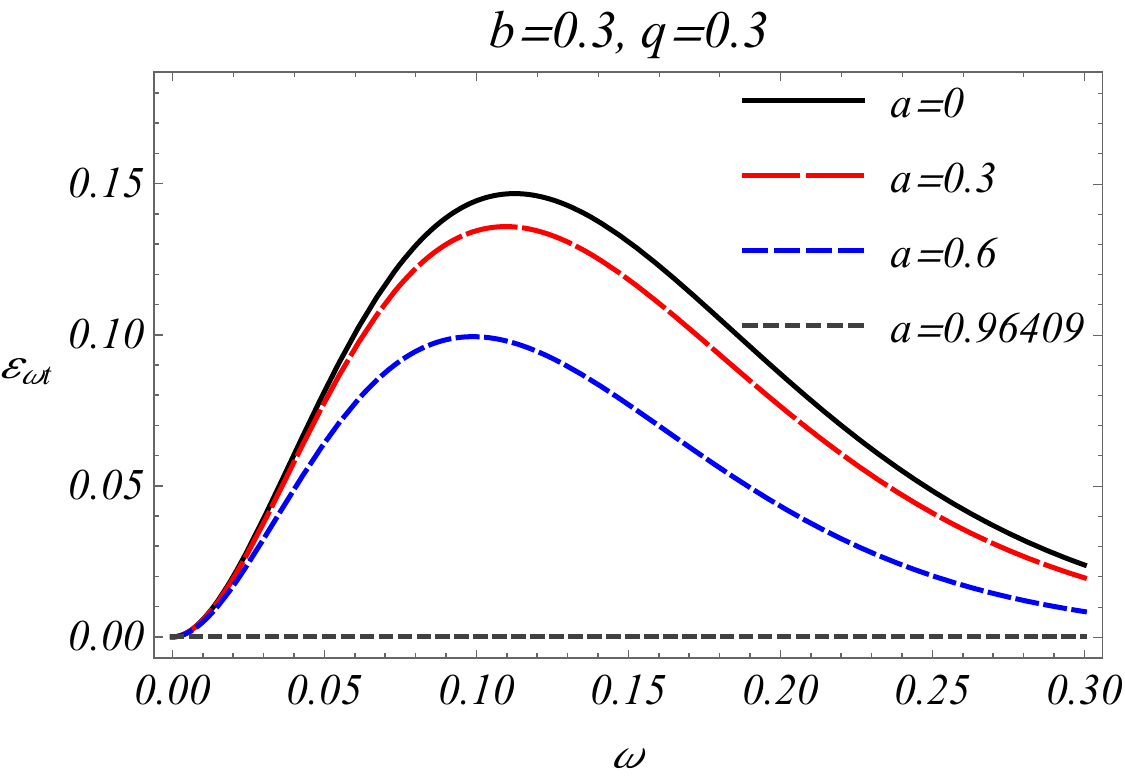}}~~
\subfigure{\includegraphics[width=0.315\textwidth]{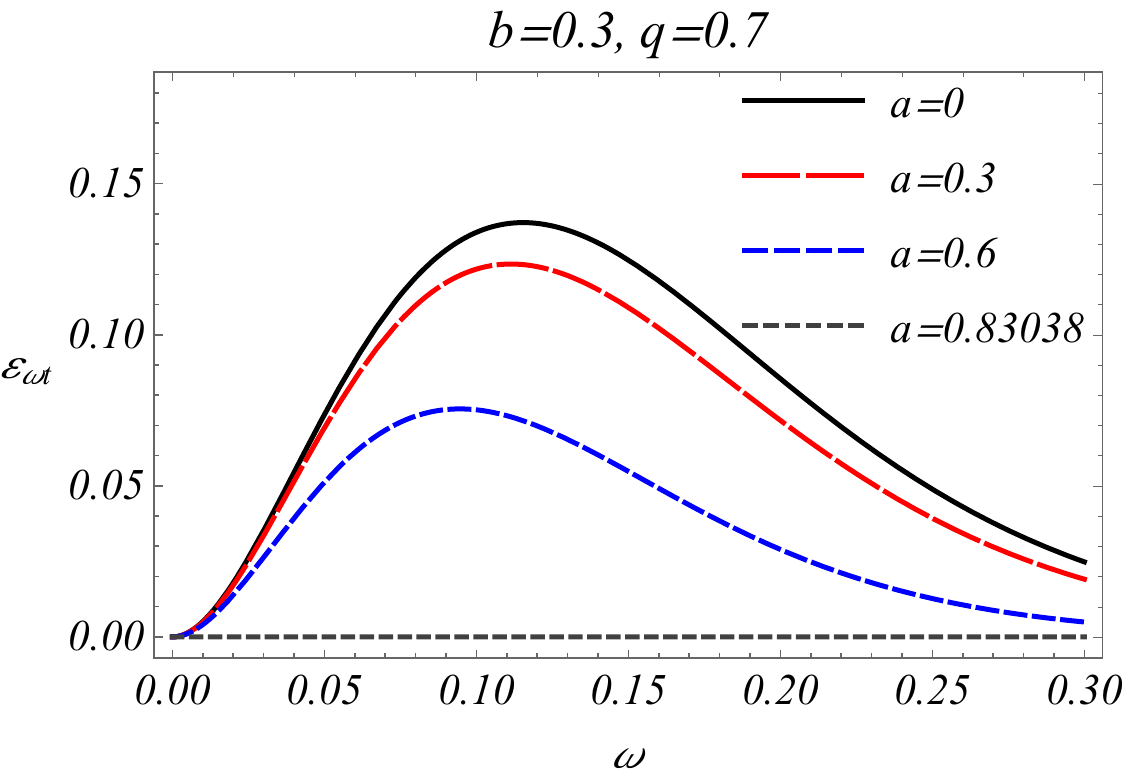}}~~
\subfigure{\includegraphics[width=0.315\textwidth]{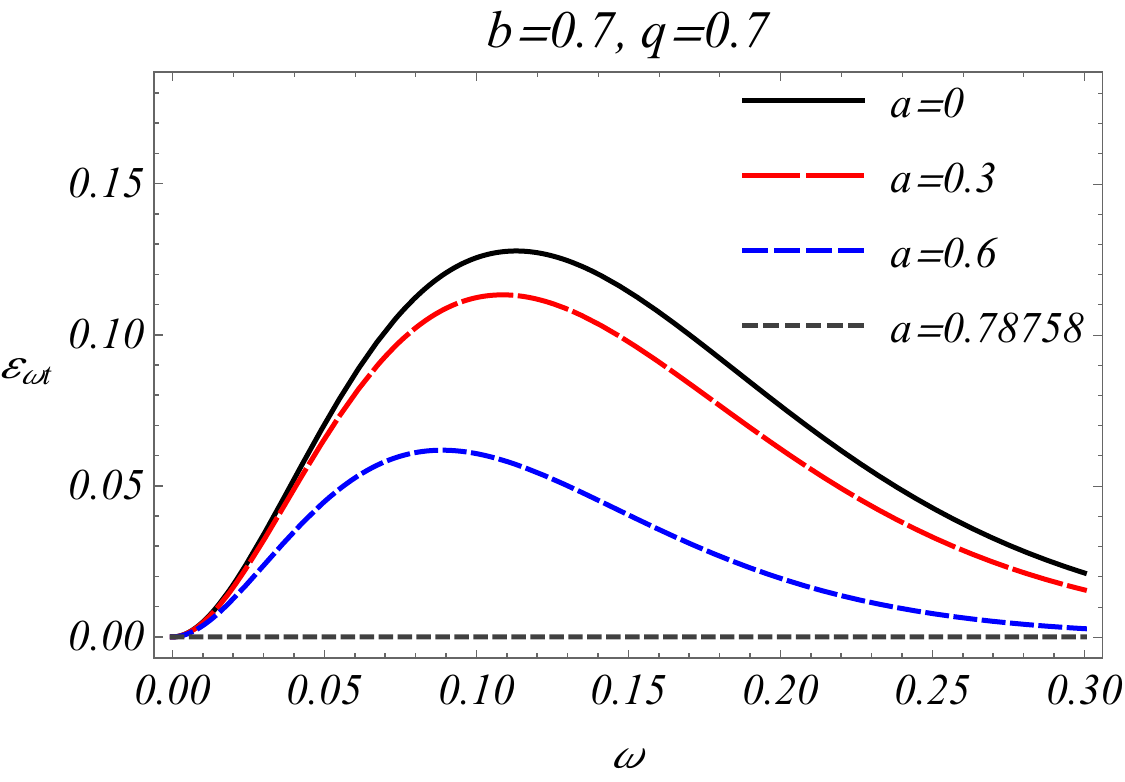}}
\end{center}
\caption{Plots showing the behavior of BH evaporation rate for different values of $b$, $q$ and $a$ corresponding to the Fig. \ref{Sh}.} \label{EER}
\end{figure*}
The energy emission rate $\mathcal{E}_{\omega t}$ is plotted versus the frequency $\omega$ in Fig. \ref{EER} for various values of the BH parameters corresponding to the shadows in Fig. \ref{Sh}. The top panel shows that the BH evaporation rate decreases by increasing the value of $q$. Furthermore, from left to right in the panel, as the value of $b$ increases, the variation in the BH evaporation rate increases. In the middle panel, the variation in the BH evaporation rate is much lower for small values of $q$, and with an increase in the value of $q$, the variation in the BH evaporation rate becomes prominent. Therefore, we can see that the BH evaporation rate decreases with increasing $b$. In the lower panel, with respect to $a$, a slow rate of evaporation of BH is observed. Furthermore, for extreme values of $a$, the BH does not evaporate. It is because the Hawking temperature is zero Kelvin for extremal BHs, and therefore, they do not radiate.

\section{Conclusion}\label{conclusion}

We focused on the non-linear effect of electrodynamics and the electric charge together with BH spin on various BH properties. The rotating BH metric is obtained by incorporating modified NJA. The real and imaginary parts of quasinormal modes have been studied related to the radius of the photon sphere for the static BH. The horizon radius is investigated for the rotating BH in terms of spin $a$. To study the shadows, we incorporated the HJ formalism, and by using Bardeen's method for an observer located at infinity, we obtained the effective potential, shadows, and distortion. We also estimated parameters using the shadow area and oblateness. The shadows are compared with the EHT observations for SMBHs Sgr A* and M87*, and the constraints on BH parameters are obtained. Lastly, the rate of energy emissions is discussed. The results presented in previous sections are summarized below:
\begin{itemize}
\item The angular velocity associated with quasinormal modes increases with an increase in $q$. For a small value of $q$, the angular velocity remains constant with respect to $b$. When the value of $q$ is increased, the angular velocity increases rapidly and then gradually becomes constant with respect to $b$. A very similar behavior of the Lyapunov exponent has been observed.
\item The event horizon of the BH decreases with an increase in spin up to its extremal value. Moreover, with respect to $b$ and $q$, the event horizon also decreases. However, the decrease with respect to $b$ is prominent for larger values of $q$.
\item The effective potential determines the unstable circular null orbits. The unstable circular null orbits are found to shrink with increasing $b$, $q$, and $a$.
\item The shadows also follow the behavior of unstable circular null orbits. That is, the shadow size decreases with respect to increases in $b$ and $q$. However, for a small value of $q$, the shadow size is almost constant with respect to $b$. The shadows shift towards the right as we increase $a$ and the maximum flatness is seen for extremal $a$.
\item The flatness measure shows that the distortion increases with respect to $q$ and $a$ with an accelerated rate and increases with an increase in $b$ with a decelerated rate.

\item The study of shadow observables in the NED for rotating charged BH parameters reveals that as the BH parameters increase, both the shadow area ($A$) and oblateness ($D$) values decrease.  With increasing values of $q$, the dependence of shadow observables on $q$ becomes more considerable. The research can serve as a useful tool for identifying BH parameters from their shadow parameters since it also illustrates in plots how the coordinates of crossings uniquely determine the two BH parameters $a$ and $q$.

\item The constraints for M87* and Sgr A* are presented, and density plots are used to study the dependence of the angular diameter on charged BH parameters in NED for specific cases.
Generally, the distribution of angular diameter over the $a$-$q$ space depends on the change in inclination angle.
For the first instance of M87* restrictions, the highest limit of the charge parameter $q$ is greater than $0.35 M$, but it drops to less than $0.35 M$ for the case of $b=0.5 M^{-1}$. The maximum limit of the spin parameter $a$ in the graphs for Sgr A* decreases as the fixed value of the parameter $b$ increases.

\item The BH evaporation rate decreases with increasing values of $q$, $b$, and $a$. It also shows that the extremal BH does not emit and thus has a zero-emission rate. Furthermore, the effect of the electric charge of the BH on the emission rate becomes stronger at higher values of $b$. Similarly, the increase of $q$ enhances the effect of $b$ on the rate.  
\end{itemize}

\section*{Acknowledgements}

J.R. acknowledges the Grants F-FA-2021-510 of the Agency of Innovative Development of Uzbekistan. J.R. also thanks the SU in Opava for its hospitality. Z.S. acknowledges the Research Centre for Theoretical Physics and Astrophysics, Institute of Physics, SU in Opava, and the GA{\v C}R 23-07043S
project.

\def\prc{Phys. Rev. C}
	\def\pre{Phys. Rev. E}
	\def\prd{Phys. Rev. D}
	\def\prl{Physical Review Letters}
	\def\jcap{Journal of Cosmology and Astroparticle Physics}
	\def\apss{Astrophysics and Space Science}
	\def\mnras{Monthly Notices of the Royal Astronomical Society}
	\def\apj{The Astrophysical Journal}
	\def\aap{Astronomy and Astrophysics}
	\def\actaa{Acta Astronomica}
	\def\pasj{Publications of the Astronomical Society of Japan}
	\def\apjl{Astrophysical Journal Letters}
	\def\pasa{Publications Astronomical Society of Australia}
	\def\nat{Nature}
	\def\physrep{Physics Reports}
	\def\araa{Annual Review of Astronomy and Astrophysics}
	\def\apjs{The Astrophysical Journal Supplement}
	\def\aapr{The Astronomy and Astrophysics Review}
	\def\procspie{Proceedings of the SPIE}

\bibliographystyle{spphys}
\bibliography{reference,gravreferences}

\end{document}